\documentclass[12pt]{article}
\usepackage{epsfig,epsf}

\newlength{\dinwidth}
\newlength{\dinmargin}
\def\lapproxeq{\lower .7ex\hbox{$\;\stackrel{\textstyle <}{\sim}\;$}}
\def\gapproxeq{\lower .7ex\hbox{$\;\stackrel{\textstyle >}{\sim}\;$}}

\def\be{\begin{equation}}
\def\ee{\end{equation}}
\def\bea{\begin{eqnarray}}
\def\eea{\end{eqnarray}}

\def\gtrsim{ \;\raisebox{-.7ex}{$\stackrel{\textstyle >}{\sim}$}\; }
\def\lesim{ \;\raisebox{-.7ex}{$\stackrel{\textstyle <}{\sim}$}\; }
\newcommand{\lsim}{ \mathop{}_{\textstyle \sim}^{\textstyle <} }

 \newcommand{\qsq}{q^2}

\def\GeV{{\rm GeV}}
\def\MeV{{\rm MeV}}
\def\eV{{\rm eV}}

\begin{document}
\titlepage

\begin{flushright}
hep-ph/0312250\\
KEK--TH--902\\
IPPP/03/41 \\
DCPT/03/82\\
CERN--TH/2003-162\\
LTH 613\\
17 Dec 2003, updated 8 April 2004\\
\end{flushright}

\vspace*{4cm}

\begin{center}
{\Large \bf Predictions for $g-2$ of the muon and 
            $\alpha_{\rm QED}(M_Z^2)$}

\vspace*{1cm} {\sc K. Hagiwara}$^a$, {\sc A.D. Martin}$^b$, 
{\sc Daisuke Nomura}$^b$,
and {\sc T. Teubner}$^{c,d}$ \\

\vspace*{0.5cm}
$^a$ {\em Theory Group, KEK, Tsukuba, Ibaraki 305-0801, Japan} \\
$^b$ {\em Department of Physics and Institute for
Particle Physics Phenomenology,\\
University of Durham, Durham DH1 3LE, U.K.}\\
$^c$ {\em Theory Division, CERN, CH-1211 Geneva 23, Switzerland}\\
$^d$ {\em Present address: Department of Mathematical Sciences,\\
 University of Liverpool, Liverpool L69 3BX, U.K.}
\end{center}

\vspace*{1cm}

\begin{abstract}
We calculate $(g-2)$ of the muon and the QED coupling $\alpha(M_Z^2)$, 
by improving the determination of the hadronic vacuum polarization 
contributions and their uncertainties.  We include the recently 
re-analysed CMD-2 data on $e^+e^-\to \pi^+\pi^-$. We carefully 
combine a wide variety of data for the $e^+e^-$ production of 
hadrons, and obtain the optimum form of 
$R(s)\equiv \sigma_{\rm had}^0(s)/\sigma_{\rm pt}(s)$, together 
with its uncertainty.  Our results for the hadronic contributions 
to $g-2$ of the muon are
$a_\mu^{\rm had, LO}=
     (692.4 \pm 5.9_{\rm exp} \pm 2.4_{\rm rad}) \times 10^{-10}$ 
and
$a_\mu^{\rm had, NLO}= 
     (-9.8 \pm 0.1_{\rm exp} \pm 0.0_{\rm rad}) \times 10^{-10}$,
and for the QED coupling
$\Delta \alpha^{(5)}_{\rm had} (M_Z^2)=  
     (275.5 \pm 1.9_{\rm exp} \pm 1.3_{\rm rad}) \times 10^{-4}$. 
These yield $(g-2)/2 = 0.00116591763(74)$, which is about 
$2.4\sigma$ below the present world average measurement,
and $\alpha(M_Z^2)^{-1}= 128.954 \pm 0.031$.
We compare our ($g-2$) value with other predictions and, 
in particular, make a detailed comparison with the latest 
determination of $(g-2)$ by Davier et~al.
\end{abstract}

\newpage

\tableofcontents

\vspace{1cm}

\section{Introduction}

Hadronic vacuum polarization effects play a key role in the prediction 
of many physical quantities.  Here we are concerned with their effect 
on the prediction of the anomalous magnetic moment of the muon,
$a_\mu\equiv(g_\mu-2)/2$, and on the running of the QED coupling 
to the $Z$ boson mass.  We explain below why it is crucial to predict
these two quantities as precisely as possible in order to test the 
Standard Model and to probe New Physics.

First, we recall that the anomalous magnetic moments of the electron 
and muon are two of the most accurately measured quantities in particle 
physics. Indeed the anomalous moment of the electron has been measured 
to a few parts per billion and is found to be completely described by 
quantum electrodynamics. This is the most precisely tested agreement 
between experiment and quantum field theory.  On the other hand, since 
the muon is some 200 times heavier than the electron, its moment is 
sensitive to small-distance strong and weak interaction effects, and
therefore depends on all aspects of the Standard Model.  The world 
average of the existing measurements of the anomalous magnetic moment 
of the muon is
\begin{eqnarray}
 a_\mu^{\rm exp} = 11 659 203(8) \times 10^{-10},
 \label{eq:BNL2001}
\end{eqnarray}
which is dominated by the recent value obtained by the Muon $g-2$ 
collaboration at Brookhaven National Laboratory~\cite{BNL2002}. 
Again, the extremely accurate measurement offers a stringent test of 
theory, but this time of the whole Standard Model.  If a statistically 
significant deviation, no matter how tiny, can be definitively 
established between the measured value $a_\mu^{\rm exp}$ and the 
Standard Model prediction, then it will herald the existence of new 
physics beyond the Standard Model. In particular the comparison offers 
valuable constraints on possible contributions from SUSY particles.

The other quantity, the QED coupling at the $Z$ boson mass, $M_Z$, is 
equally important. It is the least well known of the three parameters 
(the Fermi constant $G_\mu$, $M_Z$ and $\alpha(M_Z^2)$), which are 
usually taken to define the electroweak part of the Standard Model. 
Its uncertainty is therefore one of the major limiting factors for
precision electroweak physics. It limits, for example, the accuracy 
of the indirect estimate of the Higgs mass in the Standard Model.

The hadronic contributions to $g-2$ of the muon and to the running 
of $\alpha(s)$ can be calculated from perturbative QCD (pQCD) only 
for energies well above the heavy flavour thresholds\footnote{In some 
previous analyses pQCD has been used in certain regions between the 
flavour thresholds. With the recent data, we find that the pQCD and
data driven numbers are in agreement and not much more can be gained 
by using pQCD in a wider range.}. To calculate the important 
non-perturbative contributions from the low energy hadronic vacuum 
polarization insertions in the photon propagator we use the measured 
total cross section\footnote{
Strictly speaking we are dealing with a fully inclusive cross
section which includes final state radiation,
$e^+e^-\to{\rm hadrons} (+ \gamma)$.}
\begin{equation} 
  \sigma_{\rm had}^0(s)  \equiv  
  \sigma_{\rm tot}^0(e^+e^-\to\gamma^*\to{\rm hadrons}), 
  \label{eq:sigma_had}
\end{equation}
where the 0 superscript is to indicate that we take the bare cross 
section with no initial state radiative or vacuum polarization 
corrections, but with final state radiative corrections.  Alternatively 
we may use
\begin{equation} 
R(s) = \frac{\sigma_{\rm had}^0(s)}{\sigma_{\rm pt}(s)}\,,
\label{eq:R(s)} 
\end{equation}
where $\sigma_{\rm pt} \equiv 4\pi\alpha^2/3s$ with $\alpha =
\alpha(0)$.  Analyticity and the optical theorem then
yield the dispersion relations
\be 
a_\mu^{\rm had,LO} 
 = \left(\frac{\alpha m_\mu}{3\pi}\right)^2
   \int_{s_{\rm th}}^\infty {\rm d}s\,\frac{R(s)K(s)}{s^2} \,,
\label{eq:disprel1} 
\ee
\be 
\Delta\alpha_{\rm had}(s) = -\frac{\alpha s}{3\pi} {\rm P} 
   \int_{s_{\rm th}}^\infty {\rm d}s'\, \frac{R(s')}{s'(s'-s)}\, ,
\label{eq:disprel2} 
\ee
for the hadronic contributions to $a_\mu\equiv(g_\mu-2)/2$ and 
$\Delta\alpha(s)=1-\alpha/\alpha(s)$, respectively. The superscript
LO on $a_\mu$ denotes the leading-order hadronic contribution. There 
are also sizeable 
next-to-leading order (NLO) vacuum polarization and so-called
``light-by-light'' hadronic contributions to $a_\mu$, which we will 
introduce later. The kernel $K(s)$
in~(\ref{eq:disprel1}) is a known function (see (\ref{eq:K})), which 
increases monotonically from 0.40 at
$m_{\pi^0}^2$ (the $\pi^0\gamma$ threshold) to 0.63 at $s=4m_\pi^2$ (the
$\pi^+\pi^-$ threshold), and then to 1 as $s \rightarrow \infty$. 
As compared to (\ref{eq:disprel2}) evaluated at $s=M_Z^2$, we see that 
the integral in (\ref{eq:disprel1}) is much more dominated by 
contributions from the low energy domain.

At present, the accuracy to which these hadronic corrections can be 
calculated is the limiting factor in the
precision to which $g-2$ of the muon and $\alpha(M_Z^2)$ can be 
calculated. The hadronic corrections in turn rely
on the accuracy to which $R(s)$ can be determined from the 
experimental data, particularly in the low energy
domain. For a precision analysis, the reliance on the experimental 
values of $R(s)$ or $\sigma_{\rm had}^0(s)$ poses several problems:
\begin{itemize}
\item First, we must study how the data have been corrected for
radiative effects.  For example, to express $R(s)$ in (\ref{eq:disprel1}) 
and (\ref{eq:disprel2}) in terms of the observed hadron production 
cross section, $\sigma_{\rm had}(s)$, we have
\be 
 R(s) \equiv \frac{\sigma^0_{\rm had}(s)}{4\pi\alpha^2/3s} \simeq
  \left(\frac{\alpha}{\alpha(s)}\right)^2
  \frac{\sigma_{\rm had}(s)}{4\pi\alpha^2/3s}\,, 
  \label{eq:R(s)lowestorder}
\ee
if the data have not been corrected for vacuum polarization effects.  
The radiative correction factors, such as $(\alpha/\alpha(s))^2$ 
in (\ref{eq:R(s)lowestorder}), depend on each experiment, and we 
discuss them in detail in Section~\ref{sec:clustering}.

\item Second, below about $\sqrt{s}\sim1.5\ \GeV$, inclusive measurements
of $\sigma_{\rm had}^0(s)$ are not available, and instead a sum of the 
measurements of exclusive processes
($e^+e^-\to\pi^+\pi^-$, $\pi^+\pi^-\pi^0$, $K^+K^-,\dots$) is used.

\item To obtain the most reliable `experimental' values for $R(s)$
or $\sigma_{\rm had}^0(s)$ we have to combine carefully, in a consistent 
way, data from a variety of experiments
of differing precision and covering different energy intervals. 
In Section~\ref{sec:clustering} we show how this
is accomplished using a clustering method which minimizes a non-linear 
$\chi^2$ function.

\item In the region $1.5\lesim\sqrt s \lesim 2\ \GeV$ where both inclusive
and exclusive experimental determinations of $\sigma_{\rm had}^0(s)$ 
have been made, there appears to be some difference in the values. 
In Section~\ref{sec:QCDsumrules} we introduce QCD sum rules explicitly 
designed to resolve this discrepancy.

\item Finally, we have to decide whether to use the indirect information
on $e^+e^-\to{\rm hadrons}$ obtained for $\sqrt s < m_\tau$, via the 
Conserved-Vector-Current (CVC) hypothesis,
from precision data for the hadronic decays of $\tau$ leptons. 
However, recent experiments at Novosibirsk have
significantly improved the accuracy of the measurements of the 
$e^+e^-\to{\rm hadronic}$ channels, and reveal a sizeable discrepancy 
with the CVC prediction from the $\tau$ data; see
the careful study of~\cite{DEHZ}. Even with the re-analysed
CMD-2 data the discrepancy still remains~\cite{DEHZ03}.
This suggests that the understanding of the CVC hypothesis may be 
inadequate at the desired level of precision. It is
also possible that the discrepancy is coming from the $e^+e^-$ or 
$\tau$ spectral function data itself, e.g.~from
some not yet understood systematic effect.\footnote{The energy 
dependence of the discrepancy between $e^+e^-$ and
$\tau$ data is displayed in Fig.~2 of \cite{DEHZ03}.
One possible origin would be an unexpectedly large mass difference
between charged and neutral $\rho$ mesons, see, for example,
\cite{Ghozzi:2003yn}.}

The experimental discrepancy may be clarified by measurements of the 
radiative return\footnote{See \cite{CGKR} for
a theoretical discussion of the application of  ``radiative return'' 
to measure the cross sections for $e^+e^-\to
\pi\pi, K\bar K,\dots$ at $\phi$ and $B$ factories~\cite{Kloe,BABAR}. } 
events, that is $e^+e^- \to \pi^+\pi^-\gamma$, at
DA$\Phi$NE~\cite{Kloe} and BaBar~\cite{Davier_at_Pisa}. 
Indeed the preliminary measurements of the 
pion form factor by the KLOE Collaboration
\cite{DENIG} compare well with the recent precise CMD-2 $\pi^+\pi^-$ 
data~\cite{CMD2new,Akhmetshin:2001ig} in the energy region above 0.7
GeV, and are significantly below the values obtained, 
via CVC, from $\tau$ decays~\cite{DEHZ}.  We therefore do not
include the $\tau$ data in our analysis.
\end{itemize}

We have previously published \cite{HMNT} a short summary of our 
evaluation of (\ref{eq:disprel1}), which gave
\be 
a_\mu^{\rm had,LO} 
= (683.1\pm5.9_{\rm exp} \pm 2.0_{\rm rad}) \times10^{-10}. 
\label{eq:A} \ee
When this was combined with the other contributions to $g-2$ we found that
\be a_\mu^{\rm SM} \equiv (g-2)/2 = (11659166.9\pm7.4)\times10^{-10}, 
 \label{eq:B} \ee
in the Standard Model, which is about three standard deviations below 
the measured value given in (\ref{eq:BNL2001}). The purpose of this 
paper is threefold.  First, to describe our method of analysis in detail,
and to make a careful comparison with the contemporary evaluation of 
Ref.~\cite{DEHZ03}.  Second, the recent
 CMD-2 data for the $e^+e^- \to \pi^+\pi^-, \pi^+\pi^-\pi^0$ and
$K^0_S K^0_L$ channels~\cite{Akhmetshin:2001ig, Akhmetshin:2000ca, 
Akhmetshin:1999ym} has just been re-analysed,
and the measured values re-adjusted~\cite{CMD2new}.  We therefore 
recompute $a_\mu^{\rm had,LO}$ to see how the
values given in (\ref{eq:A}) and (\ref{eq:B}) are changed. 
Third, we use our knowledge of the data for $R(s)$ to
give an updated determination of $\Delta\alpha_{\rm had}(s)$, and 
hence of the QED coupling $\alpha(M_Z^2)$.

The outline of the paper is as follows. As mentioned above, 
Section~\ref{sec:clustering} describes how to process
and combine the data, from a wide variety of different experiments, 
so as to give the optimum form of $R(s)$,
defined in (\ref{eq:R(s)}). In Section~\ref{sec:evaluation} we 
describe how we evaluate dispersion relations
(\ref{eq:disprel1}) and (\ref{eq:disprel2}), for $a_\mu^{\rm had,LO}$ 
and $\Delta\alpha_{\rm had}$ respectively,
and, in particular, give Tables and plots to show which energy 
intervals give the dominant contributions {\em and}
dominant uncertainties. Section~\ref{sec:QCDsumrules} shows how QCD 
sum rules may be used to resolve discrepancies
between the inclusive and exclusive measurements of $R(s)$.  
Section~\ref{sec:comparison} contains a 
comparison with other predictions of $g-2$, and in
particular a contribution-by-contribution comparison with the very
recent DEHZ~03 determination~\cite{DEHZ03}. In
Section~\ref{sec:internal} we calculate the {\em internal}\footnote{
In this notation, the familiar light-by-light contributions are 
called {\em external}; see Section~\ref{sec:internal}.} hadronic 
light-by-light contributions to $a_\mu$. 
Section~\ref{sec:calculation} describes an updated
calculation of the NLO hadronic contribution, $a_\mu^{\rm had,NLO}$. 
In this Section we give our prediction for
$g-2$ of the muon.  Section~\ref{sec:determination} 
is devoted to the computation of the value of the QED coupling at 
the $Z$ boson mass, $\alpha(M_Z^2)$; comparison is made with earlier 
determinations. We also give the implications of the
updated value for the estimate of the Standard Model Higgs boson mass. 
Finally in Section~\ref{sec:conclusions} we present our conclusions.

\section{Processing the data for $e^+e^-\,\to\,$hadrons}
\label{sec:clustering}

The data that are used in this analysis for $R(s)$, in order 
to evaluate dispersion relations (\ref{eq:disprel1})
and (\ref{eq:disprel2}), are summarized in Table~\ref{tab:t1}, 
for both the individual exclusive channels ($e^+e^- \to \pi^+\pi^-, 
\pi^+\pi^-\pi^0, K^+K^-, \ldots$) and the 
inclusive process ($e^+e^-\to\gamma^*\to {\rm hadrons}$)\footnote{
A complete compilation of these data can be found in \cite{MRW}.}. 
In Sections~\ref{sec:vacpolcor}--\ref{sec:narrowres} we discuss 
the radiative corrections to the individual data sets, and then 
in Section~\ref{sec:combdatasets} we address the problem of 
combining different data sets for a given channel.

Incidentally, we need to assume that initial state radiative 
corrections (which are described by pure QED) have
been properly accounted for in all experiments.  We note that the 
interference between initial and final state
radiation cancels out in the total cross section.
\begin{table}[ht]\begin{center}
\begin{tabular}{|l|p{12cm}|}
\hline
& \\
\raisebox{1.2ex}[0ex][1ex]{Channel} & 
 \raisebox{1.2ex}[0ex][1ex]{Experiments with References}
\\ \hline\hline

$\pi^+\pi^-$ &
 OLYA~\cite{Vasserman:1979hw,Bukin:1978sq,Koop:1979},
 OLYA-TOF~\cite{Vasserman:1981xq}, 
 NA7~\cite{Amendolia:1984di},
 OLYA and CMD~\cite{Barkov:1985ac,Dolinsky:1991vq},
 DM1~\cite{Quenzer:1978qt},
 DM2~\cite{Bisello:1989hq}, BCF~\cite{Bollini:1975pn,Bernardini:1973pe},
 MEA~\cite{Esposito:1977xg,Esposito:1980bz},
 ORSAY-ACO~\cite{Cosme:1976ft},
 CMD-2~\cite{CMD2new,Akhmetshin:2001ig,Eidelman:2001ju}
\\ \hline

$\pi^0\gamma$ &
 SND~\cite{Achasov:2003ed,Achasov:2000zd}
\\
$\eta\gamma$  &
 SND~\cite{Achasov:2000zd,Achasov:1997nq},
 CMD-2~\cite{Akhmetshin:1995vz,Akhmetshin:1999zv,Akhmetshin:2001hm}
\\ \hline

$\pi^+\pi^-\pi^0$ &
 ND~\cite{Dolinsky:1991vq}, DM1~\cite{Cordier:1980qg},
 DM2~\cite{Antonelli:1992jx},
 CMD-2~\cite{CMD2new,Akhmetshin:2000ca,Akhmetshin:1995vz,Akhmetshin:1998se},
 SND~\cite{Achasov:2002ud,Achasov:2003ir}, CMD~\cite{Barkov:1989}
\\ \hline

$K^+K^-$ &
 MEA~\cite{Esposito:1977xg}, OLYA~\cite{Ivanov:1981wf},
 BCF~\cite{Bernardini:1973pe}, DM1~\cite{Delcourt:1981eq},
 DM2~\cite{Bisello:1988ez,Augustin:1983ix},
 CMD~\cite{Dolinsky:1991vq}, CMD-2~\cite{Akhmetshin:1995vz},
 SND~\cite{Achasov:2001ni}
\\ \hline

$K^0_SK^0_L$ &
 DM1~\cite{Mane:1981ep},
 CMD-2~\cite{CMD2new,Akhmetshin:1999ym,Akhmetshin:2002vj},
 SND~\cite{Achasov:2001ni}
\\ \hline

$\pi^+\pi^-\pi^0\pi^0$ &
 M3N~\cite{Paulot:1979ep}, DM2~\cite{Bisello:1990kh},
 OLYA~\cite{Kurdadze:1986tc}, CMD-2~\cite{Akhmetshin:1998df},
 SND~\cite{Achasov:2003bv}, ORSAY-ACO~\cite{Cosme:1976tf},
 $\gamma\gamma 2$~\cite{Bacci:1981zs}, MEA~\cite{Esposito:1981dv}
\\ \hline

$\omega(\to \pi^0\gamma)\pi^0$ &
 ND and ARGUS~\cite{Dolinsky:1991vq}, DM2~\cite{Bisello:1990kh},
 CMD-2~\cite{Akhmetshin:1998df,Akhmetshin:2003ag},
 SND~\cite{Achasov:2000wy,Achasov:1999wr}, ND~\cite{Dolinsky:1986kj}
\\ \hline

$\pi^+\pi^-\pi^+\pi^-$ &
 ND~\cite{Dolinsky:1991vq}, M3N~\cite{Paulot:1979ep},
 CMD~\cite{Barkov:1988gp}, DM1~\cite{Cordier:1982zn,Cordier:1979yp},
 DM2~\cite{Bisello:1990kh}, OLYA~\cite{Kurdadze:1988mu},
 $\gamma\gamma 2$~\cite{Bacci:1980ru},
 CMD-2~\cite{Akhmetshin:1998df,Akhmetshin:1999ty,Akhmetshin:2000it},
 SND~\cite{Achasov:2003bv}, ORSAY-ACO~\cite{Cosme:1976tf}
\\ \hline

$\pi^+\pi^-\pi^+\pi^-\pi^0$ &
 MEA~\cite{Esposito:1981dv}, M3N~\cite{Paulot:1979ep},
 CMD~\cite{Dolinsky:1991vq,Barkov:1988gp}, 
 $\gamma\gamma 2$~\cite{Bacci:1981zs}
\\ \hline

$\pi^+\pi^-\pi^0\pi^0\pi^0$ &
 M3N~\cite{Paulot:1979ep}
\\ \hline

$\omega(\to \pi^0\gamma)\pi^+\pi^-$ &
 DM2~\cite{Antonelli:1992jx}, CMD-2~\cite{Akhmetshin:2000wv},
 DM1~\cite{Cordier:1981zs}
\\ \hline

$\pi^+\pi^-\pi^+\pi^-\pi^+\pi^-$ &
 M3N~\cite{Paulot:1979ep}, CMD~\cite{Barkov:1988gp},
 DM1~\cite{Bisello:1981sh}, DM2~\cite{Schioppa:1981th}
\\
$\pi^+\pi^-\pi^+\pi^-\pi^0\pi^0$ &
 M3N~\cite{Paulot:1979ep}, CMD~\cite{Barkov:1988gp},
 DM2~\cite{Schioppa:1981th}, $\gamma\gamma 2$~\cite{Bacci:1981zs},
 MEA~\cite{Esposito:1981dv}
\\
$\pi^+\pi^-\pi^0\pi^0\pi^0\pi^0$ &
 isospin-related
\\ \hline

$\eta\pi^+\pi^-$ &
 DM2~\cite{Antonelli:1988fw}, CMD-2~\cite{Akhmetshin:2000wv}
\\ \hline

$K^+K^-\pi^0$ &
 DM2~\cite{Bisello:1991du,Bisello:1991kd}
\\ \hline

$K^0_S \pi K$ &
 DM1~\cite{Mane:1982si}, DM2~\cite{Bisello:1991du,Bisello:1991kd}
\\ \hline

$K^0_S X$ &
 DM1~\cite{Mane:1982th}
\\ \hline

$\pi^+ \pi^- K^+ K^-$ &
 DM2~\cite{Bisello:1991du}
\\ \hline

$p \bar{p}$ &
 FENICE~\cite{Antonelli:1998fv,Antonelli:1994kq},
 DM2~\cite{Bisello:1983at,Bisello:1990rf}, DM1~\cite{Delcourt:1979ed}
\\ \hline

$n \bar{n}$ &
 FENICE~\cite{Antonelli:1998fv,Antonelli:1993vz}
\\ \hline\hline

incl. ($<2$ GeV) &
 $\gamma\gamma 2$~\cite{Bacci:1979ab}, MEA~\cite{Esposito:1981fi},
 M3N~\cite{Cosme:1979qe}, BARYON-ANTIBARYON~\cite{Ambrosio:1980mf}
\\ \hline

incl. ($>2$ GeV) &
 BES~\cite{Bai:1999pk,Bai:2001ct},
 Crystal Ball~\cite{Jakubowski:1988cd,Osterheld:1986hw,Edwards:1990pc},
 LENA~\cite{Niczyporuk:1982ya}, MD-1~\cite{Blinov:1996fw},
 DASP~\cite{Albrecht:1982bq}, CLEO~\cite{Ammar:1998sk},
 CUSB~\cite{Rice:1982br}, DHHM~\cite{Bock:1980ag}
\\ \hline
\end{tabular}
\vspace{2mm} 
\caption{Experiments and references for the $e^+e^-$
  data sets for the different exclusive and the inclusive
  channels as used in this analysis.  The recent re-analysis from
  CMD-2 \cite{CMD2new} supersedes their previously published data for
  $\pi^+\pi^-$ \cite{Akhmetshin:2001ig}, $\pi^+\pi^-\pi^0$
  \cite{Akhmetshin:2000ca} and $K^0_S K^0_L$ \cite{Akhmetshin:1999ym}.}
\label{tab:t1} 
\end{center}
\end{table}

\subsection{Vacuum polarization corrections} 
\label{sec:vacpolcor}

The observed cross sections in $e^+e^-$ annihilation contain effects from
the $s$-channel photon vacuum polarization (VP)
corrections.  Their net effect can be expressed by replacing the 
QED coupling constant by the running effective coupling as follows:
\be \alpha^2 \to \alpha(s)^2. \ee
On the other hand, the hadronic cross section which enters the 
dispersion integral representations of the vacuum
polarization contribution in (\ref{eq:disprel1}) and 
(\ref{eq:disprel2}) should be the bare cross section. 
 We therefore need to multiply the experimental data by the factor
\be 
 C_{\rm vp} = C_{\rm vp}^A = 
 \left(\frac{\alpha}{\alpha(s)}\right)^{2}, \label{eq:c1}
\ee
if no VP corrections have been applied to the data and if the 
luminosity is measured correctly by taking into account all the VP 
corrections to the processes used for the luminosity measurement.
These two conditions are met only for some recent data.

In some early experiments (DM2, NA7), the muon-pair production process 
is used as the normalization cross section, $\sigma_{\rm norm}$. 
For these measurements, all the corrections to the photon propagator 
cancel out exactly, and the correction factor is unity:
\be  C_{\rm vp} = C_{\rm vp}^B = 1. \label{eq:c2}\ee
However, most experiments use Bhabha scattering as the normalization 
(or luminosity-defining) process.  If no VP correction has been 
applied to this normalization cross section, the correction is dominated
by the contribution to the $t$ channel photon exchange amplitudes at 
$t_{\rm min}$, since the Bhabha scattering cross section behaves as 
$d\sigma/dt \propto \alpha^2/t^2$ at small $|t|$. Thus we may 
approximate the correction factor for the Bhabha scattering cross 
section by
\be    \alpha^2 \to (\alpha(t_{\rm min}))^2. \ee
In this case, the cross section should be multiplied by the factor
\be  
C_{\rm vp} =  C_{\rm vp}^C 
  = \frac{(\alpha/\alpha(s))^2}{(\alpha/\alpha(t_{\rm min}))^2}
  = \left(\frac{\alpha(t_{\rm min})}{\alpha(s)}\right)^2,
\label{eq:c3}         \ee
where
\be   t_{\rm min} = -s (1-\cos\theta_{\rm cut})/2. \ee
If, for example, $|\cos\theta_{\rm cut}| \simeq 1$, then 
$\alpha(t_{\rm min})\simeq \alpha$, and the correction
factor (\ref{eq:c3}) would be nearer to (\ref{eq:c1}). 
On the other hand, if $|\cos\theta_{\rm cut}|\lesim 0.5$,
then $\alpha(t_{\rm min})\sim\alpha(s)$, and the correction 
(\ref{eq:c3}) would be near to (\ref{eq:c2}).

In most of the old data, the leptonic (electron and muon) contribution 
to the photon vacuum polarization function has been accounted for 
in the analysis. [This does not affect data which use $\sigma(\mu^+\mu^-)$ 
as the normalization cross section, since the correction cancels out, 
and so (\ref{eq:c2}) still applies.] However, for those
experiments which use Bhabha scattering to normalize the data, the 
correction factor (\ref{eq:c3}) should be modified to
\be   
C_{\rm vp} = C_{\rm vp}^D = 
 \frac{(\alpha_l(s)/\alpha(s))^2}
      {(\alpha_l(t_{\rm min})/\alpha(t_{\rm min}))^2}\,, 
\label{eq:c4} 
\ee
where $\alpha_l(s)$ is the running QED coupling with only the 
electron and muon  contributions to the photon vacuum polarization 
function included.  In the case of the older inclusive $R$ data, only 
the electron contribution has been taken into account, and we take 
only $l=e$ in (\ref{eq:c4}):
\be   
C_{\rm vp} = C_{\rm vp}^E = 
 \frac{(\alpha_e(s)/\alpha(s))^2}
      {(\alpha_e(t_{\rm min})/\alpha(t_{\rm min}))^2}.\, 
\label{eq:ec4} 
\ee

We summarize the information we use for the vacuum polarization
corrections in Table~\ref{tab:t2} where we partly use information
given in Table~III of \cite{SWARTZ} and in addition give corrections for
further data sets and recent experiments not covered there.  It is
important to note that the most recent data from CMD-2 for $\pi^+\pi^-$,
$\pi^+\pi^-\pi^0$ and $K^0_S K^0_L$, as re-analysed in \cite{CMD2new},
and the $K^0_S K^0_L$ data above the $\phi$ \cite{Akhmetshin:2002vj}, 
are already presented as undressed cross sections, and hence are not
further corrected by us.  The same applies to the inclusive $R$
measurements from BES, CLEO, LENA and Crystal Ball.
In the last column of Table~\ref{tab:t2} we present the ranges of
vacuum polarization correction factors $C_{\rm vp}$, if we approximate
-- as done in many analyses -- the required time-like
$\alpha(s)$ by the smooth space-like $\alpha(-s)$.  The numbers
result from applying formulae (\ref{eq:c1}), (\ref{eq:c2}),
(\ref{eq:c4}), (\ref{eq:ec4}) as specified in the second to last 
column, over the energy ranges
relevant for the respective data sets\footnote{To obtain these
  numbers we have used the parametrization of Burkhardt and
  Pietrzyk~\cite{Burkhardt:2001xp} for $\alpha(q^2)$ in the space-like 
  region, $q^2 < 0$.}.
\begin{figure}[ht]
\begin{center}
{\epsfxsize=10cm \leavevmode \epsffile[90 150 475 670]{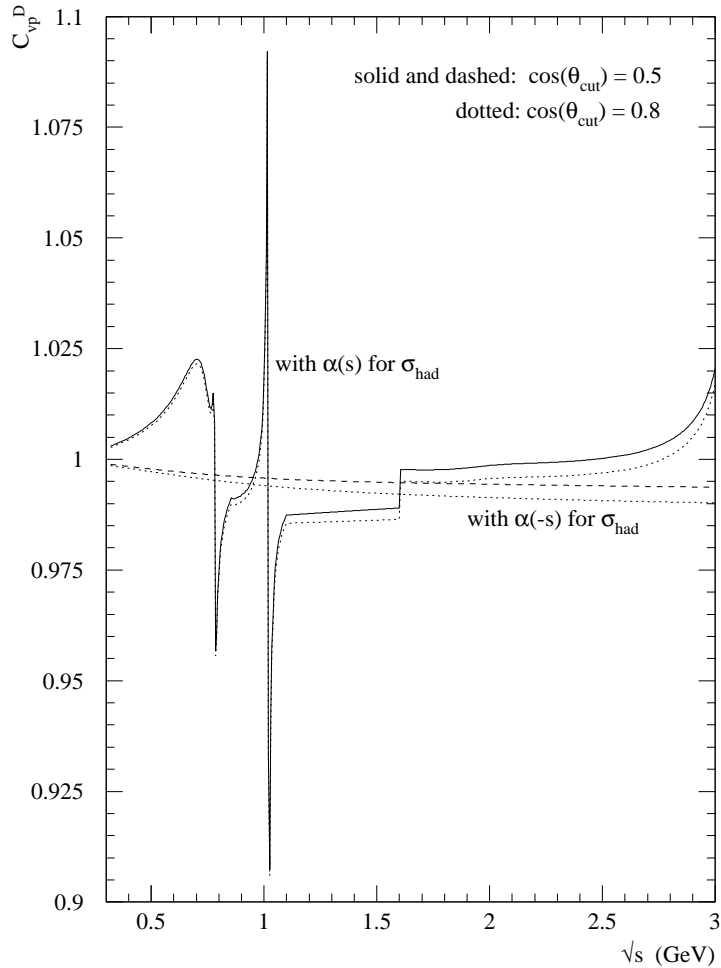}}
\end{center}
\vspace{-0.5cm}
\caption{Vacuum polarization correction factor $C_{\rm vp}^D$ in the
  low energy regime.  The continuous line is the full result as
  applied in our analysis, whereas the dashed line is obtained when
  using $\alpha(-s)$ as an approximation for $\alpha(s)$.  
  Both curves are for $\cos\theta_{\rm cut} = 0.5$ whereas
  the dotted lines are obtained for $\cos\theta_{\rm cut} = 0.8$.}
  \label{fig:rcvp} 
\end{figure}
The correction factors obtained in this way are very close to, but below,
one, decrease with increasing energy, and are very similar to the
corrections factors as given in Table~III of \cite{SWARTZ}.  However,
for our actual analysis we make use of a recent
parametrization of $\alpha$, which is also available in the time-like
regime \cite{J03}.  For the low energies around the $\omega$ and
$\phi$ resonances relevant here, the running of $\alpha$ exhibits a
striking energy dependence, and so do our correction factors $C_{\rm vp}$.
We therefore do not include them in Table~\ref{tab:t2} but display the
energy dependent factor $C_{\rm vp}^D$ in Fig.~\ref{fig:rcvp}.  For
comparison, the correction using space-like $\alpha$, $\alpha(-s)$, 
is displayed as dashed and dotted lines for 
$\cos\theta_{\rm cut} = 0.5$ and 0.8 respectively.
\begin{table}[ht]\begin{center}
\begin{tabular}{|l|l|l|l|l|l|}
\hline
Experiment & Procs.
 & Norm.
 & $\left|\cos\theta_{\rm cut}\right|$
 & Type
 & $C_{\rm vp}(\alpha \mbox{\ spacel.})$ \\ \hline\hline
NA7~\cite{Amendolia:1984di}
 & $\pi^+\pi^-$
 & $\mu\mu$
 & -- 
 & B
 & $1.000$ \\ \hline
OLYA~\cite{Vasserman:1979hw,Bukin:1978sq,Koop:1979,Barkov:1985ac,
 Dolinsky:1991vq,Ivanov:1981wf} 
 & $\pi^+\pi^-, K K$
 & $ee + \mu\mu$
 & $< 0.71$
 & D
 & $0.998 - 0.993$ \\
 \mbox{\ \ }\cite{Kurdadze:1986tc,Kurdadze:1988mu}
 & $4\pi$ & & & D & $0.995 - 0.993$ \\ \hline
CMD~\cite{Barkov:1985ac,Dolinsky:1991vq}
 & $\pi^+\pi^-, K K$
 & $ee + \mu\mu$
 & $< 0.60$
 & D
 & $0.999 - 0.994$\\
 \mbox{\ \ }\cite{Barkov:1989,Barkov:1988gp} & $3\pi, 4\pi$ & & & D &
 $0.996 - 0.994$ \\ \hline
OLYA-TOF~\cite{Vasserman:1981xq}
 & $\pi^+\pi^-$
 & $ee + \mu\mu$
 & $< 0.24$
 & D
 & $0.999 - 0.998$ \\ \hline
MEA~\cite{Esposito:1977xg}
 & $\pi^+\pi^-, K K$
 & $ee$
 & $< 0.77$
 & D
 & $0.992$ \\
 \mbox{\ \ }\cite{Esposito:1980bz}
 & $\pi^+\pi^-$
 & $\mu\mu$
 & --
 & B
 & $1.000$  \\
 \mbox{\ \ }\cite{Esposito:1981dv}
 & $4\pi$
 & $ee$
 & $< 0.77$
 & D
 & $0.993 - 0.992$  \\ \hline
DM1~\cite{Quenzer:1978qt,Delcourt:1981eq,Mane:1981ep}
 & $\pi^+\pi^-, K K$
 & $ee$
 & $< 0.50$
 & D
 & $0.998 - 0.994$ \\
 \mbox{\ \ }\cite{Cordier:1980qg,Cordier:1982zn,Cordier:1979yp} &
 $3\pi, 4\pi$ & & & D & $0.998 - 0.994$ \\ \hline
DM2~\cite{Bisello:1989hq,Bisello:1988ez,Augustin:1983ix}
 & $\pi^+\pi^-, K K$
 & $\mu\mu$
 & --
 & B
 & $1.000$  \\
 \mbox{\ \ }\cite{Antonelli:1992jx,Bisello:1990kh} & $3\pi, 4\pi$ & $ee$
 & unknown & -- & no corr. appl. \\ \hline
SND~\cite{Achasov:2003ed,Achasov:2000zd,Achasov:2001ni}
 & $\pi^0\gamma, K K$
 & $ee$
 & ($< 0.89$)
 & A
 & $0.974 - 0.967$  \\
 \mbox{\ \ }\cite{Achasov:2002ud,Achasov:2003ir,Achasov:2003bv} & $3\pi,
 4\pi$ & & & A & $0.973 - 0.963$ \\ \hline
CMD-2~\cite{Akhmetshin:1999ym,Akhmetshin:1995vz}
 & $K K$
 & $ee$
 & ($< 0.64$)
 & A
 & $0.968 - 0.967$  \\
 \mbox{\ \ }\cite{Akhmetshin:2000ca,Akhmetshin:1995vz,Akhmetshin:1998se,Akhmetshin:1998df,Akhmetshin:1999ty,Akhmetshin:2000it} &
 $3\pi, 4\pi$ & &
 & A
 & $0.972 - 0.963$ \\ \hline
$\gamma\gamma 2$~\cite{Bacci:1979ab}
 & $R$
 & $ee$
 & $< 0.64$
 & E
 & $0.992 - 0.991$ \\ \hline
DASP~\cite{Albrecht:1982bq}
 & $R$
 & $ee$
 & $< 0.71$
 & E
 & $0.985$ \\ \hline
DHHM~\cite{Bock:1980ag}
 & $R$
 & $ee$
 & $< 0.70$
 & D
 & $0.990 - 0.989$ \\ \hline
BES~\cite{Bai:1999pk,Bai:2001ct}
 & $R$
 & $ee$
 & ($< 0.55$)
 & B
 & $1.000$ \\ \hline
Crystal Ball~\cite{Jakubowski:1988cd,Osterheld:1986hw,Edwards:1990pc}
 & $R$
 & $ee$
 & 
 & B
 & $1.000$ \\ \hline
LENA~\cite{Niczyporuk:1982ya}
 & $R$
 & $ee$
 & 
 & B
 & $1.000$ \\ \hline
CLEO~\cite{Ammar:1998sk}
 & $R$
 & $ee$
 & various
 & B
 & $1.000$ \\ \hline
\end{tabular}
\vspace{2mm} \caption{Information about vacuum
  polarization correction factors for different data sets as explained
  in the text. The letters A, B, D, E indicate that the correction factor
  is given by (\ref{eq:c1}), (\ref{eq:c2}), (\ref{eq:c4}), (\ref{eq:ec4})
  respectively.  The $\pi^+\pi^-$ and most recent $\pi^+\pi^-\pi^0$
  and $K^0_L K^0_S$ data from CMD-2, as well as the $R$ measurements
  from BES are given as undressed quantities and are already
  corrected for vacuum polarization effects. According to their
  publications also the $R$ data from CLEO, LENA and Crystal Ball have
  leptonic and hadronic VP corrections applied both in the Bhabha and
  the hadronic cross sections. The correction factors of type A, D and E 
  displayed in the last column are obtained using  $\alpha(-s)$ as an
  approximation to $\alpha(s)$.  However in 
  the actual analysis we evaluate the corrections using $\alpha(s)$,
  see Fig.\ \ref{fig:rcvp}.}
\label{tab:t2} 
\end{center}
\end{table}

\begin{table}\begin{center}
\begin{tabular}{|c||c|c|c|c|c|}
\hline
channel & 
 $\pi^+ \pi^-$ &
 $\pi^+ \pi^- \pi^0$ &
 $\pi^+ \pi^- \pi^+\pi^-$ &
 $\pi^+ \pi^- \pi^0 \pi^0$ & \\ \hline
$\Delta a_{\mu}^{\rm vp} \times 10^{10}$ &
 +1.77 & $-0.68$ & $-0.10$ & $-0.28$ &  \\ \hline
$\Delta (\Delta \alpha_{\rm had}(M_Z^2))^{\rm vp} \times 10^{4}$ &
 $+0.06$ & $-0.07$ & $-0.02$ & $-0.05$ &  \\ \hline\hline
channel & 
 $K^+ K^-$ &
 $K^0_S K^0_L$ &
 $\pi^0\gamma$ &
 incl. ($< 2$ GeV) &
 incl. ($> 2$ GeV) \\ \hline
$\Delta a_{\mu}^{\rm vp} \times 10^{10}$ &
 $-1.05$ & $-0.17$ & $-0.16$ & $-0.54$ & $-0.07$ \\ \hline
$\Delta (\Delta \alpha_{\rm had}(M_Z^2))^{\rm vp} \times 10^{4}$ &
 $-0.14$ & $-0.02$ & $-0.01$ & $-0.18$ & $-0.54$ \\ \hline
\end{tabular}
\end{center}
\caption{Shifts of the contributions to $a_{\mu}$ and 
    $\Delta\alpha_{\rm had}(M_Z^2)$ from
    the different channels due to the application of the appropriate 
    vacuum polarization corrections to the various data sets.
    The values $\Delta a_{\mu}^{\rm vp}$
    are derived as the difference of $a_{\mu}$ calculated with and
    without VP corrections.}
\label{tab:vpshifts} \end{table}
For all exclusive data sets not mentioned in Table~\ref{tab:t2} no
corrections are applied. In most of these cases the possible effect is
very small compared to the large systematic errors or even included
already in the error estimates of the experiments.
For all inclusive data sets not cited in Table~\ref{tab:t2} (but used
in our analysis as indicated in Table~\ref{tab:t1}) we assume, in line
with earlier analyses, that only electronic VP corrections have been
applied to the quoted hadronic cross section values.  We therefore do
correct for missing leptonic ($\mu, \tau$) and hadronic contributions,
using a variant of (\ref{eq:c1}) without the electronic
corrections:
\begin{eqnarray}
 C_{\rm vp} = C_{\rm vp}^F  
  = \frac{(\alpha/\alpha(s))^2}{(\alpha/\alpha_e(s))^2}
  = \left( \frac{\alpha_e(s)}{\alpha(s)} \right)^2 \, .  
\end{eqnarray}
This may, as is clear from the discussion above, lead to an
overcorrection due to a possible cancellation between corrections to
the luminosity defining and hadronic cross sections, in which case
either $C_{\rm vp}^B$ (if $\sigma_{\rm norm}=\sigma_{\mu\mu}$)
or $C_{\rm vp}^E$ (if $\sigma_{\rm norm}=\sigma_{ee}$) should be
used.  However, those
corrections turn out to be small compared to the error in the
corresponding energy regimes.
In addition, we conservatively include these uncertainties in the
estimate of an extra error $\delta a_{\mu}^{\rm vp}$, as discussed below.

The application of the strongly energy dependent VP corrections
leads to shifts $\Delta a_{\mu}^{\rm vp}$ of the contributions to
$a_{\mu}$ as displayed in Table~\ref{tab:vpshifts}.
Note that these VP corrections are significant and of the order of the
experimental error in these channels.  In view of this, the large
positive shift for the leading $\pi^+\pi^-$ channel --- expected from
the correction factor as displayed in Fig.~\ref{fig:rcvp} --- is still
comparably small.  This is due to the dominant role of the CMD-2 data
which do not require correction, as discussed above.  Similarly, for the
inclusive data (above 2 GeV), the resulting VP corrections would
be larger without the important recent data from BES which are more
accurate than earlier measurements and  have been corrected appropriately
already.

To estimate the uncertainties in the treatment of VP corrections, we
take half of the shifts for all channels summed in
quadrature\footnote{ For data sets with no correction applied,
the shifts $\Delta a_{\mu}^{\rm vp}$ are obviously zero. To be
consistent and conservative for these sets (CLEO, LENA
and Crystal Ball) we assign vacuum polarization corrections,
but just for the error estimate.  This results in a total shift of
the inclusive data of $\Delta a_{\mu}^{\rm vp, incl.} = - 0.94 \times
  10^{-10}$, rather than the $- (0.54+0.07) \times 10^{-10}$ implied
by Table~\ref{tab:vpshifts}.}.
The total error due to VP is then given by
\be
\delta a_{\mu}^{\rm vp, excl+incl} =  \frac{1}{2} \left(
\sum^{{\rm all\ channels\ }i} 
 \left( \Delta a_{\mu}^{{\rm vp, }i} \right)^2
\right)^{1/2} = 1.20 \times 10^{-10}\,.
\label{eq:deltamuvp}
\ee
Alternatively, we may assume these systematic uncertainties are highly
correlated and prefer to add the shifts linearly.  For $a_{\mu}$ this
results in a much smaller error due to cancellations of the VP
corrections, and we prefer to take the more conservative result
(\ref{eq:deltamuvp}) as our estimate of the additional
uncertainty.  However, for $\Delta\alpha_{\rm had}$, no
significant cancellations are found to take place 
between channels, so adding the shifts linearly gives the
bigger effect.
Hence for $\Delta\alpha_{\rm had}$ we estimate the error from VP as
\be
\delta \Delta\alpha_{\rm had}^{\rm vp, excl+incl} =  \frac{1}{2}
\sum^{{\rm all\ channels\ }i} \Delta (\Delta\alpha_{\rm had})^{{\rm
    vp, }i} = 1.07 \times 10^{-4}\,.
\label{eq:deltadalphavp}
\ee

\subsection{Final state radiative corrections}

For all the $e^+e^-\to \pi^+\pi^-$ data (except
  CMD-2 \cite{Akhmetshin:2001ig}, whose values for
  $\sigma^0_{\pi\pi(\gamma)}$ already contain final state photons) and
  $e^+e^-\to K^+K^-$ data, we correct for the final state radiation
  effects by using the theoretical formula
\be  C_{\rm fsr} = 1 + \eta(s)\,\alpha/\pi \,,\ee
where $\eta(s)$ is given e.g.\ in~\cite{HGJ}\footnote{For the $\pi^+\pi^-$
  contribution very close to threshold, which is computed in chiral
  perturbation theory, we apply the exponentiated correction
  formula (47) of~\cite{HGJ}.  For a detailed discussion of FSR
  related uncertainties in $\pi^+\pi^-$ production see
  also~\cite{Gluza:2002ui}.}.
In the expression for $\eta(s)$, we take $m=m_\pi$ for $\pi^+\pi^-$, 
and $m=m_K$ for $K^+K^-$ production. Although the formula assumes 
point-like charged scalar bosons, the effects of $\pi$ and $K$ 
structure are expected to be small at energies not too far away 
from the threshold, where the cross section is significant. The above 
factor corrects the experimental data for the photon radiation effects, 
including both real emissions and virtual photon effects.  Because 
there is not sufficient information available as to how the various 
sets of experimental data are corrected for final state photon 
radiative effects, we include 50\% of the correction factor with a 50\% 
error.  That is, we take
\be    
C_{\rm fsr} = 
\left(1 + 0.5\,\eta(s)\,\frac{\alpha}{\pi}\right) 
      \pm  0.5\,\eta(s)\,\frac{\alpha}{\pi}\,, 
\ee
so that the entire range, from omitting to including the correction, 
is spanned.  The estimated additional
uncertainties from final state photon radiation in these two channels 
are then numerically 
$\delta a_{\mu}^{{\rm fsr}, \pi^+\pi^-} = 0.68 \times 10^{-10}$ 
and $\delta a_{\mu}^{{\rm fsr}, K^+K^-} = 0.42 \times 10^{-10}$,
and for $\Delta\alpha_{\rm had}$,   
$\delta \Delta\alpha_{\rm had}^{{\rm fsr}, \pi^+\pi^-} 
= 0.04 \times 10^{-4}$ and
$\delta \Delta\alpha_{\rm had}^{{\rm fsr}, K^+K^-} 
 = 0.06 \times 10^{-4}$.
For all other exclusive modes we do not apply final state radiative 
corrections, but assign an additional 1\% error to the
contributions of these channels in our estimate of the uncertainty 
from radiative corrections. This means that we effectively take
\be    C_{\rm fsr} = 1 \pm 0.01 \ee
for the other exclusive modes such as
$3\pi,\pi^0\gamma,\eta\gamma,4\pi,5\pi,K\bar Kn\pi$, etc., which gives
\begin{eqnarray}
 \delta a_\mu^{\rm fsr,~other} &=& 0.81\times 10^{-10}, \\  
 \delta \Delta \alpha_{\rm had}^{\rm fsr,~other} 
  &=& 0.10 \times 10^{-4}.
\end{eqnarray}

\subsection{Radiative corrections for the narrow 
($J/\psi, \psi^\prime, \Upsilon$) resonances} 
\label{sec:narrowres}

The narrow resonance contributions to the dispersion integral 
are proportional to the leptonic widths $\Gamma(V\to e^+e^-)$. 
The leptonic widths tabulated in \cite{PDG2002} contain photon 
vacuum polarization corrections, as well as final state photon 
emission corrections.  We remove those corrections to obtain the 
bare leptonic width
\be    \Gamma_{ee}^0 = C_{\rm res} \Gamma(V\to e^+e^-) \ee
where
\begin{eqnarray}
  C_{\rm res} = \frac{(\alpha/\alpha(m_V^2))^2}
                          {1 + (3/4)\alpha/\pi}\,. 
\label{eq:radcorr_narrowres}
\end{eqnarray}
Since a reliable evaluation of $\alpha(m_V^2)$ for the very narrow
$J/\psi, \psi'$ and $\Upsilon$ resonances is not available, we use
$\alpha(-m_V^2)$ in the place of $\alpha(m_V^2)$ in 
(\ref{eq:radcorr_narrowres}).  The correction factors obtained 
in this way are small, namely $C_{\rm res}= 0.95$ for $J/\psi$ and 
$\psi^\prime$, and 0.93 for $\Upsilon$ resonances, in agreement with 
the estimate given in~\cite{minireviewRPP}.  A more precise evaluation 
of the correction factor (\ref{eq:radcorr_narrowres}) will be discussed 
elsewhere \cite{HMNTprep}.

To estimate the uncertainty in the treatment of VP corrections,
we take half of the errors summed linearly over all the narrow
resonances.  In this way we found 
\begin{eqnarray}
 \delta a_\mu^{\rm vp, res}  
  &=& \frac12 \sum_{V=J/\psi,\psi^\prime, \Upsilon} 
     \delta a_\mu^{{\rm vp}, V} \nonumber\\
  &=& (0.15 + 0.04 + 0.00) \times 10^{-10} \label{eq:a_mu-vpres}   \\
  &=& 0.19 \times 10^{-10}, \\
 \delta \Delta \alpha_{\rm had}^{\rm vp, res}  
  &=& \frac12 \sum_{V=J/\psi,\psi^\prime, \Upsilon} 
     \delta \Delta \alpha_{\rm had}^{{\rm vp}, V}  \nonumber \\
  &=&( 0.17 + 0.06 + 0.02 + 0.00 ) \times 10^{-4} \label{eq:alf-vpres}  \\
  &=& 0.25 \times 10^{-4}, 
\end{eqnarray}
where the three  numbers in (\ref{eq:a_mu-vpres}) mean the
contributions from $J/\psi, \psi^\prime$ and $\Upsilon(1S-6S)$,
respectively.  Similarly, the four numbers in (\ref{eq:alf-vpres}) 
are the contributions from $J/\psi, \psi^\prime, \Upsilon(1S)$ and 
$\Upsilon(2S-6S)$.

\subsection{Combining data sets} 
\label{sec:combdatasets}

To evaluate dispersion integrals (\ref{eq:disprel1}) and 
(\ref{eq:disprel2}) and their uncertainties, we need to
input the function $R(s)$ and its error. It is clearly desirable to 
make as few theoretical assumptions as possible on the shape and the 
normalization of $R(s)$. Two typical such assumptions are the use 
of Breit--Wigner shapes for resonance contributions and the use of 
perturbative QCD predictions in certain domains of $s$. If we
adopt these theoretical parameterizations of $R(s)$, then it becomes 
difficult to estimate the error of the integral. Therefore, we do not 
make any assumptions on the shape of $R(s)$, and use the trapezoidal
rule for performing the integral up to $\sqrt s = 11.09~\GeV$, beyond 
which we use the most recent perturbative QCD estimates, including the 
complete quark mass corrections up to order $\alpha_S^2$, see 
e.g.~\cite{CKK}. This approach has been made possible because of the 
recent, much more precise, data on $2\pi,3\pi, K\bar K, \pi^0\gamma, 
\eta\gamma$ channels in the $\omega$ and $\phi$ resonant 
regions\footnote{The $J/\psi, \psi'$ and the $\Upsilon$ 
resonances are still treated in the zero-width approximation.}. 
Although this procedure is free from theoretical prejudice, we still 
have to address the problem of combining data from different experiments
(for the same hadronic channel), each with their individual
uncertainties.  If we would perform the dispersion integrals 
(\ref{eq:disprel1}), (\ref{eq:disprel2}) for each data set from each 
experiment separately and then average the resulting contributions to 
$a_{\mu}$ (or $\Delta\alpha_{\rm had}$), this, in general, would lead to a
loss of information resulting in unrealistic error estimates (as 
discussed e.g. in \cite{ADH}), and is, in addition, impracticable in 
the case of data sets with very few points.  On the other hand, a 
strict point-to-point integration over all data points from different 
experiments in a given channel would clearly lead to an
overestimate of the uncertainty because the weighting of precise data 
would be heavily suppressed by nearby data points of lower quality. 
The asymmetry of fluctuations in poorly measured multi-particle final 
states and in energy regions close to the thresholds could in addition 
lead to an overestimate of the mean values of $a_{\mu}$
and $\Delta\alpha_{\rm had}$.

For these reasons, data should be combined before the integration 
is performed.  As different experiments give data points in different 
energy bins, obviously some kind of `re-binning' has to be 
applied\footnote{Another
  possibility to `combine' data, is to fit them
  simultaneously to a function with enough free parameters, typically
  polynomials and Breit--Wigner shapes for continuum and resonance
  contributions, see e.g.~\cite{SWARTZ}.  We decided to avoid any such
  prejudices about the shape of $R$ and possible problems of
  separating continuum and resonance contributions.}.
The bin-size of the combined data will depend, of course, on the 
available data and has to be much smaller in resonance as compared 
to continuum regimes (see below).  For the determination of the mean 
$R$ value, within a bin, the $R$ measurements from different 
experiments should contribute according to their weight.

\begin{figure}
\begin{center}
{\epsfxsize=13cm \leavevmode \epsffile[80 270 490 550]{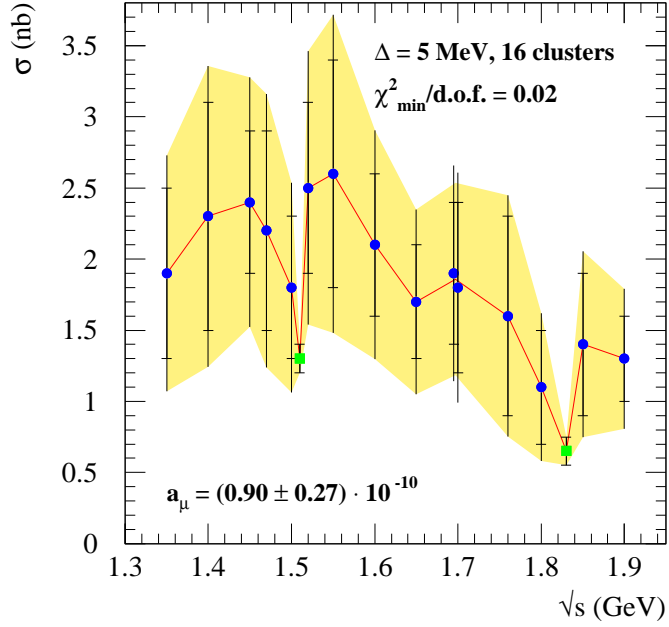}}
{\epsfxsize=13cm \leavevmode \epsffile[80 270 490 550]{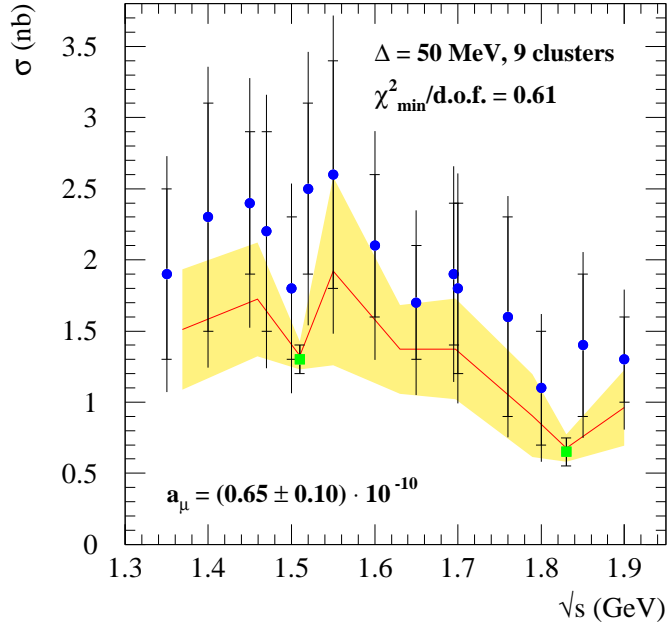}}
\end{center}\caption{Two toy data sets chosen to illustrate the
  problems of combining precise with less precise data. The upper plot
  shows the result obtained with a very small `cluster' size. The
  lower shows the data clustered in 50~MeV bins, which allows
  renormalization of the data within their systematic errors.  Here
  the (much less precise) points of set 1 are renormalized by
  $1/1.35$ whereas the two precise points of set two are nearly
  unchanged ($1/0.9995$).  The length of the error bars give the
  statistical plus systematic errors added in quadrature for each data
  point.  The small horizontal lines in the bars indicate the size of
  the statistical errors.  The error band of the clustered data is
  defined through the diagonal elements of the covariance matrix.}
\label{fig:data_5}
\end{figure}

The problem that the weight of accurate, but sparse, data may become 
lower than inaccurate, but densely-populated, data is well illustrated 
by the toy example shown in Fig.~\ref{fig:data_5}. The plots show two 
hypothetical sets of $R$ data. The set shown by circles has many data 
points with large statistical and a 30\% systematic error.  The second 
set has only two data points, shown by squares, but has small
statistical and only a 1\% systematic error. (The length of the error 
bars of each point is given by the statistical and systematic error
added in quadrature, whereas the little horizontal line inside the bar
indicates the size of the statistical error alone.)  Two alternative 
ways of treating the data are shown in Fig.~\ref{fig:data_5}, together 
with the respective contribution to $a_\mu$, which follows from the 
trapezoidal integration. In the first plot, the impact of the two 
accurate data points is local (with a 5 MeV cluster size no combination 
with the the other set takes place and only two of the less accurate
points around 1.7 GeV are combined), and we see that the integral has a 
30\% error. In the second plot, we have assumed that $R(s)$ does not
change much in a 50~MeV interval, and hence have combined data points 
which lie in 50~MeV `clusters'. In this clustering process, the overall 
normalization factors of the two data sets are allowed to vary within their
uncertainties.  In the toy example, this means that in the upper plot no 
renormalization adjustment takes place, as there is no cluster with 
points from both data sets.  In the lower plot, however, the points of 
the more accurate set 2 are binned together in the clusters with mean 
energies $1.51$ and $1.83$ GeV and lead to a renormalization of all the 
points of the less accurate set by a factor $1/1.35$. (Vice versa the 
adjustment of set 2 is marginal, only ($1/0.9995$), due to its small 
errors.) It is through this renormalization procedure that the
sparse, but very accurate, data can affect the integral.  As a result, 
in the example shown, the value of the integral is reduced by about 
30\% and the error is reduced from 30\% to 15\%. The goodness of the 
fit can be judged by the $\chi_{\rm min}^2$ per degree of freedom, 
which is 0.61 in this toy example. We find that by increasing the
cluster size, that is by strengthening our theoretical assumption 
about the piecewise constant nature of $R$, the error of the integral 
decreases (and the $\chi_{\rm min}^2$ per degree of freedom rises). 
Note that the `pull down' of the mean $R$ values observed in our toy 
example is {\em not} an artefact of the statistical treatment (see 
the remark below) but a property of the data. 

More precisely, to combine all data points for the same channel which 
fall in suitably chosen (narrow) energy bins, we determine the mean 
$R$ values and their errors for all clusters by minimising the 
non-linear $\chi^2$ function
\begin{equation}
\chi^2(R_m, f_k) =
 \sum_{k=1}^{N_{\rm exp}} \left[\left(1-f_k\right)/
 {\rm d}f_k\right]^2 + \sum_{m=1}^{N_{\rm clust}}\ \sum_{i=1}^{N_{\{k,m\}}}
 \left[\left(R_i^{\{k,m\}} - f_k  R_m\right)/
{\rm d}R_i^{\{k,m\}}\right]^2. \label{eq:chisquaredef}
\end{equation}
Here $R_m$ and $f_k$ are the fit parameters for the mean $R$ value
of the $m^{\rm th}$ cluster and the overall normalization factor
of the $k^{\rm th}$ experiment, respectively.  $R_i^{\{k, m\}}$
and ${\rm d}R_i^{\{k, m\}}$ are the $R$ values and errors from
experiment $k$ contributing to cluster $m$. For ${\rm d}R_i^{\{k,
m\}}$ the statistical and, if given, point-to-point systematic errors
are added in quadrature, whereas ${\rm d}f_k$ is the overall
systematic error of the $k^{\rm th}$ experiment. Minimization of
(\ref{eq:chisquaredef}) with respect to the $(N_{\rm exp} + N_{\rm
clust})$ parameters, $f_k$ and $R_m$, gives our best estimates for
these parameters together with their error correlations.

In order to parameterize $R(s)$ in terms of $R_m$, we need a 
prescription to determine the location of the cluster, $\sqrt s = E_m$. 
We proceed as follows. When the original data points, which contribute 
to the cluster $m$, give
\be R(\sqrt s = E_i^{\{k,m\}})\ =\ R_i^{\{k,m\}} \pm \sqrt{\left({\rm
    d}R_i^{\{k,m\}}\right)^2 + \left({\rm d}f_k\right)^2} \label{eq:tt1} \ee
from the $k$th experiment, we calculate the cluster energy $E_m$ by
\be E_m = \left[\sum_k\sum_{i=1}^{N_{\{k,m\}}}
  \frac{1}{\left({\rm d}R_i^{\{k,m\}}\right)^2 + \left({\rm d}f_k\right)^2}E_i^{\{k,m\}}\right]
\Bigg/ \left[\sum_k\sum_{i=1}^{N_{\{k,m\}}} \frac{1}{\left({\rm
  d}R_i^{\{k,m\}}\right)^2 + \left({\rm d}f_k\right)^2}\right], \label{eq:tt2} \ee
where the sum over $k$ is for those experiments whose data points
contribute to the cluster $m$.  Here we use the point-to-point errors,
${\rm d}R_i^{\{k,m\}}$, added in quadrature with the systematic error,
${\rm d}f_k$, to weight the contribution of each data point to the cluster
energy $E_m$. 
Alternatively, we could use just the statistical errors to 
determine the cluster energies $E_m$.  We have checked that the
results are only affected very slightly by this change for our
chosen values for the cluster sizes.

The minimization of the non-linear $\chi^2$ function with respect to
the free parameters $R_m$ and $f_k$ is performed numerically in an
iterative procedure\footnote{Our
  non-linear definition (\ref{eq:chisquaredef}) of the $\chi^2$
  function avoids the pitfalls of simpler definitions without rescaling
  of the errors which would allow
  for a linearized solution of the minimization problem, see
  e.g.~\cite{dAgostini,Takeuchi}.}
and we obtain the following parameterization of $R(s)$:
\be 
 R(s=E_m^2)\ \equiv\ R_m\ =\ \overline R_m \pm {\rm d}R_m,
 \label{eq:tt3} 
\ee
where the correlation between the errors ${\rm d}R_m$ and ${\rm d}R_n$,
\be \rho_{\rm corr}(m,n) = V(m,n)/({\rm d}R_m)({\rm d}R_n), \label{eq:tt4} \ee
with $V(m,m) = ({\rm d}R_m)^2$, is obtained from the covariance 
matrix $V(m,n)$ of the fit, that is
\be 
\chi^2 = \chi^2_{\rm min} 
+ \sum_{m=1}^{N_{\rm clust}} 
  \sum_{n=1}^{N_{\rm clust}} 
  (R_m - \overline R_m) V^{-1}(m,n) (R_n-\overline R_n). \label{eq:tt5} 
\ee
Here the normalization uncertainties are integrated out. We keep the 
fitted values of the normalization factors $f_k$
\be f_k = \bar f_k. \ee
The $\chi^2$ function takes its minimum value $\chi^2_{\rm min}$ 
when $R_m = \overline R_m$ and $f_k = \bar f_k$.  The goodness of 
the fit can be judged from
\be 
 \frac{\chi^2_{\rm min}}{{\rm d.o.f.}}\ 
 =\ \frac{\chi^2_{\rm min}}{\sum_k(N_k-1)-N_{\rm cluster}},
\label{eq:tt6} \ee
where $\sum_kN_k$ stands for the total number of data points and 
$\sum_k(-1)$ stands for the overall normalization uncertainty per 
experiment. Once a good fit to the function $R(s)$ is obtained, 
we may estimate any integral and its error as follows. Consider the 
definite integral
\be 
I(a,b)\ =\ \int_{a^2}^{b^2} {\rm d}s\,R(s)K(s)\ 
        =\ 2\int_a^b {\rm d}E\, E R(E^2)K(E^2) \ 
        =\ \bar I\pm\Delta I.
\label{eq:tt7} \ee
When $a = E_m < E_n = b$, the integral $I$ is estimated by the 
trapezoidal rule to be
\bea 
\bar I & = & 
  2\left(  \frac{E_{m+1} - E_m}{2}E_m R_m K_m  
         + \frac{E_n - E_{n-1}}{2}E_n R_n K_n  
         + \sum_{k=m+1}^{n-1}\frac{E_{k+1}-E_{k-1}}{2}E_k R_k K_k 
   \right),~~~ \label{eq:tt8} 
\eea
where $K_k = K(E_k^2)$, and its error $\Delta I$ is determined, 
via the covariance matrix $V$, to be
\bea 
(\Delta I)^2 & = & \sum_{k=m}^n~ \sum_{l=m}^n 
  \frac{\partial\bar I}{\partial R_k} V(k,l)
  \frac{\partial\bar I}{\partial R_l} \\
 & = & 
    \sum_{k,l=m}^n
     \left((E_{k+1}-E_{k-1})E_k\;K_k\;\right) V(k,l)
     \left((E_{l+1}-E_{l-1})E_lK_l\right), \label{eq:tt9} 
\eea
where $E_{m-1} = E_m$ and $E_{n+1}= E_n$ at the edges, according 
to (\ref{eq:tt8}). When the integration boundaries do not match a 
cluster energy, we use the trapezoidal rule to interpolate between 
the adjacent clusters.


We have checked that for all hadronic channels we find a stable value 
and error for $a_{\mu}^{\rm had,LO}$,
together with a good\footnote{However, there are three channels
for which $\chi^2_{\rm min}/{\rm d.o.f.} > 1.2$, indicating that the 
data sets are mutually incompatible.  These are the
$e^+e^-\to\pi^+\pi^-\pi^+\pi^-, \pi^+\pi^-\pi^0, \pi^+\pi^-\pi^0\pi^0$ 
channels with $\chi^2_{\rm min}/{\rm d.o.f.} = 2.00,\ 1.44,\ 1.28$ 
respectively.  For these cases the error is enlarged by a
factor of $\sqrt{\chi^2_{\rm min}/{\rm d.o.f.}}$.  Note that for the
four pion channel a re-analysis from CMD-2 is under way which is expected to
bring CMD-2 and SND data in much better agreement~\cite{Fedotovitch}. 
If we were to use the same procedure, but now enlarging the errors of
the data sets with $\chi_{\rm min}^2/{\rm d.o.f.} > 1$, then we find
that the experimental error on our determination of $a_{\mu}^{\rm
  had,LO}$ is increased by less than 3\% from the values given in
(\ref{eq:amuhadloincl}) and (\ref{eq:amuhadloexcl}) below.} 
$\chi^2$ fit if we vary the minimal cluster size around our chosen
default values (which are typically about 0.2~MeV for a narrow
resonance and about 10~MeV or larger for the continuum).
\begin{figure}
\begin{center}
{\epsfxsize=12cm \leavevmode \epsffile[90 115 460 670]{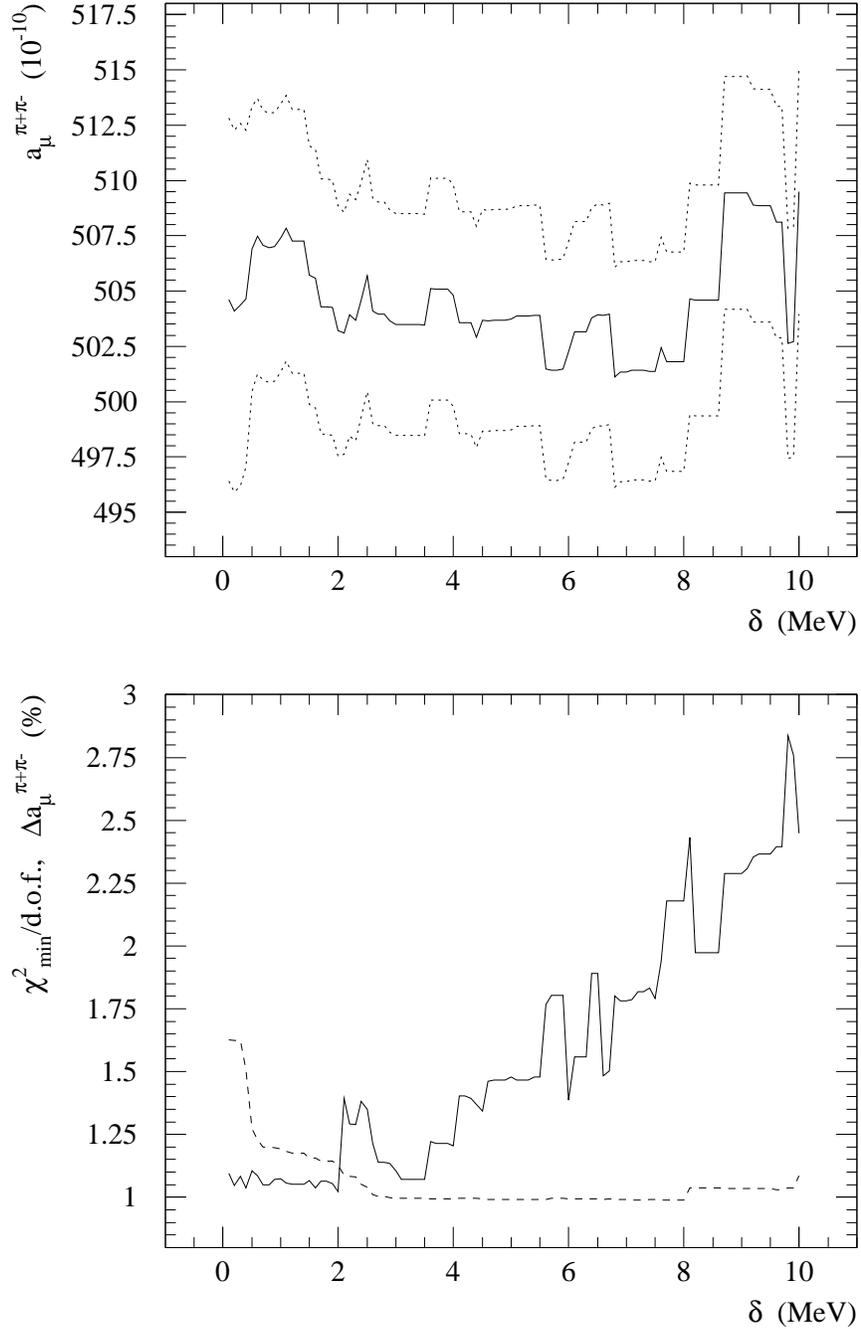}}
\vspace{-6.0ex}
\end{center}
\caption{Dependence of the fit on the cluster size parameter $\delta$
  in the case of the $\pi^+ \pi^-$ channel: the band in the upper plot
  shows the contribution to $a_{\mu}$ and its errors for different
  choices of the cluster size.  The three lines show $\bar{a}_\mu$ 
  (solid), $\bar{a}_\mu +\Delta a_\mu$ and $\bar{a}_\mu-\Delta 
   a_\mu$ (dotted), respectively.  The lower plot displays the
  $\chi^2_{\rm min}/{\rm d.o.f.}$ (continuous line) together with 
  the error size $\Delta a_{\mu}$ in~\% (dashed line).} 
\label{fig:chisqscan}
\end{figure}
For the most important $\pi^+\pi^-$ channel we show in 
Fig.~\ref{fig:chisqscan} the behaviour of the contribution 
to $a_{\mu}$, its error and the quality of the fit expressed 
through $\chi^2_{\rm min}/{\rm d.o.f.}$ as a function of the 
typical cluster size $\delta$. It is clear that very large values 
of $\delta$, even if they lead to a satisfactory $\chi^2_{\rm min}$, 
should be discarded as the fit would impose too much theoretical 
prejudice on the shape of $R(s)$.  Thus, in practice, we also have 
to check how the curve of the clustered data, and its errors, describe 
the data.  One would, in general, try to avoid combining together 
too many data points in a single cluster.

\begin{table}\begin{center}
\begin{tabular}{|l||c|c|c|c|c|c|c|}
\hline
channel & data range & $\delta$ (MeV) & $\chi^2_{\rm min}/{\rm d.o.f.}$ 
& range used &
$a_{\mu}$ & $\Delta a_{\mu}$ & w/o fit \\ \hline\hline

$\pi^+ \pi^-$ & $ 0.32 - 3$ & $3.5$ & $1.07$ & $ 0.32 - 1.425$ &
$502.76$ & $5.01$ & $500.10$ \\ \hline
$\pi^+ \pi^- \pi^0$ & $ 0.483 - 2.4$ & $20, 0.6, 0.6$ & $2.11$ & $ 0.66
- 1.425$ & $46.05$ & $0.63$ & $46.54$ \\
  & & $20, 0.2, 0.2$ & $1.44$ & $ 0.66 - 1.425$ & $46.42$ & $0.76$ &
$47.38$ \\ 
\hline
$\pi^+ \pi^- \pi^+ \pi^-$ & $ 0.765 - 2.245$ & $11$ & $2.00$ & $ 0.765
- 1.432$ & $6.18$ & $0.23$ & $5.70$ \\ \hline
$\pi^+ \pi^- \pi^0 \pi^0$ & $0.915 - 2.4$ & $10$ & $1.28$ & $0.915 -
1.438$ & $9.89$ & $0.57$ & $9.44$ \\ \hline
$K^+ K^-$ & $1.009 - 2.1$ & $5, 0.6$ & $1.00$ & $1.009 - 1.421$ &
$21.58$ & $0.76$ & $21.31$ \\ \hline
$K^0_S K^0_L$ & $1.004 - 2.14$ & $10, 0.1$ & $0.86$ & $1.004 - 1.442$
& $13.16$ & $0.16$ & $13.11$ \\ \hline
inclusive & $1.432 - 3.035$ & $20$ & $0.28$ & $1.432 - 2.05$ & $32.95$ 
          & $2.58$\ & $31.99$ \\
          & $2 - 11.09$ & $20$ & $0.74$ & $2 - 11.09$ & $42.02$ & $1.14$ 
          & $41.51$ \\ \hline
\end{tabular} \end{center}
\caption{Details of the clustering and fit for the
    dominant channels as described in the text. The values of 
    $a_\mu$ and its error have been multiplied by $10^{10}$
    and energy ranges are given in GeV.
    For the $\pi^+\pi^-\pi^0$ channel the bands of clustered data 
    for $\omega$ and $\phi$ displayed in Fig.~\ref{fig:3pi} were
    obtained using a clustering size of 0.6 MeV, which leads to 
    a slightly worse $\chi^2_{\rm min}$, but a better eyeball fit, 
    than for the 0.2 MeV clustering.  For the numerics we have used
    the 0.2 MeV clustering size.  The differences are small.}
\label{tab:clus}
\end{table}
In Table \ref{tab:clus} we give the details of the clustering and
non-linear fit for the most relevant channels.  The fits take into
account data as cited in Table \ref{tab:t1} with energy ranges as
indicated in the second column of Table \ref{tab:clus}.  
We use clustering sizes $\delta$ as
displayed in the third column.  In the $\pi^+\pi^-\pi^0$, $K^+K^-$
and $K_S^0 K_L^0$ channels the binning has to be very fine in
the $\omega$ and $\phi$ resonance regimes;  the
respective values of the clustering sizes in the continuum, ($\omega$
and) $\phi$ regions are given in the Table.  
The $\chi^2_{\rm min}/{\rm d.o.f.}$ displayed in the fourth
column is always good, apart from the three channels
$\pi^+\pi^-\pi^+\pi^-,\ \pi^+\pi^-\pi^0$ and $\pi^+ \pi^- 2\pi^0$, 
in which we inflate the error as mentioned above.  In most cases 
the fit quality and result is amazingly stable with respect to 
the choice of the cluster size, indicating
that no information is lost through the clustering.
Table \ref{tab:clus} also gives information about the contribution of
the leading channels to $a_{\mu}$ within the given ranges.  
For comparison, the last column shows the contributions to $a_{\mu}$ 
obtained by combining data without allowing for
renormalization of individual data sets through the fit parameters 
$f_k$. In this case, we 
use the same binning as in the full clustering, but calculate the 
mean values $R_m$ just as the weighted average of the $R$ data within a 
cluster:
\be 
  R_m \equiv \tilde R_m 
 = \left[ \sum_k\sum_{i=1}^{N_{\{k,m\}}}
   \frac{1}{\left({\rm d}R_i^{\{k,m\}}\right)^2 
            + \left({\rm d}f_k\right)^2}
   R_i^{\{k,m\}}\right]
  \Bigg/ \left[\sum_k\sum_{i=1}^{N_{\{k,m\}}} 
    \frac{1}{\left({\rm d}R_i^{\{k,m\}}\right)^2 
              + \left({\rm d}f_k\right)^2} \right]\,.
\label{eq:rmstart} \ee
(These $\tilde R_m$ values are actually used as starting values for 
our iterative fit procedure.)  The point-to-point trapezoidal integration 
(\ref{eq:tt8}) with these $\tilde R_m$ values from (\ref{eq:rmstart}) 
without the 
fit neglects correlations between different energies.  As is clear from 
the comparison of columns six and eight of Table \ref{tab:clus}, such 
a procedure leads to wrong results, especially in the most important 
$\pi^+\pi^-$ channel.

As explained above, the dispersion integrals (\ref{eq:disprel1}) and
(\ref{eq:disprel2}) are evaluated by integrating (using the
trapezoidal rule (\ref{eq:tt8}) for the mean value and
(\ref{eq:tt9}) for the error and thus including correlations) over the
clustered data directly for all hadronic channels, including the
$\omega$ and $\phi$ resonances. Thus we avoid possible problems due to
missing or double-counting of non-resonant backgrounds. Moreover
interference effects are taken into account automatically. As an
example we display in Fig.~\ref{fig:rhoomega} the most important
$\pi^+ \pi^-$ channel, together with an enlargement of the region of
$\rho$--$\omega$ interference.  As in Fig.~\ref{fig:data_5}, the error
band is given by the diagonal elements of the covariance matrix of our
fit, indicating the uncertainty of the mean values.  Data points are
displayed (here and in the following) after application of radiative
corrections.  The error bars show the statistical and systematic
errors added in quadrature and the horizontal markers inside the error
bars indicate the size of the statistical error alone.

\begin{figure}
\begin{center}
{\epsfxsize=13cm \leavevmode \epsffile[80 270 490 550]{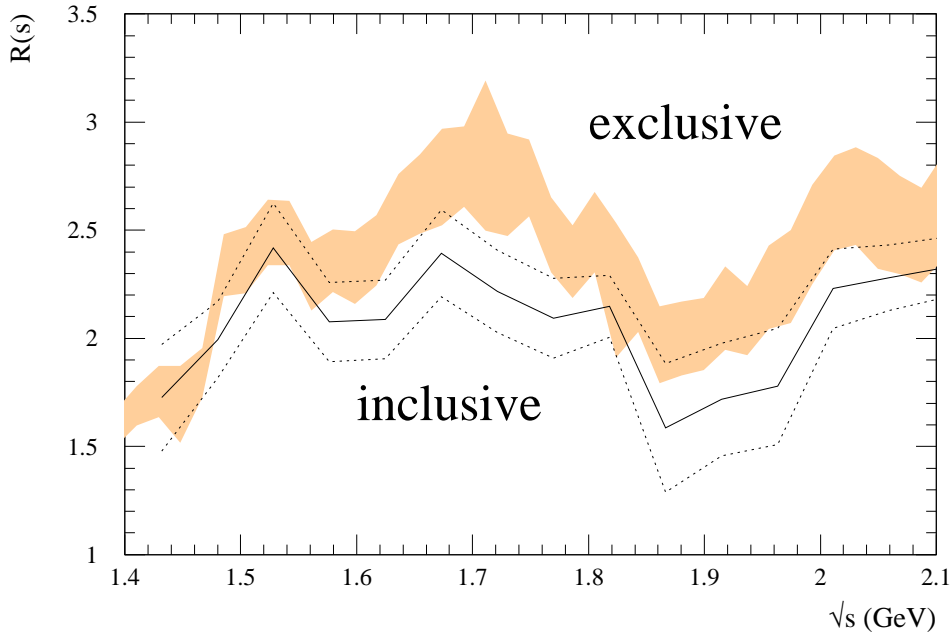}}
\end{center}
\caption{
  The behaviour of $R$ obtained from inclusive data and
  from the sum of exclusive channels, after clustering and fitting
  the various data sets.
  Note the suppressed zero of the vertical scale. }
\label{fig:incl-excl}
\end{figure}

In the region between $1.43$ and $\sim 2$ GeV we have the choice 
between summing up the exclusive channels or
relying on the inclusive measurements from the $\gamma\gamma 2$, 
MEA, M3N and ADONE experiments 
\cite{Bacci:1979ab}--\cite{Ambrosio:1980mf}. 
 Two-body final states were not included 
in these analyses.  Therefore we correct the $R$ data from $\gamma\gamma 
2$, MEA and ADONE for missing contributions from $\pi^+\pi^-$, $K^+ K^-$ 
and $K^0_S K^0_L$, estimating them from our exclusive data 
compilation.\footnote{We do not correct the data from M3N as they quote 
an extra error of 15\% for the missing channels which is taken into 
account in the analysis.}
The corrections are small 
compared to the large statistical and systematic errors
and energy dependent, ranging from up to 7\% at 
1.4 GeV down to about 3\% at 2 GeV.  In addition, we add some purely 
neutral modes to the inclusive data, see below.  Surprisingly, even after 
having applied these corrections, the sum of exclusive channels overshoots 
the inclusive data.
The discrepancy is shown in Fig.~\ref{fig:incl-excl}, where we display
the results of our clustering algorithm for the inclusive and the sum of
exclusive data including error bars defined by the diagonal elements
of the covariance matrices (errors added in quadrature for the
exclusive channels).  We study the problem of this exclusive/inclusive
discrepancy in detail in Section~\ref{sec:QCDsumrules}.

\section{Evaluation of the dispersion relations 
for $a_\mu^{\rm had,LO}$ and $\Delta\alpha_{\rm had}$}
\label{sec:evaluation}

Here we use dispersion relations  (\ref{eq:disprel1}) and  
(\ref{eq:disprel2}) to determine $a_\mu^{\rm had,LO}$
and $\Delta\alpha_{\rm had}(M_Z^2)$ respectively\footnote{It is 
conventional to compute $\Delta\alpha_{\rm had}$
for 5 quark flavours, and to denote it by 
$\Delta\alpha_{\rm had}^{(5)}$.  For simplicity of presentation we often
omit the superscript~(5), but make the notation explicit when we add 
the contribution of the top quark in Section~\ref{sec:determination}.}, 
which in turn we will use to predict $g-2$ of the muon (in
Section~\ref{sec:calculation}) and the QED coupling $\alpha(M_Z^2)$ (in
Section~\ref{sec:determination}). The dispersion
relation (\ref{eq:disprel1}) has the form
\begin{eqnarray}
 a_\mu^{\rm had,LO}
=
 \frac{1}{4\pi^3} \int_{s_{\rm th}}^\infty {\rm d}s\ 
 \sigma_{\rm had}^0(s)\left(\frac{m_\mu^2}{3s}\ K(s)\right),
 \label{eq:dispersion_rel}
\end{eqnarray}
where $\sigma_{\rm had}^0(s)$ is the total cross section for 
$e^+e^-\to {\rm hadrons}\,(+\gamma)$ at centre-of-mass energy 
$\sqrt{s}$, as defined in (\ref{eq:sigma_had}). For 
$s>4m_\mu^2$ the kernel function $K(s)$ is given by~\cite{BKK}
\begin{eqnarray}
 K(s>4m_\mu^2) = 
   \frac{3s}{m_\mu^2}\left\{\frac{x^2}{2}(2-x^2) 
        + \frac{(1+x^2)(1+x)^2}{x^2}
        \left( \ln ( 1 + x ) - x + \frac{x^2}{2} \right)
        + \frac{1+x}{1-x} x^2 \ln x\right\}, \label{eq:K}
\end{eqnarray}
with $x \equiv (1-\beta_\mu)/(1+\beta_\mu)$ where 
$\beta_\mu \equiv \sqrt{1-4m_\mu^2/s}$; while for $s<4m_\mu^2$ the
form of the kernel can be found in~\cite{AK}, and is used to 
evaluate the small $\pi^0\gamma$ contribution to
$a_\mu^{\rm had,LO}$.  Dispersion relation~(\ref{eq:disprel2}), 
evaluated at $s=M_Z^2$, may be written in the form
\begin{equation} 
  \Delta \alpha_{\rm had}(M_Z^2) 
 = -\frac{M_Z^2}{4\pi^2\alpha} 
  {\rm P} \int_{s_{\rm th}}^\infty {\rm d} s\,
  \frac{\sigma_{\rm had}^0(s)}{s-M_Z^2}\,. 
\label{eq:210703-1} 
\end{equation}
To evaluate (\ref{eq:dispersion_rel}) and (\ref{eq:210703-1}) 
we need to input the function $\sigma_{\rm had}^0(s)$ and its 
error. Up to $\sqrt s \sim 2~\GeV$ we can calculate $\sigma_{\rm had}^0$ 
from the sum of the cross sections for all the exclusive channels 
$e^+e^-\to\pi^+\pi^-,\pi^+\pi^-\pi^0$, etc.  On the other hand for
$\sqrt s \gtrsim 1.4~\GeV$ the value of $\sigma_{\rm had}^0$ can be 
obtained from inclusive measurements of
$e^+e^-\to {\rm hadrons}$.  Thus, as mentioned above, there is an 
`exclusive, inclusive overlap' in the interval 
$1.4\lesim\sqrt s\lesim 2~\GeV$, which allows a comparison of the 
two methods of determining $\sigma_{\rm had}^0$
from the data. As we have seen, the two determinations do not agree, 
see Fig.~\ref{fig:incl-excl}. It is worth
noting that the data in this interval come from older experiments. 
The new, higher precision, Novosibirsk data on
the exclusive channels terminate at $\sqrt s \sim 1.4~\GeV$, and 
the recent inclusive BES data~\cite{Bai:1999pk,Bai:2001ct} start only 
at $\sqrt s \sim 2~\GeV$. Thus in Table~\ref{tab:A} we show the
contributions of the individual channels to $a_\mu^{\rm had,LO}$ and 
$\Delta\alpha_{\rm had}(M_Z^2)$ using first
inclusive data in the interval $1.43<\sqrt s<2~\GeV$, and then 
replacing them by the sum of the exclusive channels.

\begin{table}\begin{center}
\begin{tabular}{|l||c|c||c|c|}
\hline
channel & \multicolumn{2}{c||}{inclusive (1.43,2~GeV)} & 
\multicolumn{2}{c|}{exclusive (1.43,2~GeV)} \\
& $a_\mu^{\rm had,LO}$ & $\Delta\alpha_{\rm had}(M_Z^2)$ & 
$a_\mu^{\rm had,LO}$ & $\Delta\alpha_{\rm had}(M_Z^2)$
\\ \hline
$\pi^0 \gamma$ (ChPT) & 
  $0.13 \pm 0.01$ & $0.00\pm0.00$ & $0.13 \pm 0.01$ & $0.00\pm0.00$
\\ \hline
$\pi^0 \gamma$ (data) & 
  $4.50 \pm 0.15$ & $0.36\pm0.01$ & $4.50 \pm 0.15$ & $0.36\pm0.01$
\\ \hline
$\pi^+ \pi^-$ (ChPT) & 
  $2.36 \pm 0.05$ & $0.04 \pm0.00$ & $2.36 \pm 0.05$ & $0.04\pm0.00$
\\ \hline
$\pi^+ \pi^-$ (data) & 
  $502.78 \pm 5.02$ & $34.39\pm0.29$ & $ 503.38 \pm 5.02$ & $34.59\pm0.29$
\\ \hline
$\pi^+ \pi^- \pi^0$ (ChPT) & 
  $0.01 \pm 0.00$ & $0.00 \pm 0.00$ & $0.01 \pm 0.00$ & $0.00 \pm 0.00$
\\ \hline
$\pi^+ \pi^- \pi^0$ (data) & 
  $46.43 \pm 0.90$ & $4.33\pm 0.08$ & $ 47.04 \pm 0.90$ & $4.52\pm0.08$
\\ \hline
$\eta \gamma$ (ChPT) & 
  $ 0.00 \pm 0.00 $ & $0.00 \pm 0.00$ & $0.00 \pm 0.00$ & $0.00 \pm 0.00$
\\ \hline
$\eta \gamma$ (data) & 
  $ 0.73 \pm 0.03 $ & $0.09\pm0.00$ & $0.73 \pm 0.03$ & $0.09\pm0.00$
\\ \hline
$K^+ K^-$ & 
  $ 21.62 \pm 0.76 $ & $3.01\pm0.11$ & $22.35 \pm 0.77$ & $3.23\pm0.11$
\\ \hline
$K_S^0 K_L^0$ & 
  $ 13.16 \pm 0.31 $ & $1.76\pm0.04$ & $13.30 \pm 0.32$ & $1.80\pm0.04$
\\ \hline
$2 \pi^+ 2\pi^-$ & 
  $ 6.16 \pm 0.32 $ & $1.27\pm0.07$ & $14.77 \pm 0.76$ & $4.04\pm0.21$
\\ \hline
$\pi^+ \pi^- 2\pi^0$ & 
  $ 9.71 \pm 0.63 $ & $1.86\pm0.12$ & $20.55 \pm 1.22$ & $5.51\pm0.35$
\\ \hline
$2 \pi^+ 2 \pi^- \pi^0$ & 
  $ 0.26 \pm 0.04 $ & $0.06\pm 0.01$ & $2.85 \pm 0.25$ & $0.99\pm0.09$
\\ \hline
$ \pi^+ \pi^- 3 \pi^0$ & 
  $ 0.09 \pm 0.09 $ & $0.02\pm0.02$ & $1.19 \pm 0.33$ & $0.41\pm0.10$
\\ \hline
$ 3\pi^+ 3\pi^-$ & 
  $ 0.00 \pm 0.00 $ & $0.00\pm0.00$ & $0.22 \pm 0.02$ & $0.09\pm 0.01$
\\ \hline
$ 2\pi^+ 2\pi^- 2\pi^0$ & 
  $ 0.12 \pm 0.03 $ & $0.03\pm0.01$ & $ 3.32 \pm 0.29$ & $1.22\pm0.11$
\\ \hline
$ \pi^+ \pi^- 4\pi^0$ (isospin)& 
  $ 0.00 \pm 0.00 $ & $0.00\pm0.00$ & $0.12 \pm 0.12$ & $0.05\pm0.05$
\\ \hline
$ K^+ K^- \pi^0$ & 
  $ 0.00 \pm 0.00 $ & $0.00\pm0.00$ & $0.29 \pm 0.07$ & $0.10\pm 0.03$
\\ \hline
$ K_S^0 K_L^0 \pi^0$ (isospin) & 
  $ 0.00 \pm 0.00 $ & $0.00\pm0.00$ & $0.29 \pm 0.07$ & $0.10\pm 0.03$
\\ \hline
$ K_S^0 \pi^{\mp} K^{\pm}$ & 
  $  0.05 \pm 0.02 $ & $0.01\pm0.00$ & $1.00 \pm 0.11$ & $0.33\pm 0.04$
\\ \hline
$ K_L^0 \pi^{\mp} K^{\pm}$ (isospin) & 
  $  0.05 \pm 0.02 $ & $0.01\pm0.00$ & $1.00 \pm 0.11$ & $0.33\pm 0.04$
\\ \hline
$ K \bar{K} \pi \pi$ (isospin) & 
  $  0.00 \pm 0.00 $ & $0.00\pm0.00$ & $3.63 \pm 1.34$ & $1.33 \pm 0.48$
\\ \hline
$ \omega (\to \pi^0\gamma) \pi^0$ & 
  $ 0.64 \pm 0.02 $ & $0.12\pm0.00$ & $0.83 \pm 0.03$ & $0.17 \pm 0.01$
\\ \hline
$ \omega (\to \pi^0\gamma) \pi^+ \pi^-$ & 
  $ 0.01 \pm 0.00 $ & $0.00\pm0.00$ & $ 0.07 \pm 0.01$ & $0.02 \pm 0.00$
\\ \hline
$ \eta (\to \pi^0\gamma) \pi^+ \pi^-$ & 
  $ 0.07 \pm 0.01 $ & $0.02\pm0.00$ & $0.49 \pm 0.07$ & $0.15 \pm 0.02$
\\ \hline
$ \phi(\to {\rm unaccounted}) $ & 
  $ 0.06 \pm 0.06 $ & $0.01\pm0.01$ & $0.06 \pm 0.06$ & $0.01 \pm 0.01$
\\ \hline
$ p\bar{p} $ & 
  $ 0.00 \pm 0.00 $ & $0.00\pm0.00$ & $0.04 \pm 0.01$ & $0.02 \pm 0.00$
\\ \hline
$ n\bar{n} $ & 
  $ 0.00 \pm 0.00 $ & $0.00\pm0.00$ & $0.07 \pm 0.02$ & $0.03 \pm 0.01$
\\ \hline
$J/\psi, \psi^\prime$ &
  $ 7.30 \pm 0.43 $ & $8.90\pm0.51$ & $7.30 \pm 0.43$ & $8.90\pm0.51$
\\ \hline
$\Upsilon(1S-6S)$ &
  $ 0.10\pm 0.00$ & $1.16\pm0.04$ & $0.10 \pm 0.00$ & $1.16\pm0.04$
\\ \hline
inclusive $R$ & 
 $ 73.96 \pm 2.68$ & $ 92.75\pm1.74$ & $42.05 \pm 1.14$ & $81.97 \pm 1.53$
\\ \hline
pQCD & 
 $ 2.11 \pm 0.00$ & $125.32\pm0.15$ & $2.11\pm0.00$ & $125.32\pm0.15$
\\ \hline \hline
sum & 
 $692.38 \pm 5.88$ & $275.52\pm1.85$ & $ 696.15 \pm 5.68$ & $276.90\pm1.77$
\\ \hline
\end{tabular}
\end{center}
\caption{Contributions to the dispersion 
  relations (\ref{eq:disprel1}) and (\ref{eq:disprel2}) from the
  individual channels.}
\label{tab:A}
\end{table}

Below we describe, in turn, how the contributions of each channel 
have been evaluated. First we note that narrow
$\omega$ and $\phi$ contributions to the appropriate channels are 
obtained by integrating over the (clustered)
data using the trapezoidal rule. We investigated the use of parametric 
Breit--Wigner forms by fitting to the data
over various mass ranges. We found no significant change in the 
contributions if the resonant parameterization was
used in the region of the $\omega$ and $\phi$ peaks, but that the 
contributions of the resonance tails depend a
little on the parametric form used. The problem did not originate 
from a bias due to the use of the linear
trapezoidal rule in a region where the resonant form was concave, but 
rather was due to the fact that different
resonant forms fitted better to different points in the tails. For 
this reason we believe that it is more reliable
to rely entirely on the data, which are now quite precise in the 
resonance regions.

\subsection{$\pi^0\gamma$ channel}
\label{subsec:eval_pi0gamma}

\begin{figure}
\begin{center}
{\epsfxsize=12cm \leavevmode \epsffile[90 115 460 670]{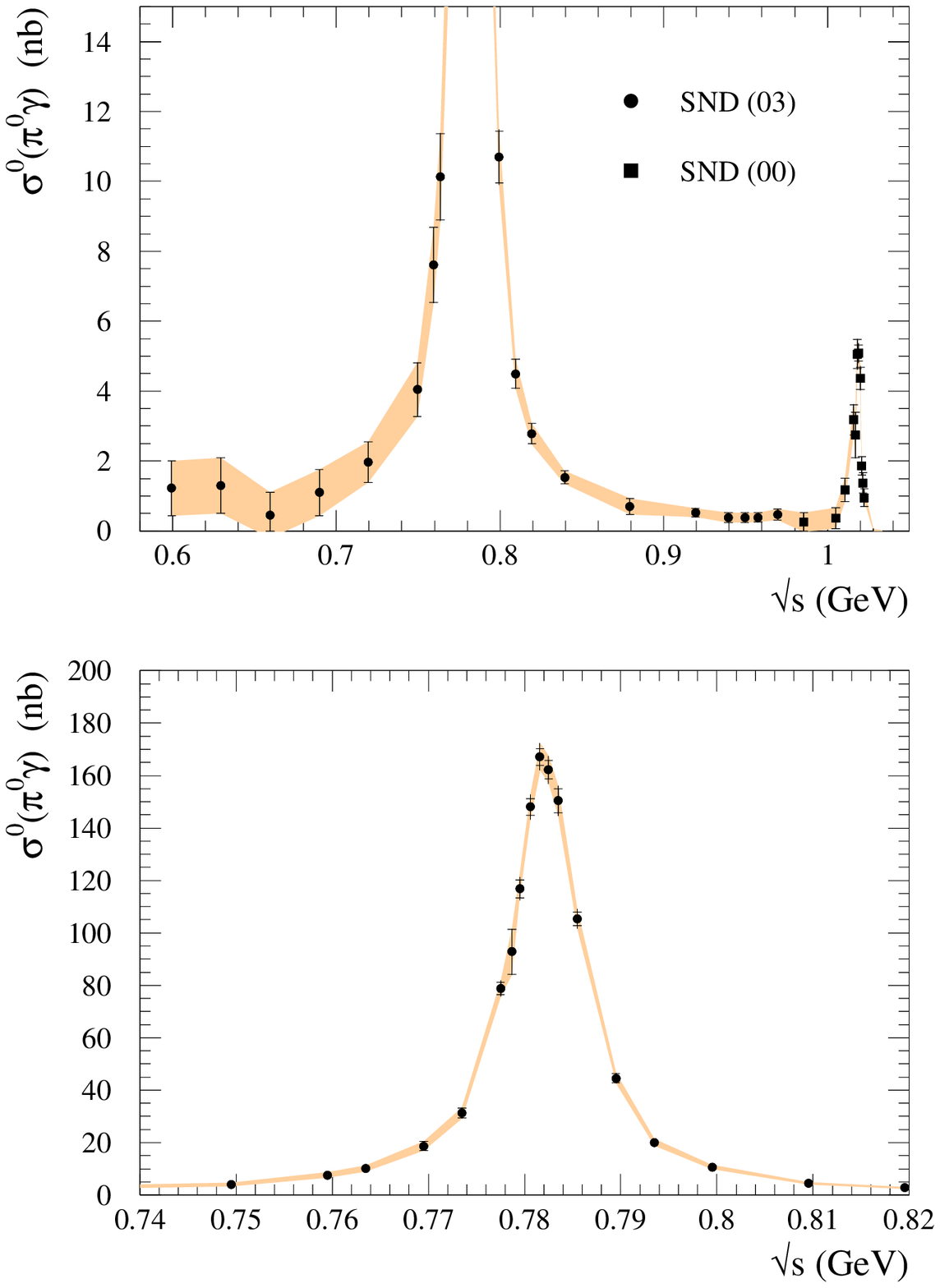}}
\vspace{-6.0ex}
\end{center}
\caption{Data for $\sigma^0(e^+e^-\to \pi^0\gamma)$. 
  The shaded band shows the behaviour of the cross section
  after clustering and fitting the data.
  The second plot is an enlargement in the region of the $\omega$ 
  resonance.}
\label{fig:pi0gamma_total}
\end{figure}

The contribution of the $e^+e^-\to\pi^0\gamma$ channel defines the 
lower limit, $\sqrt s_{\rm th}=m_\pi$, of the
dispersion integrals. There exist two data sets 
\cite{Achasov:2003ed,Achasov:2000zd} for this channel, which cover the
interval $0.60<\sqrt s<1.03~\GeV$ (see Fig.~\ref{fig:pi0gamma_total}).
After clustering, a trapezoidal rule integration over this 
$\pi^0\gamma$ energy interval gives a contribution
\begin{eqnarray}
a_\mu (\pi^0\gamma,\,0.6<\sqrt s<1.03~\GeV) 
= (4.50 \pm 0.15)\times 10^{-10} 
\end{eqnarray}
and
\be \Delta \alpha_{\rm had} (\pi^0\gamma,\,0.6<\sqrt s<1.03~\GeV) 
= (0.36 \pm 0.01)\times 10^{-4}.\ee
In Fig. \ref{fig:pi0gamma_total} we show an overall picture 
of the $e^+ e^- \to \pi^0\gamma$ data and a blow up around the
$\rho$-$\omega$ region.

\begin{figure} \begin{center}
{\epsfxsize=13cm \leavevmode \epsffile[80 270 490 550]{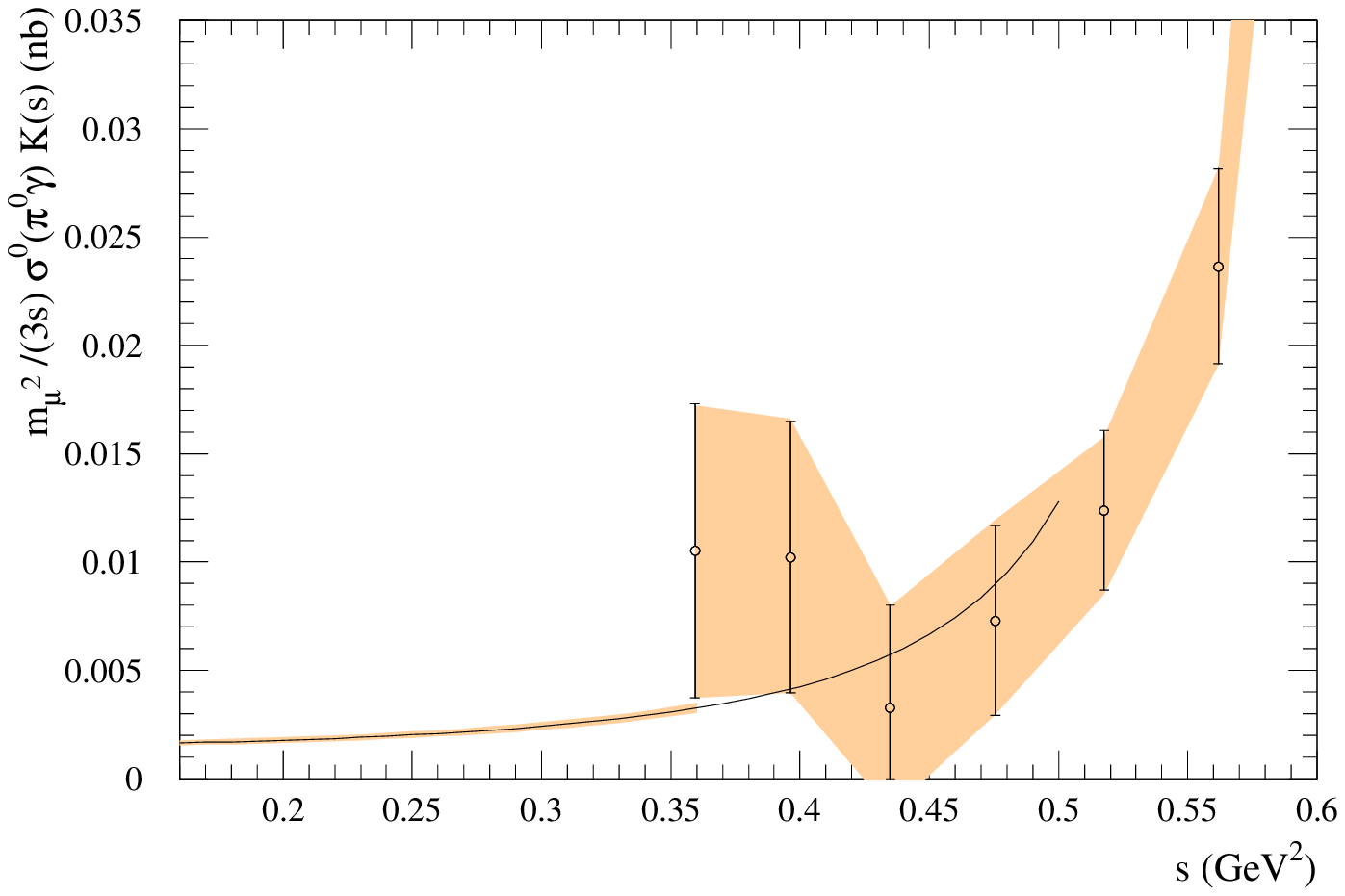}}
\end{center}
\caption{Predictions for $\sigma^0(e^+e^-\to\pi^0\gamma)$ from ChPT 
  compared with low energy 
  experimental data from the SND collaboration~\cite{Achasov:2003ed}.
  In the figure we have multiplied $m^2_\mu /(3s) K(s)$
  by the cross section so that the area below the data is proportional 
  to the contribution to $a_\mu$.
  The continuous curve, which is obtained assuming Vector Meson ($\omega$)
  Dominance (VMD), is used for $s < 0.36~{\rm GeV}^2$.} 
\label{fig:sigma-pi0gamma_omega}
\end{figure}

The use of the trapezoidal rule for the interval 
$m_\pi<\sqrt s < 0.6~\GeV$ would overestimate the contribution,
since the cross section is not linear in $\sqrt s$. In this region 
we use chiral perturbation theory (ChPT), based
on the Wess--Zumino--Witten (WZW) local interaction for the 
$\pi^0\gamma\gamma$ vertex,
\begin{eqnarray}
 {\cal L}_{WZW} = - \frac{\alpha}{8\pi f_\pi} \pi^0
             \epsilon^{\mu\nu\lambda\sigma} F_{\mu\nu} F_{\lambda\sigma},
\end{eqnarray}
with $f_\pi \simeq 93$~MeV, which yields
\begin{eqnarray}
 \sigma(e^+e^-\to \pi^0\gamma) = \sigma_{\rm pt} \equiv
  \frac{8\alpha\pi\Gamma(\pi^0\to2\gamma)}{3m_\pi^3} 
  \left( 1 - \frac{m_\pi^2}{s} \right)^3.
\label{eq:pi0gammaLO1}
\end{eqnarray}
Since the electromagnetic current couples to $\pi^0\gamma$ via
$\omega$ meson exchange, the low-energy cross section can be
improved by assuming the $\omega$-meson dominance~\cite{AK}, 
which gives
\begin{eqnarray}
 \sigma_{\rm VMD}(e^+e^-\to \pi^0\gamma) 
= \sigma_{\rm pt} (e^+e^-\to \pi^0\gamma)
  \left(\frac{m_\omega^2}{m_\omega^2 - s}\right)^2.
\label{eq:pi0gammaLO2}
\end{eqnarray}
We find
\begin{eqnarray}
 a_\mu(\pi^0\gamma, \sqrt{s}<0.6~\GeV)
= (  0.13 \pm 0.01 ) \times 10^{-10},
\end{eqnarray}
while the contribution to $\Delta\alpha_{\rm had}$ is
less than $10^{-6}$.
The agreement of the prediction of (\ref{eq:pi0gammaLO2}) 
for the $\pi^0\gamma$ cross
section with the SND data just above 0.6~GeV is shown 
in Fig.~\ref{fig:sigma-pi0gamma_omega}.

\subsection{$\pi^+\pi^-$ channel}
\label{pipichannel}

\begin{figure}
\begin{center}
{\epsfxsize=13cm \leavevmode \epsffile[80 270 490 550]{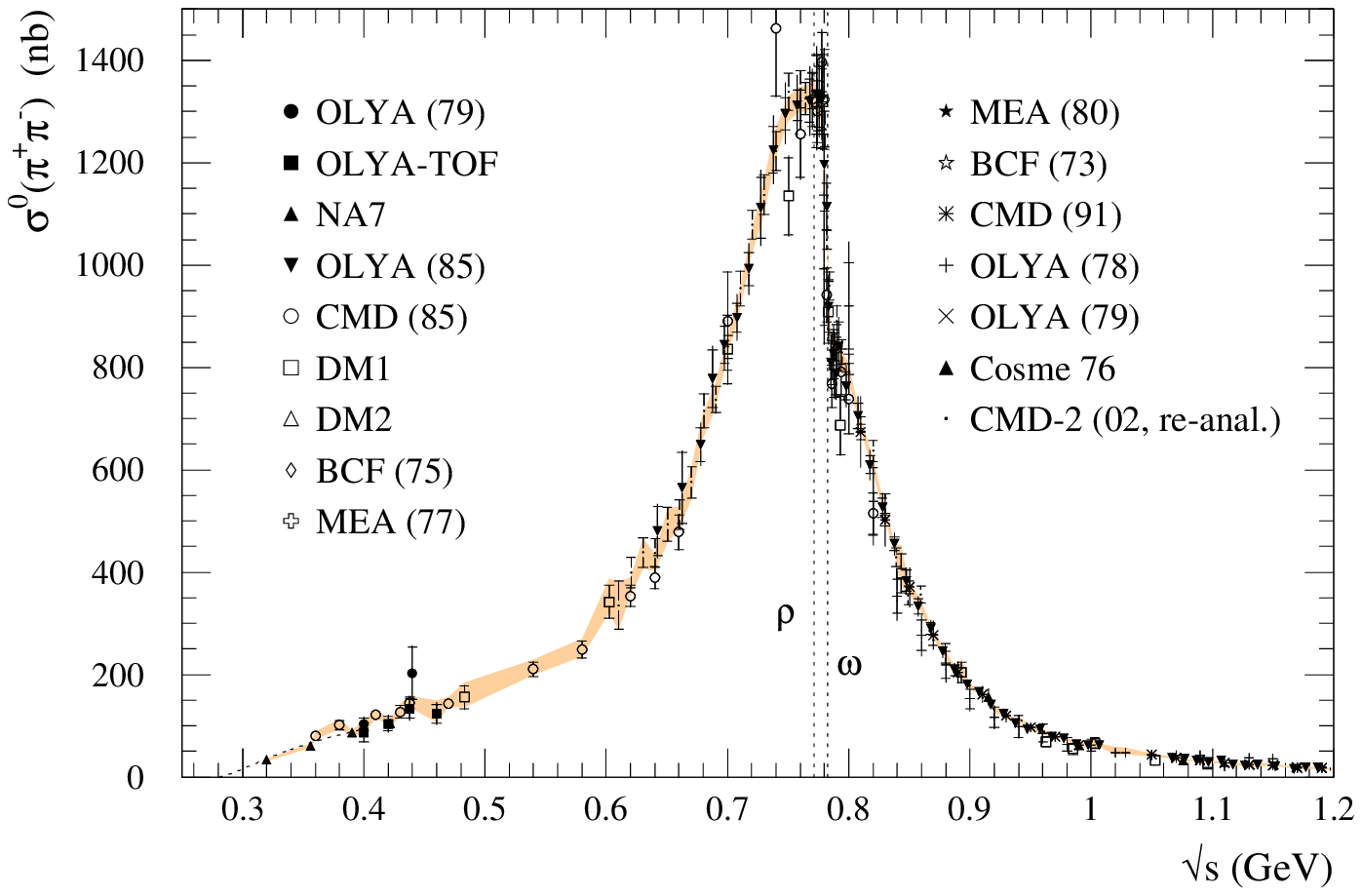}}
{\epsfxsize=13cm \leavevmode \epsffile[65 230 490 555]{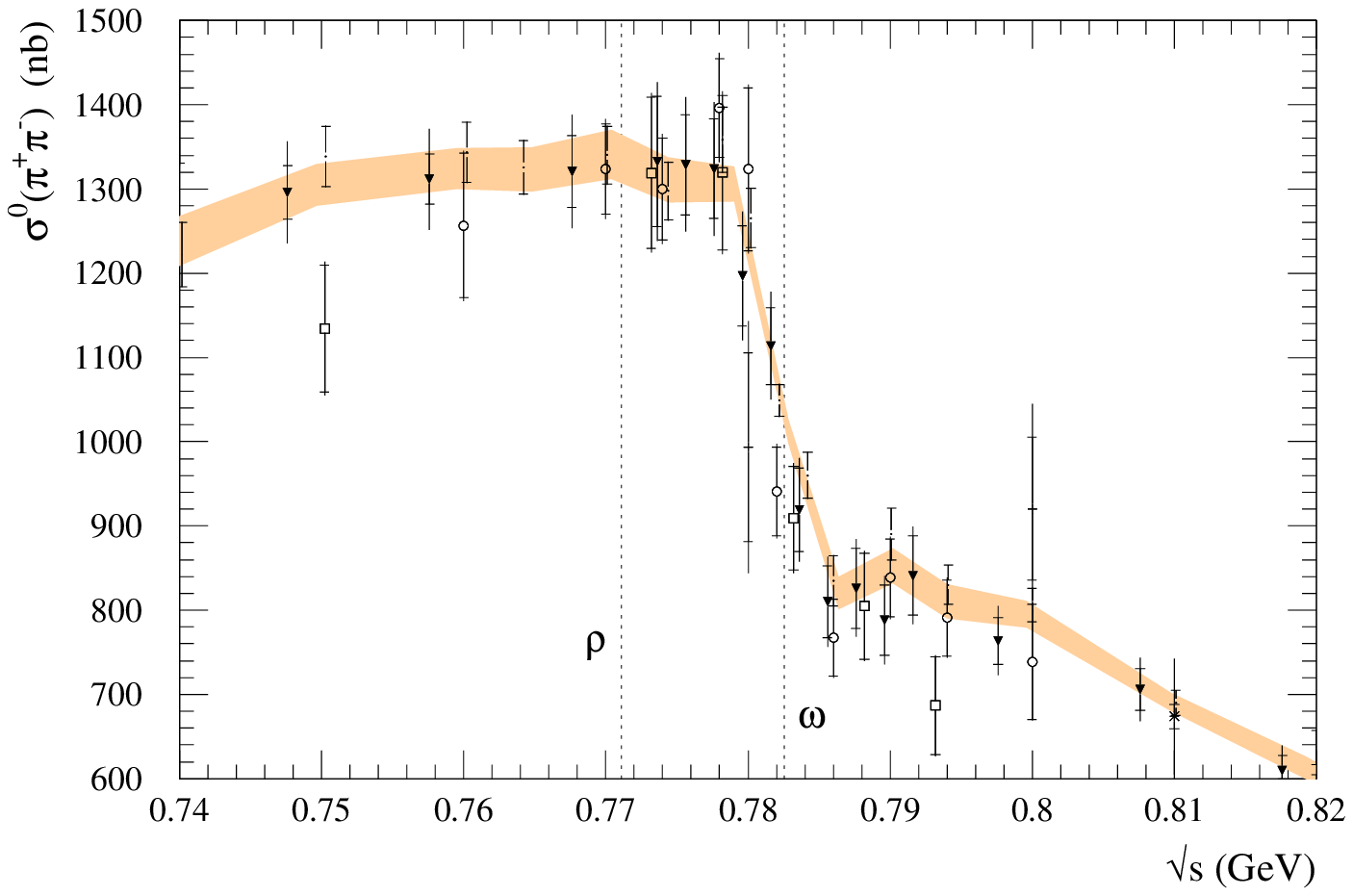}}
\end{center}
\vspace{-7.5ex}
\caption{$e^+e^-\to\pi^+\pi^-$ data up to 1.2~GeV, 
  after radiative corrections, where the shaded
  band shows the result, $\sigma^0(\pi^+\pi^-)$ (obtained from $R_m$ of
  (\ref{eq:chisquaredef})), of our fit after clustering. The width of
  the band indicates the error on the $\sigma^0(\pi^+\pi^-)$ values, 
  obtained from the diagonal elements of the full covariance matrix.  The
  second plot is an enlargement of the $\rho$-$\omega$ interference region.}
\label{fig:rhoomega}
\end{figure}

We use 16 data sets for $e^+e^-\to \pi^+\pi^-$ 
\cite{CMD2new}, \cite{Vasserman:1979hw}--\cite{Eidelman:2001ju} 
which cover the energy range $0.32<\sqrt s < 3.0~\GeV$. 
Some older data with very large errors are omitted.
In Fig.\ \ref{fig:rhoomega},
we show the region around $\rho$, which gives the most important 
contribution to $g-2$ of the muon.

The $\pi^+\pi^-$ 
contributions\footnote{If we were to leave out
the dominant $\pi^+\pi^-$ data from CMD-2 altogether, 
we find $491.33 \pm 8.47$, instead of $503.38 \pm 5.02$, for
the $\pi^+\pi^-$ contribution from the interval 
$0.32 <\sqrt s < 2~\GeV$.  (The $\chi^2_{\rm min}/{\rm d.o.f.}$ 
of the fit which clusters
the data would be even slightly better, 1.00 instead of 1.07.)  This 
means that after re-analysis the CMD-2 data
dominate the error but do not pull down the contribution, but 
rather push it up!} to $a_\mu^{\rm had,LO}$ and
$\Delta\alpha_{\rm had}(M_Z^2)$, obtained by integrating clustered 
data over various energy intervals, are shown
in Table~\ref{tab:new180703}.
\begin{table}
\begin{center}
\begin{tabular}{|r @{--} l|c|r @{$~\pm$~} l|r @{$~\pm$~} l|}
\hline  \multicolumn{2}{|c|}{$\sqrt s~(\GeV)$} 
& \rule[-2ex]{0ex}{5ex} comment &
\multicolumn{2}{c|}{$a_\mu^{\rm had,LO}\times10^{10}$} & 
\multicolumn{2}{c|}{$\Delta\alpha_{\rm had}(M_Z^2)\times10^4$}\\ \hline
 0.32 & 1.43 &             & 502.78 & 5.02 & 34.39 & 0.29 \\
(0.32 & 1.43 & `old' CMD-2 & 492.66 & 4.93 & 33.65 & 0.28) \\
 0.32 & 2    &             & 503.38 & 5.02 & 34.59 & 0.29 \\
 \multicolumn{2}{|c|}{} && \multicolumn{2}{|c|}{} & 
\multicolumn{2}{|c|}{} \\
0 & 0.32 & ChPT & 2.36 & 0.05 & 0.04 & 0.00 \\ \hline
\end{tabular}
\end{center}
\caption{$\pi^+\pi^-$ contributions to $a_\mu^{\rm had,LO}$ 
  and $\Delta\alpha_{\rm had}(M_Z^2)$ from various energy intervals. 
  The entries in brackets give the contributions
  obtained using the CMD-2 data before re-analysis.}
\label{tab:new180703}
\end{table}
As seen from the Table, if we integrate over the data up to
1.43 GeV, we obtain
\begin{eqnarray}
 a_\mu(\pi^+\pi^-, 0.32< \sqrt{s}<1.43~\GeV)
&=& (  502.78 \pm 5.02 ) \times 10^{-10},  
\label{eq:Table4_1strow} \\
 \Delta \alpha_{\rm had}(\pi^+\pi^-, 0.32< \sqrt{s}<1.43~\GeV)
&=& (  34.39 \pm 0.29 ) \times 10^{-4}.  
\end{eqnarray}
If we integrate up to 2 GeV, instead of 1.43 GeV, we obtain 
\begin{eqnarray}
 a_\mu(\pi^+\pi^-, 0.32< \sqrt{s} < 2~\GeV)
&=& (  503.38 \pm 5.02 ) \times 10^{-10}, 
\label{eq:Table4_3rdrow} \\
 \Delta \alpha_{\rm had}(\pi^+\pi^-, 0.32< \sqrt{s}< 2~\GeV)
&=& (  34.59 \pm 0.29 ) \times 10^{-4}.  
\end{eqnarray}
The contribution of the $\pi^+\pi^-$ channel is dominated by 
the $\rho$-meson, and hence the differences between
Eqs.~(\ref{eq:Table4_1strow}) and (\ref{eq:Table4_3rdrow}) 
is small.  If we use the CMD-2 data before the recent 
re-analysis~\cite{CMD2new}, we have
\begin{eqnarray}
 a_\mu(\pi^+\pi^-, 0.32< \sqrt{s} < 1.43~\GeV, 
{\rm ~old~CMD\mbox{-}2~data})
&=& (  492.66 \pm 4.93 ) \times 10^{-10}, 
\label{eq:Table4_2ndrow} \\
 \Delta \alpha_{\rm had}(\pi^+\pi^-, 0.32< \sqrt{s}< 1.43~\GeV,
{\rm ~old~CMD\mbox{-}2~data})
&=& (  33.65 \pm 0.28 ) \times 10^{-4}.  
\end{eqnarray}
The comparison of (\ref{eq:Table4_1strow}) and 
(\ref{eq:Table4_2ndrow}) shows the
effect of the re-analysis of the recent CMD-2 data,
which is an upward shift of the central value by roughly 2\%
in this interval.

It is interesting to quantify the prominent role of these most precise 
CMD-2  $e^+ e^- \to \pi^+ \pi^-$ data, which have a systematic error of 
only 0.6\%.  If we were to omit these CMD-2 data in the central $\rho$ 
regime altogether, the contribution of this channel to $a_{\mu}$ would 
{\em decrease} by roughly $12.1 \times 10^{-10}$,
i.e., by $\sim 2.4\%$, whereas the error would {\em increase} by 
about $3.4 \times 10^{-10}$, i.e., by $\sim 68\%$ in the
interval $0.32 < \sqrt{s} < 1.43$ GeV.

\begin{figure}
\begin{center}
{\epsfxsize=13cm \leavevmode \epsffile[80 270 490 550]{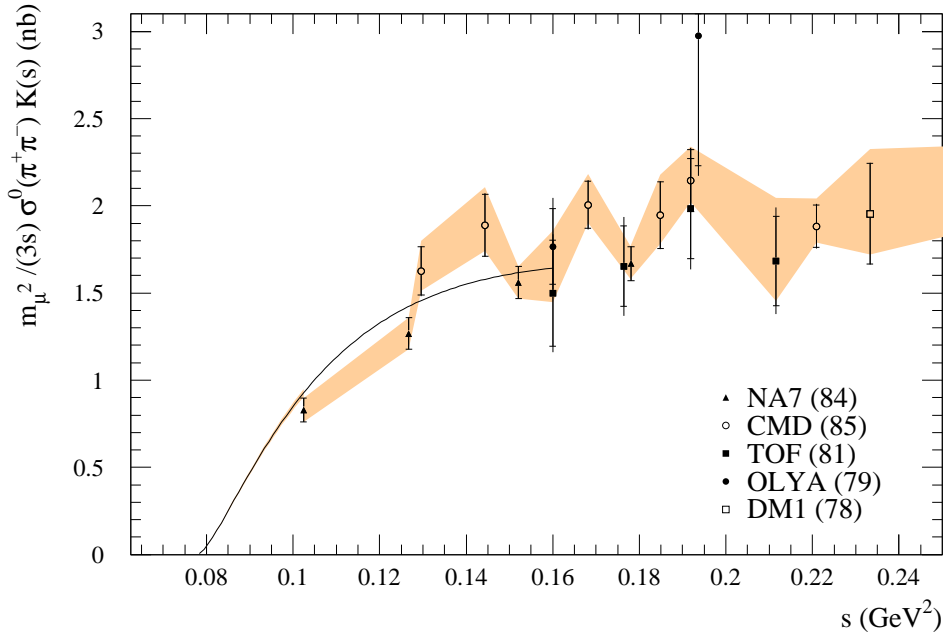}}
\end{center}
\caption{The $e^+e^-\to \pi^+\pi^-$ data near the threshold. 
  In the figure we have multiplied $m^2_\mu /(3s) K(s)$
  by the cross section so that the area below the data is proportional 
  to the contribution to $a_\mu$.  The theoretical curve obtained 
  from chiral perturbation theory is also shown, and is used up to 
  $\sqrt s = 0.32~\GeV$ ($s=0.10~\GeV^2$).} 
\label{fig:pi+pi-_thre}
\end{figure}

In the threshold region, below 0.32~GeV, we use chiral 
perturbation theory, due to the lack of
$\pi^+\pi^-$ experimental data. The pion form factor $F_\pi(s)$ 
is written as
\begin{eqnarray}
 F_\pi(s) = 1 + \frac16 \langle r^2 \rangle_\pi s + c_\pi s^2
  + {\cal O}(s^3) ,
\end{eqnarray}
with coefficients determined to be~\cite{Colangelo:1996hs}
\begin{eqnarray}
  \langle r^2 \rangle_\pi = 0.431 \pm 0.026 ~({\rm fm}^2), ~~~~
  c_\pi    = 3.2 \pm  1.0 ~({\rm GeV}^{-4}),
\end{eqnarray}
by fitting to space-like pion scattering data~\cite{Amendolia:1986wj}. 
Fig.~\ref{fig:pi+pi-_thre} compares the
prediction with the (time-like) experimental data which exist 
for $\sqrt s \geq 0.32~\GeV$.
The contributions from the threshold region are 
\begin{eqnarray}
 a_\mu( \pi^+\pi^-, \sqrt{s} < 0.32~{\rm GeV}) &=& 
     (2.36 \pm 0.05) \times 10^{-10}, \\
\Delta\alpha_{\rm had}( \pi^+\pi^-, \sqrt{s} < 0.32~{\rm GeV}) &=& 
     (0.04 \pm 0.00) \times 10^{-4}, 
\end{eqnarray}
and are also listed in the last row of Table~\ref{tab:new180703}.  
Though these contributions are small, for $a_\mu$ it is 
non-negligible.

In the calculation of the contribution from the threshold region,
we have included the effect from final state (FS) radiative
corrections. In Ref.~\cite{HGJ} both the ${\cal O}(\alpha)$ correction 
and the exponentiated formula for the FS corrections are given.
If we do not apply the FS correction, we would obtain
\begin{eqnarray}
 a_\mu( \pi^+\pi^-, \sqrt{s} < 0.32 {\rm ~GeV}) 
 = (2.30 \pm 0.05) \times 10^{-10}.
\end{eqnarray}
However, if we include the FS corrections, we have
\begin{eqnarray}
 a_\mu( \pi^+\pi^-, \sqrt{s} < 0.32 {\rm ~GeV}, 
        {\cal O}(\alpha) {\rm ~FS~corr.}) 
&=& (2.36 \pm 0.05) \times 10^{-10}. 
\end{eqnarray}
We obtain the same contribution if we use the exponentiated formula, 
which we have used in all the tables in the paper.
The effect of final state radiation is to increase the 
contribution by about 3 \%, whether the ${\cal O}(\alpha)$  
or the exponentiated form is used.  
Similarly, the contribution from this region to 
$\Delta\alpha_{\rm had}(M_Z^2)$ is given by
\begin{eqnarray}
 \Delta\alpha_{\rm had}
 ( \pi^+\pi^-, \sqrt{s} < 0.32 {\rm ~GeV}, {\rm exponentiated~FS~corr.}) 
&=& (0.04 \pm 0.00) \times 10^{-4}, 
\end{eqnarray}
so here the contribution from the threshold region is 
totally negligible.

\subsection{$\pi^+\pi^-\pi^0$ channel}

\begin{figure}
\begin{center}
{\epsfxsize=12cm \leavevmode \epsffile[80 270 460 550]{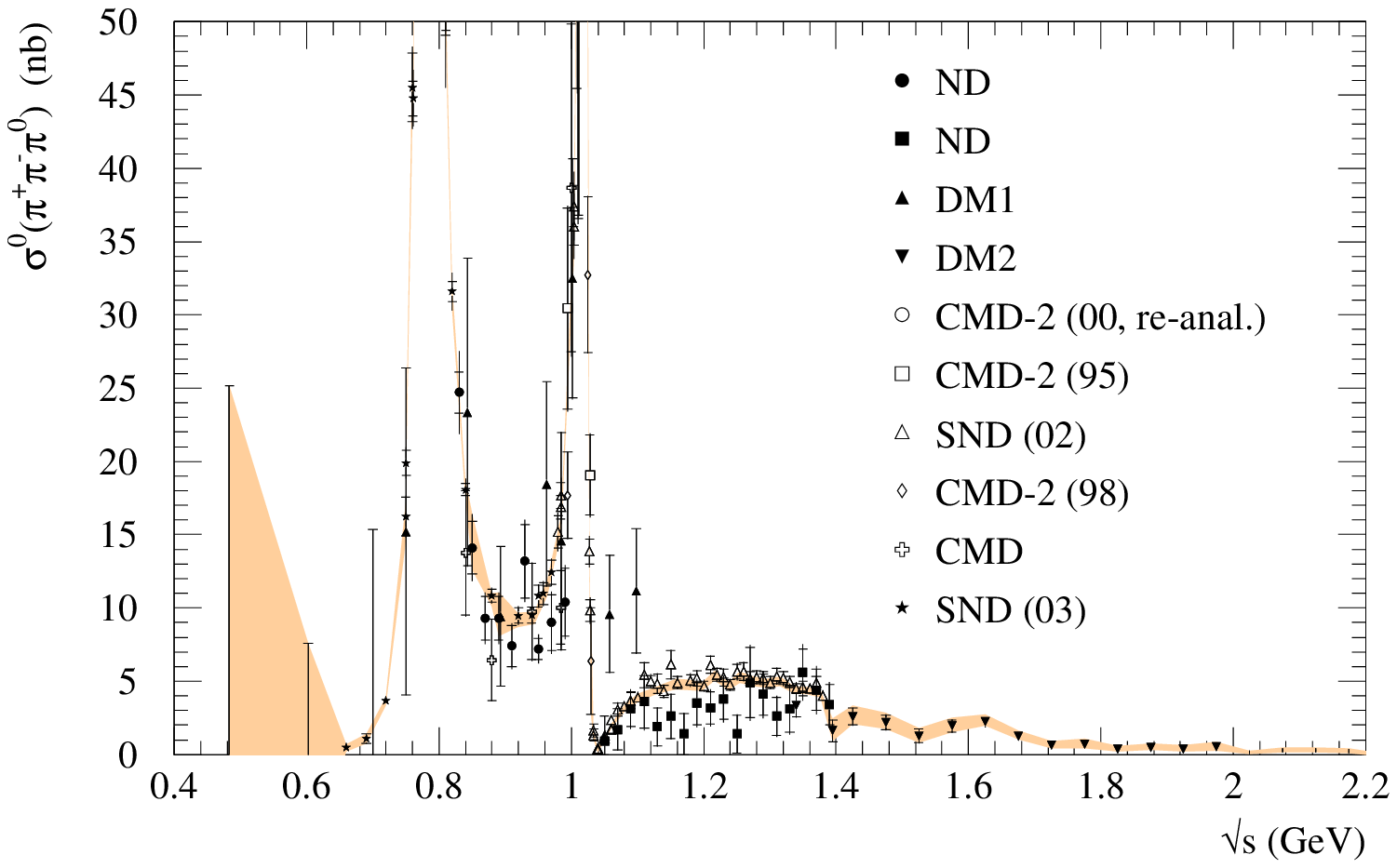}}
{\epsfxsize=12cm \leavevmode \epsffile[80 270 460 550]{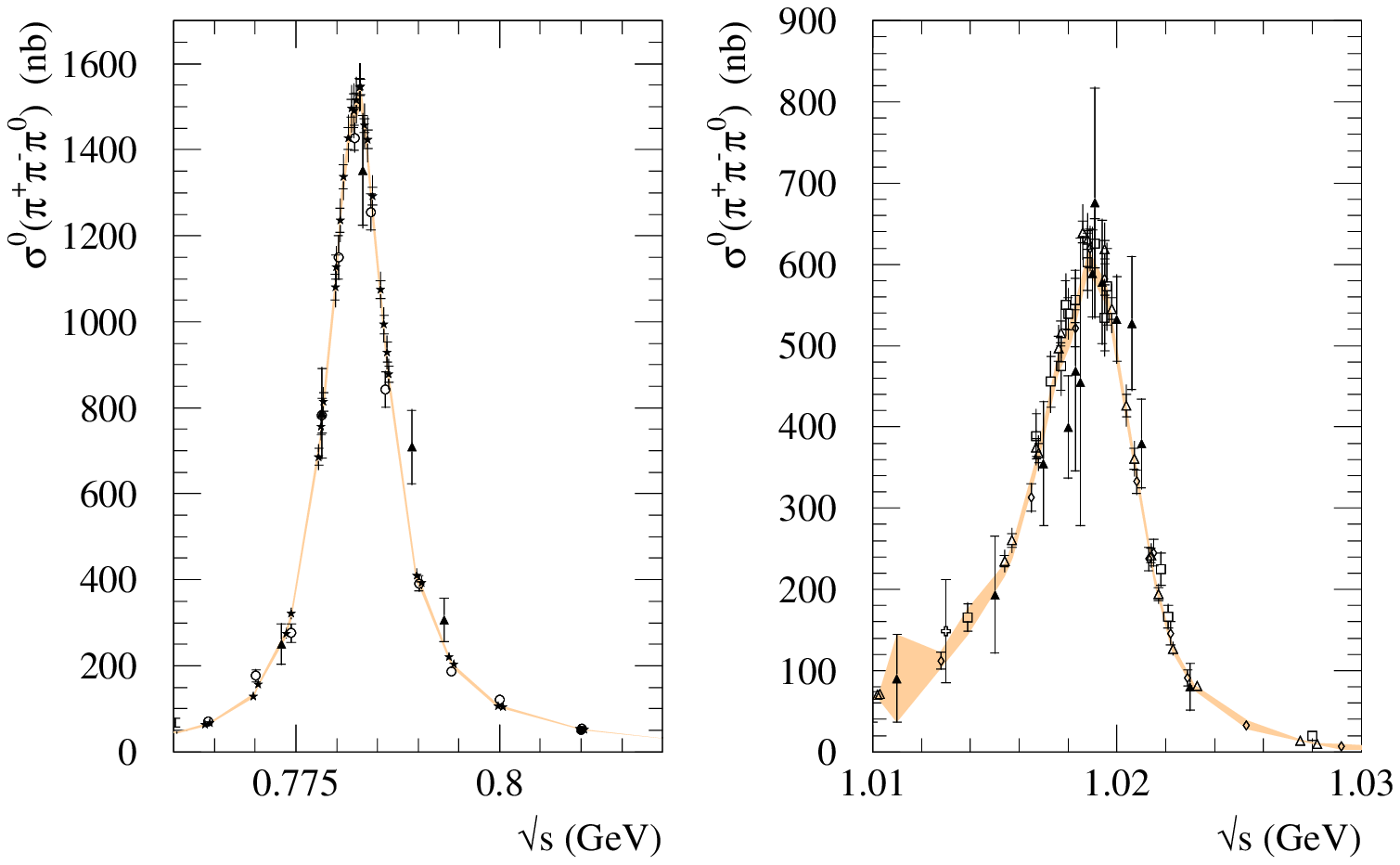}}
\end{center}
\vspace{1.0ex}
\caption{
  The data for $\sigma^0(e^+e^-\to\pi^+\pi^-\pi^0)$ 
  together with an expanded version in the $\omega$ and $\phi$ 
  resonance regions.  The shaded band shows the result of our fit 
  after clustering.  In the analysis we do not use the first two data
  points, below 0.66 GeV, but use chiral perturbation theory as shown
  in Fig.~\ref{fig:pi+pi-pi0_thre}.} 
\label{fig:3pi}
\end{figure}

\begin{figure}
\begin{center}
{\epsfxsize=13cm \leavevmode \epsffile[80 270 490 550]{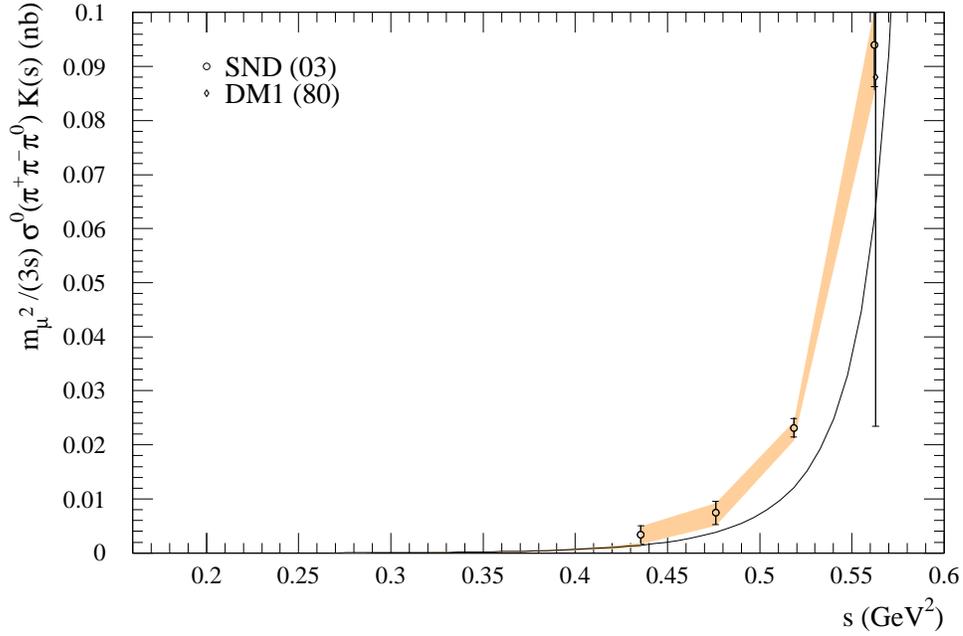}}
\vspace{-2.0ex}
\end{center}
\caption{$e^+e^-\to \pi^+\pi^-\pi^0$ data 
  \cite{Cordier:1980qg, Achasov:2003ir} 
  near the threshold compared with the predictions of chiral
  perturbation theory.  Three measurements \cite{Cordier:1980qg} of 
  zero cross section with very large errors are not shown.
  In the figure we have multiplied $m^2_\mu /(3s) K(s)$
  by the cross section so that the area below the data is proportional 
  to the contribution to $a_\mu$.  The theoretical curve obtained 
  from chiral perturbation theory is used up to 
  $\sqrt s = 0.66~\GeV$ ($s=0.44$~GeV$^2$).}
\label{fig:pi+pi-pi0_thre}
\end{figure}

We use ten experimental data sets for the $\pi^+\pi^-\pi^0$ 
channel~\cite{CMD2new, Akhmetshin:2000ca, Dolinsky:1991vq, 
Akhmetshin:1995vz}, \cite{Cordier:1980qg}--\cite{Barkov:1989}, 
 which extend up to 2.4~GeV, although the
earlier experiments have large errors, see Fig.\ \ref{fig:3pi}. 
Since the data for this channel are not very good, we inflate 
the error by a factor of $\sqrt{\chi^2_{\rm min}/{\rm d.o.f.}}$,
which is 1.20 for this channel.
(We inflate the error by a factor of 
$\sqrt{\chi^2_{\rm min}/{\rm d.o.f.}}$ whenever 
$\chi^2_{\rm min}/{\rm d.o.f.}>1.2$,
as discussed in Section~\ref{sec:combdatasets}, see Table~\ref{tab:clus}.)
We discard the data points below 0.66~GeV, in favour of the predictions 
of chiral perturbation theory~\cite{Kuraev:1995hc, Ahmedov:2002tg},
see Fig.~\ref{fig:pi+pi-pi0_thre}. 
The contributions to $a_\mu^{\rm had,LO}$ and
$\Delta\alpha_{\rm had}(M_Z^2)$ are
\begin{eqnarray}
 a_\mu(\pi^+\pi^-\pi^0,
  0.66{\rm ~GeV}<\sqrt{s}< 1.43{\rm ~GeV}, {\rm ~data})
&=& (46.43 \pm 0.90)  \times 10^{-10} , \\
 \Delta \alpha_{\rm had}(\pi^+\pi^-\pi^0,
  0.66{\rm ~GeV}<\sqrt{s}< 1.43{\rm ~GeV}, {\rm ~data})
&=& (4.33 \pm 0.08)  \times 10^{-4},
\end{eqnarray}
respectively.

In the threshold region, below 0.66~GeV, we use chiral 
perturbation theory~\cite{Kuraev:1995hc, Ahmedov:2002tg}, 
due to the lack of good $\pi^+\pi^-\pi^0$ experimental data, see 
Figs.~\ref{fig:3pi} and \ref{fig:pi+pi-pi0_thre}.
The contributions to $a_\mu^{\rm had,LO}$ and
$\Delta\alpha_{\rm had}(M_Z^2)$ from the threshold region are
\begin{eqnarray}
 a_\mu(\pi^+\pi^-\pi^0, \sqrt{s}< 0.66~{\rm GeV},~{\rm ChPT})
&=& (0.01 \pm 0.00)  \times 10^{-10} , \\
 \Delta \alpha_{\rm had}
 (\pi^+\pi^-\pi^0, \sqrt{s}<0.66~{\rm GeV},~{\rm ChPT})
&=& (0.00 \pm 0.00)  \times 10^{-4}.
\end{eqnarray}
There is a tendency that the ChPT prediction with the 
$\omega$-dominance undershoots the lowest-energy data points.
Because of the smallness of the threshold contribution, we do not
attempt further improvement of the analysis.

\subsection{$\eta\gamma$ channel}
\label{subsec:eval_etagamma}

We use five data sets from SND~\cite{Achasov:2000zd,Achasov:1997nq} 
and CMD-2~\cite{Akhmetshin:1995vz,Akhmetshin:1999zv,Akhmetshin:2001hm}. 
We divide the data set given in Ref.~\cite{Akhmetshin:2001hm} into
two parts at 0.95 GeV since it has different systematic errors 
below and above this energy.

\begin{figure}
\begin{center}
{\epsfxsize=13cm \leavevmode \epsffile[80 270 490 550]{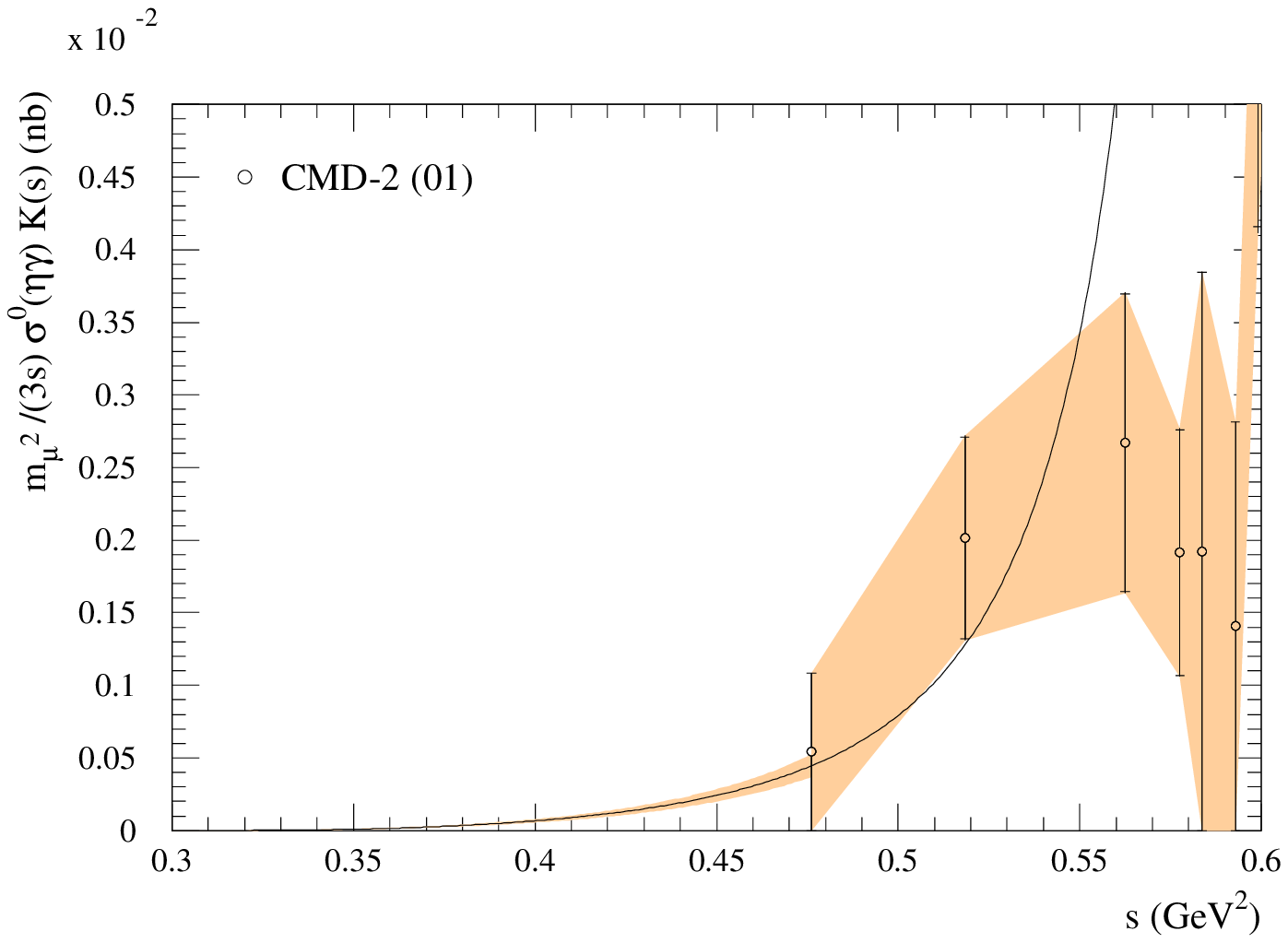}}
\end{center}
\vspace{-2.0ex}
\caption{$e^+e^-\to \eta \gamma$ data near the threshold
  compared with the predictions of chiral perturbation theory.}
\label{fig:etagamma_thre}
\end{figure}

Since the lowest data point starts only at 690 MeV, we use
ChPT at the threshold region up to the
lowest-energy data point.  We summarize our method in Appendix A,
according to which the contribution from the region to 
$a_\mu^{\rm had,LO}$ is less than $10^{-12}$, which can be safely 
neglected.
The contribution to $\Delta\alpha_{\rm had}$ is also small, and
less than $10^{-7}$.
In Fig.~\ref{fig:etagamma_thre} we show the threshold region 
of the $\eta\gamma$ production cross section and our prediction 
from ChPT.

\begin{figure}
\begin{center}
{\epsfxsize=12cm \leavevmode \epsffile[90 115 460 670]{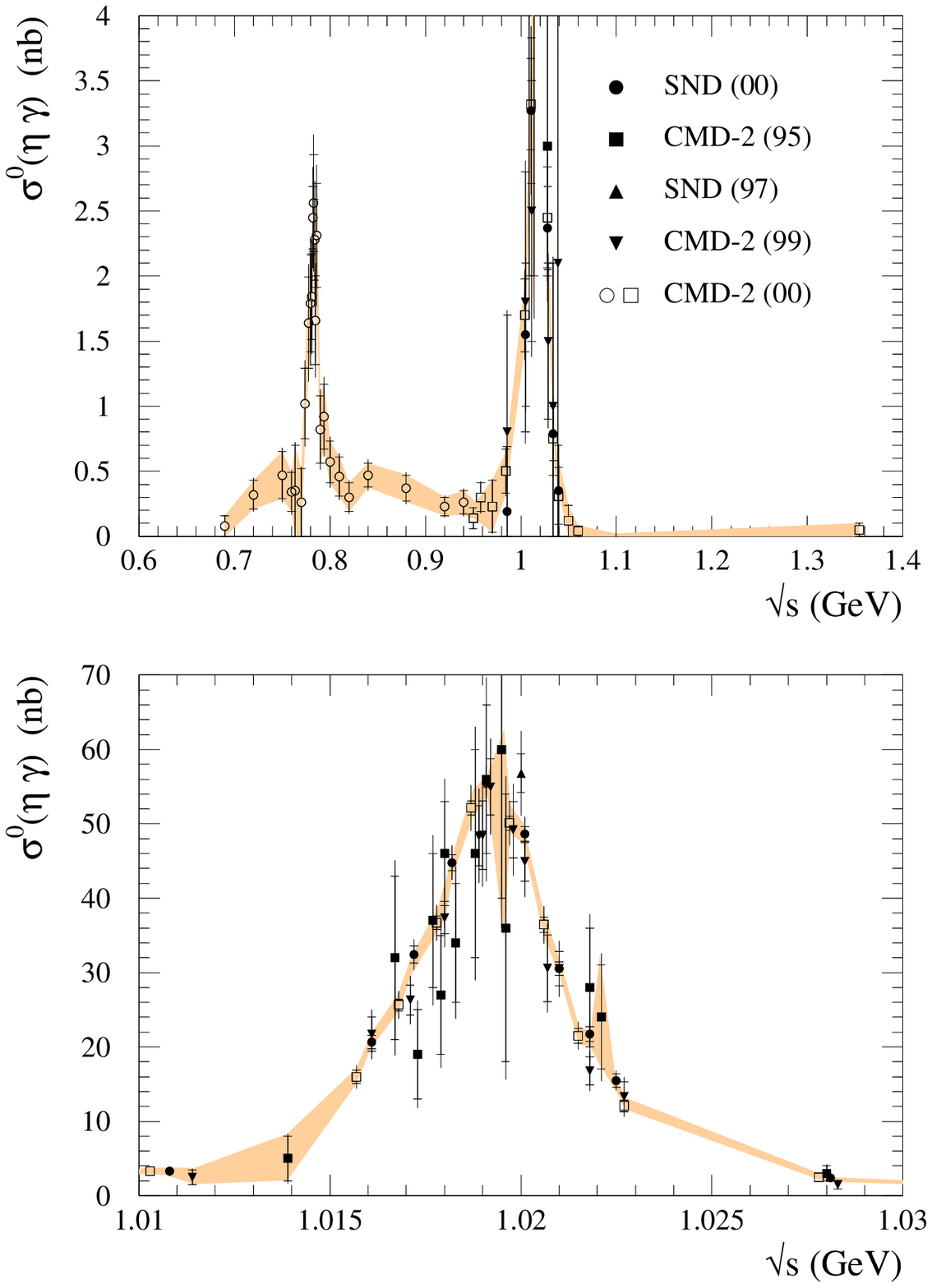}}
\vspace{-6.0ex}
\end{center}
\caption{An overall picture of the $e^+e^-\to \eta \gamma$ data
  together with an enlargement in the region of the $\phi$ resonance.}
\label{fig:etagamma_total}
\end{figure}

Above the lowest-energy data point we integrate over the data.
In Fig.~\ref{fig:etagamma_total} we show the overall
picture of the $\eta\gamma$ production cross section and our result of
the clustering.
After integrating over $0.69<\sqrt{s}<1.43$ GeV we obtain
\begin{eqnarray}
   a_\mu(\eta\gamma, 0.69<\sqrt{s}<1.43~{\rm GeV}) 
 &=& (0.73 \pm 0.03) \times 10^{-10}, \\
   \Delta\alpha_{\rm had}(\eta\gamma, 0.69<\sqrt{s}<1.43~{\rm GeV}) 
 &=& (0.09 \pm 0.00) \times 10^{-4}. 
\end{eqnarray}

\subsection{$4\pi,5\pi,6\pi$ and $\eta\pi^+\pi^-$ channels}
\label{subsec:npi}

For the $4\pi$ channel, we have data for the $2\pi^+2\pi^-$
and the $\pi^+\pi^-2\pi^0$ final states.  (The reaction
$e^+e^- \to \gamma^\ast \to 4\pi^0$ is forbidden from charge
conjugation symmetry.)  

\begin{figure}
\begin{center}
{\epsfxsize=8.2cm \leavevmode \epsfbox[138 270 432 550]{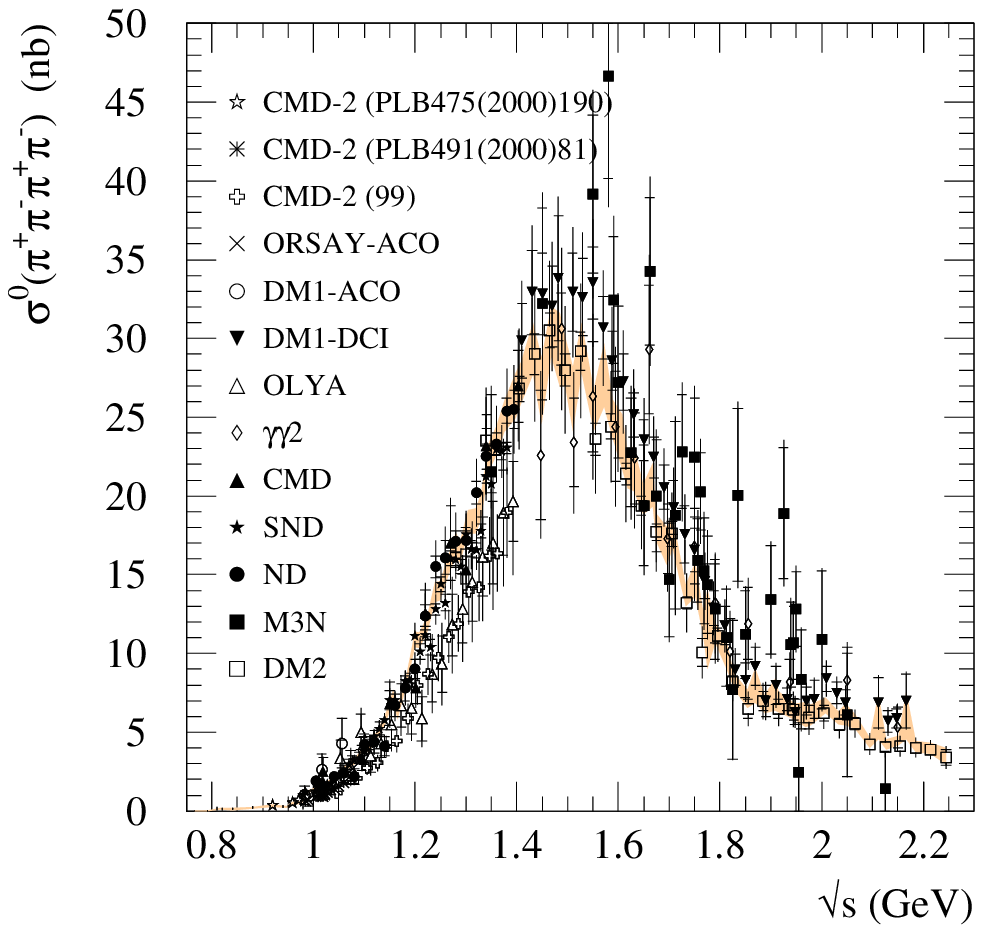}}
{\epsfxsize=8.2cm \leavevmode \epsfbox[138 270 432 550]{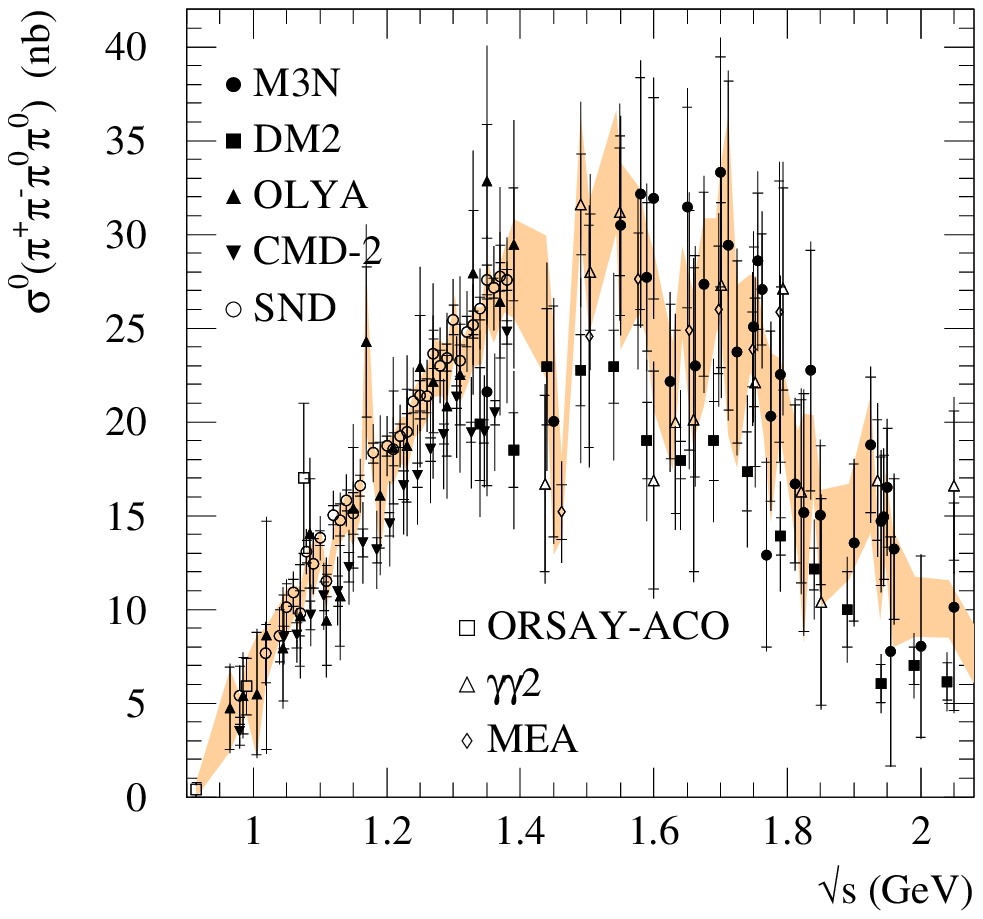}}
\end{center}
\vspace{-3.0ex}
\caption{The data for $\sigma^0(e^+e^-\to 2\pi^+2\pi^-)$ (left) 
  and $\sigma^0(e^+e^-\to \pi^+\pi^-2\pi^0)$ (right).}
\label{fig:4pi}
\end{figure}

For the $2\pi^+2\pi^-$ channel, we use thirteen data 
sets~\cite{Dolinsky:1991vq,Paulot:1979ep,Bisello:1990kh},
\cite{Akhmetshin:1998df}--\cite{Cosme:1976tf}, 
\cite{Barkov:1988gp}--\cite{Akhmetshin:2000it}, 
see Fig.\ \ref{fig:4pi}.
Since the data for this channel are not very consistent to
each other, we inflate 
the error by a factor of $\sqrt{\chi^2_{\rm min}/{\rm d.o.f.}}=1.41$.
We note, in particular, that the compatibility between 
the data from SND and CMD-2 is poor.  This may improve
after the re-analysis of the CMD-2 data for this channel 
is completed~\cite{Fedotovitch}. 
The contribution from this channel is
\begin{eqnarray}
 a_{\mu}(2\pi^+2\pi^-, \sqrt{s}< 1.43~{\rm GeV},~{\rm data}) &=&
  (6.16 \pm 0.32) \times 10^{-10},
\\
 \Delta \alpha_{\rm had} (2\pi^+2\pi^-, \sqrt{s}< 1.43~{\rm GeV},~{\rm data})
&=&
 (1.27 \pm 0.07) \times 10^{-4}.
\end{eqnarray}

For the $\pi^+\pi^-2\pi^0$ channel, we use eight data 
sets~\cite{Paulot:1979ep}--\cite{Esposito:1981dv},
see Fig.\ \ref{fig:4pi}, which contribute
\begin{eqnarray}
 a_{\mu}(\pi^+\pi^-2\pi^0, \sqrt{s}< 1.43~{\rm GeV},~{\rm data}) &=&
  (9.71 \pm 0.63) \times 10^{-10},
\\
 \Delta \alpha_{\rm had} 
(\pi^+\pi^-2\pi^0, \sqrt{s}< 1.43~{\rm GeV},~{\rm data})
&=&
 (1.86 \pm 0.12) \times 10^{-4}.
\end{eqnarray}
For the $\pi^+\pi^-2\pi^0$ channel we have inflated the
error by $\sqrt{\chi^2_{\rm min}/{\rm d.o.f.}}= 1.13$ as discussed
in Section~\ref{sec:combdatasets}.

For the $5\pi$ channel, there exist data for the $2\pi^+ 2\pi^- \pi^0$ 
and the $\pi^+\pi^- 3\pi^0$ final states.
(The reaction $e^+e^- \to \gamma^\ast \to 5\pi^0$ is forbidden 
from charge conjugation symmetry.)  We use
five data sets for the $2\pi^+ 2\pi^- \pi^0$ channel~\cite{
Dolinsky:1991vq,Paulot:1979ep,Bacci:1981zs,Esposito:1981dv,Barkov:1988gp}, 
and one data set for the $\pi^+\pi^- 3\pi^0$ 
channel~\cite{Paulot:1979ep}. We
integrate over the clustered data, which gives
\begin{eqnarray}
 a_{\mu}(2\pi^+2\pi^-\pi^0, \sqrt{s}< 1.43~{\rm GeV},~{\rm data}) &=&
  (0.26 \pm 0.04) \times 10^{-10},
\\
 \Delta \alpha_{\rm had} 
(2\pi^+2\pi^-\pi^0, \sqrt{s}< 1.43~{\rm GeV},~{\rm data})
&=&
 (0.06 \pm 0.01) \times 10^{-4},
\end{eqnarray}
and
\begin{eqnarray}
 a_{\mu}(\pi^+\pi^-3\pi^0, \sqrt{s}< 1.43~{\rm GeV},~{\rm data}) &=&
  (0.09 \pm 0.09) \times 10^{-10},
\\
 \Delta \alpha_{\rm had} 
  (\pi^+\pi^-3\pi^0, \sqrt{s}< 1.43~{\rm GeV},~{\rm data})
&=&
 (0.02 \pm 0.02) \times 10^{-4},
\end{eqnarray}
respectively.  For the $5\pi$ channels we do not inflate the error 
since the $\chi^2_{\rm min}/{\rm d.o.f.}$ values are 
\begin{eqnarray}
 \chi^2_{\rm min}/{\rm d.o.f.} (2\pi^+ 2\pi^- \pi^0) = 0.90,      \\
 \chi^2_{\rm min}/{\rm d.o.f.} (\pi^+ \pi^- 3\pi^0) = 1.07.
\end{eqnarray}

For the $6\pi$ channel, there are data for the $3\pi^+3\pi^-$ and the 
$2\pi^+2\pi^-2\pi^0$ final states, but not
for the $\pi^+\pi^-4\pi^0$ final state.  For the $\pi^+\pi^-4\pi^0$ 
channel we estimate the contribution to $a_\mu$ and 
$\Delta \alpha_{\rm had}$ by using an isospin relation. The 
reaction $e^+e^- \to \gamma^\ast \to 6\pi^0$ is forbidden from charge 
conjugation.

We use four data sets for the $3\pi^+3\pi^-$ channel 
\cite{Paulot:1979ep, Barkov:1988gp, Bisello:1981sh, Schioppa:1981th}. 
M3N~\cite{Paulot:1979ep} provides the lowest data point which is 
at 1.35 GeV, which we do not use since it has unnaturally large cross
section with a large error, $(1.56 \pm 1.11)$ nb, compared with the 
next data point from the same experiment,
$(0.10 \pm 0.31)$ nb at 1.45 GeV. The first data points from 
CMD~\cite{Barkov:1988gp} and DM1~\cite{Bisello:1981sh} contain data 
with vanishing cross section with a finite error, which result in 
points with zero cross section even after clustering.  We do not
use such points when integrating over the data. Thus the first data 
point after clustering is at 1.45 GeV. Our
evaluation of the contribution from the $3\pi^+3\pi^-$ channel 
from the region $\sqrt{s}< 1.43$ GeV 
is zero for both $a_\mu$ and $\Delta \alpha_{\rm had}$.

For the $2\pi^+ 2\pi^- 2\pi^0$ channel we use five data 
sets~\cite{Paulot:1979ep,Bacci:1981zs,Esposito:1981dv,Barkov:1988gp,
Schioppa:1981th}, 
which cover the energy interval from 1.32 GeV to
2.24 GeV. The trapezoidal integration gives us
\begin{eqnarray}
 a_{\mu}(2\pi^+ 2\pi^- 2\pi^0, \sqrt{s}< 1.43~{\rm GeV},~{\rm data}) &=&
  (0.12 \pm 0.03 ) \times 10^{-10},
\\
 \Delta \alpha_{\rm had} (2\pi^+ 2\pi^- 2\pi^0, 
                       \sqrt{s}< 1.43~{\rm GeV},~{\rm data})
&=&
 (0.03 \pm 0.01) \times 10^{-4}.
\end{eqnarray}

For the $\pi^+\pi^-4\pi^0$ channel we use the multipion isospin
decompositions \cite{PAIS,ROUGE} of both the $e^+e^-\to 6\pi$
channel and the $\tau\to 6\pi\nu_\tau$ decays, which are summarised
in the Appendix of Ref.~\cite{BARATE}. Then using the measured
ratio~\cite{CLEOtau} of $\tau^- \to 2\pi^-\pi^+3\pi^0\nu_\tau$ and
$\tau^-\to 3\pi^-2\pi^+\pi^0\nu_\tau$ decays, and the observed
$\omega$ dominance of final states of $\tau\to6\pi\nu_\tau$
decays~\cite{BARATE}, we find
\be 
\sigma(\pi^+\pi^-4\pi^0)\ =\ 
0.031 ~ \sigma(2\pi^+2\pi^-2\pi^0) +
0.093 ~ \sigma(3\pi^+3\pi^-). \label{eq:C} 
\ee
Hence we obtain the small $\pi^+\pi^-4\pi^0$ contribution\footnote{
Relation (\ref{eq:C}) was not used in our previous analysis 
\cite{HMNT}. As a consequence, the (weaker) isospin bound then gave 
a larger contribution for the $\pi^+\pi^- 4\pi^0$ channel. However 
DEHZ~\cite{DEHZ} did use the observed information of $\tau\to
6\pi\nu_\tau$ decays to tighten the isospin bound.} shown in 
Table~\ref{tab:A}.  We assign a 100 \% error to the cross
section computed in this way.  For $a_\mu^{\rm had, LO}$ and 
$\Delta\alpha_{\rm had}$ they are less than $10^{-12}$ and $10^{-6}$,
respectively, when integrated up to 1.43 GeV.

For the $\eta\pi^+ \pi^-$ channel, we use two data 
sets~\cite{Akhmetshin:2000wv,Antonelli:1988fw}.
The entry for the $\eta\pi^+\pi^-$ channel in Table~\ref{tab:A} shows
the contribution of $\sigma(e^+e^-\to\eta\pi^+\pi^-)$ multiplied
by $(1-B(\eta\to 3\pi^0) - B(\eta\to\pi^+\pi^-\pi^0))\simeq 0.448$,
since these $\eta$ decay modes are already included in the contribution
of the $5\pi$ channels.  The contributions to the muon $g-2$
and $\Delta\alpha_{\rm had}$ are
\begin{eqnarray}
 a_{\mu}(\eta (\to\pi^0\gamma)\pi^+\pi^-, \sqrt{s}< 1.43~{\rm GeV}) 
&=& (0.07 \pm 0.01) \times 10^{-10},        \\
 \Delta \alpha_{\rm had} 
(\eta (\to\pi^0\gamma)\pi^+\pi^-, \sqrt{s}< 1.43~{\rm GeV})  
&=& (0.02 \pm 0.00) \times 10^{-4}.
\end{eqnarray}

\subsection{$K^+K^-$ and $K_S K_L$ contributions}
\label{subsec:2K}

\begin{figure}
\begin{center}
{\epsfxsize=12cm \leavevmode \epsffile[90 115 460 670]{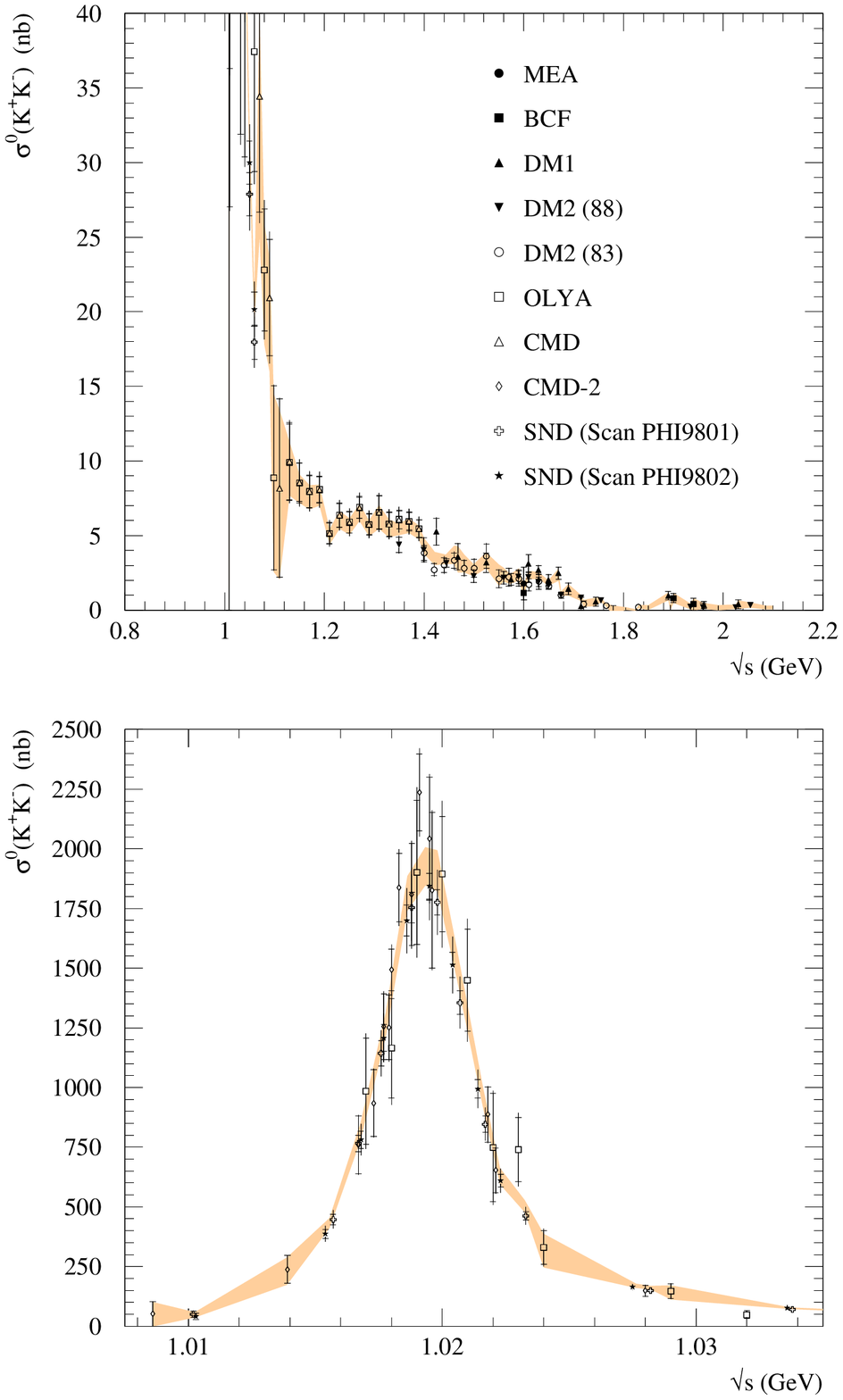}}
\vspace{3.0ex}
\end{center}
\caption{
  The data for $\sigma^0(e^+e^-\to K^+K^-)$ 
  together with an enlargement of the region of the $\phi$ 
  resonance.
  The shaded band shows the result of our fit after clustering.} 
\label{fig:k+k-}
\end{figure}

For the $K^+K^-$ channel, we use ten data sets~\cite{Dolinsky:1991vq,
Bernardini:1973pe,Esposito:1977xg,Akhmetshin:1995vz}, 
\cite{Ivanov:1981wf}--\cite{Achasov:2001ni}, which extend from 
1.0 GeV to 2.1 GeV, see Fig.~\ref{fig:k+k-}.  When integrated, this
channel contributes to the muon $g-2$ and $\Delta \alpha_{\rm had}$ 
an amount
\begin{eqnarray}
 a_{\mu}(K^+K^-, \sqrt{s}< 1.43~{\rm GeV},~{\rm data}) &=&
  (21.62 \pm 0.76) \times 10^{-10},
\\
 \Delta \alpha_{\rm had} 
(K^+K^-, \sqrt{s}< 1.43~{\rm GeV},~{\rm data}) &=&
 (3.01 \pm 0.11) \times 10^{-4}.
\end{eqnarray}

\begin{figure}
\begin{center}
{\epsfxsize=12cm \leavevmode \epsffile[90 115 460 670]{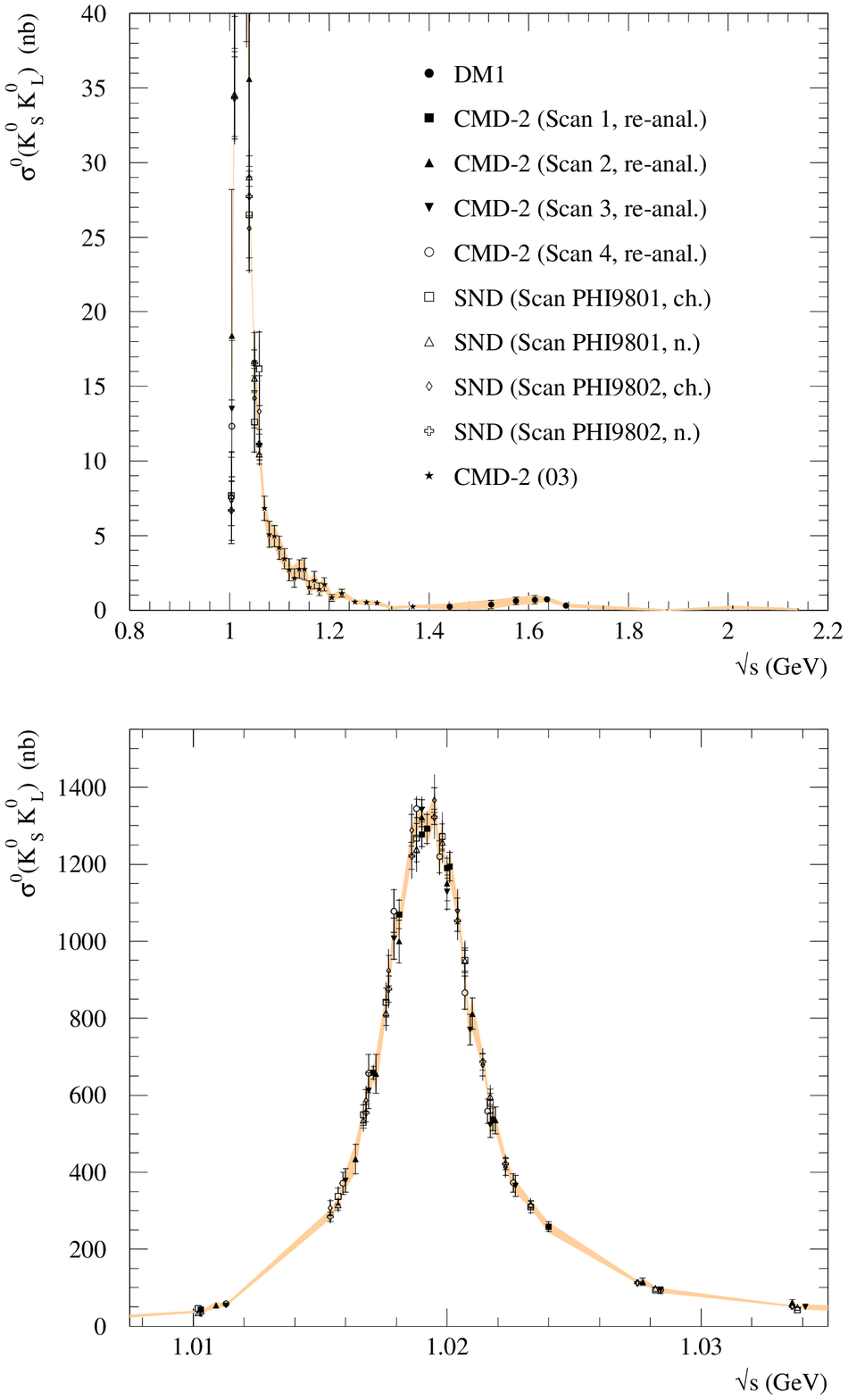}}
\vspace{3.0ex}
\end{center}
\caption{The data for $\sigma^0(e^+e^-\to K_S^0K_L^0)$ 
  together with an enlargement of the region of the $\phi$ 
  resonance.
  The shaded band shows the result of our fit after clustering;
  however, the errors on the contribution of this channel to $a_\mu$
  and $\Delta\alpha_{\rm had}$ are increased to allow for the lack of
  data in certain regions of the $\phi$ resonance tails, see the
  discussion in the text.} 
\label{fig:kskl}
\end{figure}

For the $K_S^0K_L^0$ channel, we use ten data 
sets~\cite{CMD2new,Akhmetshin:1999ym,Achasov:2001ni,Mane:1981ep,
Akhmetshin:2002vj}, which also 
extend from 1.0~GeV to  2.1~GeV, see Fig.~\ref{fig:kskl}.  Using the
trapezoidal rule, the channel gives a contribution to the muon 
$g-2$ and $\Delta \alpha_{\rm had}$ of
\begin{eqnarray}
 a_{\mu}(K^0_SK^0_L, \sqrt{s}< 1.43~{\rm GeV},~{\rm data}) &=&
  (13.16 \pm 0.31) \times 10^{-10},
\label{eq:a_mu_KsKl}
\\
 \Delta \alpha_{\rm had} (K^0_SK^0_L, \sqrt{s}< 1.43~{\rm GeV},~{\rm data})
 &=& (1.76 \pm 0.04) \times 10^{-4}.
\label{eq:alpQED_KsKl}
\end{eqnarray}
This channel is the one case where the use of the trapezoidal rule 
may overestimate the resonance contribution, due to the lack of data
in certain regions of the $\phi$ resonance tails, see Fig.\
\ref{fig:kskl}.  We find that the use of a smooth resonance form in
the tails decreases the contributions to $a_\mu$ and 
$\Delta\alpha_{\rm had}$ by about $0.15\times 10^{-10}$ and
$0.02\times 10^{-4}$ respectively.  We have therefore increased the
error in (\ref{eq:a_mu_KsKl}) and (\ref{eq:alpQED_KsKl}) to include 
this additional uncertainty.

\subsection{$K\bar K + n\pi$ contributions}

We take into account the $K\bar K + n\pi$ final states for 
$n=1$ and 2. 

For the $K\bar K \pi$, in addition to the data for the 
$K_S^0\pi^\pm K^\mp$~\cite{Bisello:1991du,Bisello:1991kd,Mane:1982si} 
and $K^+K^-\pi^0$~\cite{Bisello:1991du,Bisello:1991kd} channels, 
we use the equalities
$\sigma(K_L^0\pi K) = \sigma(K_S^0 \pi K)$ and 
$\sigma(K_S^0K_L^0\pi^0) = \sigma(K^+K^-\pi^0)$, which follow
directly from isospin.  The contribution from the 
$K_S^0 \pi^\pm K^\mp + K^0_L \pi^\pm K^\mp$ channel is
\begin{eqnarray}
 a_{\mu}
  (K_S^0 \pi^\pm K^\mp + K^0_L \pi^\pm K^\mp, \sqrt{s}< 1.43~{\rm GeV},
 ~{\rm data~and~isospin}) =
  (0.10 \pm 0.04) \times 10^{-10},&&
\\
 \Delta \alpha_{\rm had} 
  (K_S^0 \pi^\pm K^\mp + K^0_L \pi^\pm K^\mp, \sqrt{s}< 1.43~{\rm GeV},
 ~{\rm data~and~isospin}) =
  (0.02 \pm 0.00) \times 10^{-4}.&&
\end{eqnarray}
For the $K^+ K^- \pi^0 + K^0_SK^0_L \pi^0$ channel, the contribution
from the region $\sqrt{s}< 1.43$ GeV is taken to be zero since the first 
data point is at 1.44 GeV.

To evaluate the $K\bar K\pi\pi$ contribution we use the inclusive 
data for $K_SX$~\cite{Mane:1982th}, 
together with the cross section relation
\bea 2K_SX & = & K_SX + K_LX  \nonumber \\
& = & 2K_SK_L + 2(K_SK_L + K_SK_S + K_LK_L)(\pi + \pi\pi) \nonumber\\
& & \qquad\qquad\qquad\qquad\qquad\qquad + (K_S + K_L)(K\pi + K\pi\pi), 
\label{eq:crosssectionrelation} \eea
where $2K_SX$ stands for 2$\sigma^0(e^+e^- \to K_S X)$ and 
similarly for the other abbreviations.
On the right-hand-side $\pi\pi$ stands for $\pi^+\pi^-$ or 
$\pi^0\pi^0$, $K\pi$ for $K^+\pi^-$ or $K^-\pi^+$,
and $K\pi\pi$ for $K^+\pi^-\pi^0$ or $K^-\pi^+\pi^0$. On the other 
hand, the $K\bar K\pi\pi$ cross section is given by
\bea K\bar K\pi\pi & = & (K_SK_L + K_SK_S + K_LK_L)(\pi\pi) + (K_S +
K_L)(K\pi\pi) + (K^+K^-)(\pi\pi) \nonumber \\
 & = & 2K_SX - 2K_SK_L - (K_SK_L + K_SK_S + K_LK_L)(2\pi + \pi\pi)
 \nonumber \\
& & \qquad\qquad\qquad\qquad\qquad\qquad\qquad - 2K_S(K\pi) +
(K^+K^-)(\pi\pi) \nonumber \\
& = & 2(K_SX - K_SK_L - K^+K^-\pi - K_S(K\pi)), \label{eq:KKpipi} \eea
where to obtain the second equality we have used 
(\ref{eq:crosssectionrelation}). In other words, the total $K\bar
K \pi\pi$ contribution is obtained from twice the inclusive 
$K_SX$ cross section by subtracting the appropriate $K\bar
K$ and $K\bar K \pi$ contributions.
For this channel, the contribution
from the region $\sqrt{s}< 1.43$ GeV is also taken to be zero since 
the data of the $K_S^0 X$ final state start from 1.44 GeV.

\subsection{Unaccounted modes}

We still have to take into account contributions from the reactions 
$e^+e^- \to \omega \pi^0$ and $e^+e^-\to \omega \pi^+ \pi^-$, in 
which the $\omega$ decays radiatively into $\pi^0\gamma$. 
We used seven data sets for the $e^+e^- \to \omega \pi^0$
channel~\cite{Dolinsky:1991vq,Bisello:1990kh,Akhmetshin:1998df},
\cite{Akhmetshin:2003ag}--\cite{Dolinsky:1986kj},
and three data sets~\cite{Antonelli:1992jx,Akhmetshin:2000wv,
Cordier:1981zs} for the $e^+e^- \to \omega \pi^+ \pi^-$ channel.  
Note that the contributions from the $\omega (\to\pi^+ \pi^-\pi^0)
\pi^0$ and $\omega (\to \pi^+ \pi^-) \pi^0$ channels are already
included as a part of the multi-pion channels.  We therefore need 
simply to multiply the original cross section 
$\sigma(e^+e^- \to \omega\pi^0)$ by the 
branching ratio $B(\omega\to \pi^0\gamma) = 0.087$~\cite{PDG2002}. 
The same comments apply for the $\omega \pi^+\pi^-$ channel. 
The two channels give contributions
\begin{eqnarray}
  a_{\mu}(\omega (\to \pi^0 \gamma) \pi^0,
\sqrt{s}< 1.43~{\rm GeV}) &=&
  (0.64 \pm 0.02) \times 10^{-10},
\\
 \Delta \alpha_{\rm had} (\omega(\to \pi^0\gamma ) \pi^0,
\sqrt{s}< 1.43~{\rm GeV}) &=&
 (0.12 \pm 0.00) \times 10^{-4},
\end{eqnarray}
and
\begin{eqnarray}
  a_{\mu}(\omega (\to \pi^0 \gamma) \pi^+ \pi^-,
\sqrt{s}< 1.43~{\rm GeV}) &=&
  (0.01 \pm 0.00 ) \times 10^{-10},
\\
 \Delta \alpha_{\rm had} (\omega(\to \pi^0\gamma ) \pi^+ \pi^-,
\sqrt{s}< 1.43~{\rm GeV}) &=&
 (0.00 \pm 0.00) \times 10^{-4},
\end{eqnarray}
respectively.

Purely neutral contributions from the direct decays of $\rho$ and $\omega$
to $\pi^0 \pi^0 \gamma$ can be safely neglected, as the branching fractions
are of the order $5 \times 10^{-5}$ and $7 \times 10^{-5}$ respectively
\cite{PDG2002,Akhmetshin:2003rg,Achasov:2002jv}, and are 
suppressed compared to the decays into $\pi^0 \gamma$.

For the $\phi$ resonance we have so far accounted for the 
$\phi\to K^+K^-$, $K_S^0K_L^0$, $3\pi$, $\eta\gamma$ and
$\pi^0\gamma$ channels.  Since the branching fractions of these final states 
add up to 99.8\%~\cite{PDG2002}, we must allow for the 0.2\% from
the remaining final states.  To do this, we first note that the
contribution to $a_\mu^{\rm had,LO}$ from the $K^+K^-$ channel in 
the $\phi$ region is 
\begin{eqnarray}
 a_\mu(\phi\to K^+K^-; 
2m_{K^+} <\sqrt{s}< 1.03~{\rm GeV}) =
  16.15 \times 10^{-10}.
\end{eqnarray}
Using this, we estimate that the total contribution from the $\phi$ to be 
$$a_\mu(\phi)=a_\mu(\phi\to K^+K^-)/B(\phi\to K^+K^-) 
= 32\times 10^{-10}.$$  
Hence we include the small residual contribution 
\begin{eqnarray}
a_\mu(\phi \to {\rm remaining~channels})
=a_\mu(\phi)\times 0.002=0.06\times 10^{-10}, 
\end{eqnarray}
and assign to it a 100\% error.  
In a similar way the contribution  
$\Delta\alpha_{\rm had}(\phi\to K^+K^-)=2.12\times10^{-4}$ 
is used to estimate
\begin{eqnarray}
\Delta\alpha_{\rm had}(\phi \to {\rm remaining~channels})
=0.01\times 10^{-4},
\end{eqnarray}
to which we again assign a 100\% error.  

\subsection{Baryon-pair contribution}

If we are to integrate up to high enough energy to pair-produce 
baryons, we have to take into account the $p\bar{p}$ and $n\bar{n}$ 
final states.  The data come from the FENICE~
\cite{Antonelli:1998fv,Antonelli:1994kq}, 
DM1~\cite{Delcourt:1979ed} and DM2~\cite{Bisello:1983at,Bisello:1990rf}
collaborations for the $p\bar{p}$ channel, and from the 
FENICE collaboration~\cite{Antonelli:1998fv,Antonelli:1993vz}  
for the $n\bar{n}$ channel.  They do not contribute when 
we integrate over the exclusive channels only up to 1.43 GeV, but if 
we integrate up to 2.0 GeV, the $p\bar{p}$ channel gives a contribution of
\begin{eqnarray}
  a_{\mu}(p\bar{p}, \sqrt{s}< 2.0~{\rm GeV}) &=&
 (0.04 \pm 0.01 )  \times 10^{-10},  \\
 \Delta \alpha_{\rm had} (p\bar{p}, \sqrt{s}< 2.0~{\rm GeV})
&=&
 (0.02 \pm 0.00) \times 10^{-4},
\end{eqnarray}
while the $n\bar{n}$ channel gives
\begin{eqnarray}
  a_{\mu}(n\bar{n}, \sqrt{s}< 2.0~{\rm GeV}) &=&
 (0.07 \pm 0.02 )  \times 10^{-10}, \\
 \Delta \alpha_{\rm had} (n\bar{n}, \sqrt{s}< 2.0~{\rm GeV})
&=&
 (0.03 \pm 0.01) \times 10^{-4}.
\end{eqnarray}

\subsection{Narrow resonance ($J/\psi,\psi',\Upsilon$) contributions}

We add the contributions from the narrow resonances,
$J/\psi, \psi^\prime$ and $\Upsilon(1S - 6S)$.
We treat them in the zero-width approximation, in which
the total production cross section of a vector meson $V$
$(V=J/\psi, \psi^\prime, \Upsilon)$ is
\begin{eqnarray}
 \sigma(e^+e^- \to V) = 12 \pi^2 \frac{\Gamma^0_{ee}}{M_V}
                        \delta(s-M_V^2).
\end{eqnarray}
Here $\Gamma^0_{ee}$ is the bare leptonic width of $V$,
\begin{eqnarray}
 \Gamma^0_{ee} = C_{\rm res} \Gamma(V \to e^+ e^-),
\end{eqnarray}
where
\begin{eqnarray}
 C_{\rm res} = \frac{(\alpha/\alpha(m_V^2))^2}{1+(3/4)\alpha/\pi},
\end{eqnarray}
which is about 0.95 for $J/\psi$ and $\psi^\prime$, and about
0.93 for the six $\Upsilon$ resonances~\cite{minireviewRPP}. 
We use the values compiled in RPP for the leptonic widths, 
$\Gamma(V \to e^+e^-)$, and obtain the contributions
\begin{eqnarray}
 a_\mu(J/\psi)      &=& (5.89 \pm 0.41 )\times 10^{-10}, \\
 a_\mu(\psi^\prime) &=& (1.41 \pm 0.12 )\times 10^{-10}, \\
 a_\mu(\Upsilon(1S))    &=& (0.05 \pm 0.00 )\times 10^{-10}, \\
 a_\mu(\Upsilon(2S-6S)) &=& (0.05 \pm 0.00 )\times 10^{-10},
\end{eqnarray}
and
\begin{eqnarray}
 \Delta \alpha_{\rm had}(J/\psi)      &=& (6.65 \pm 0.47 )\times 10^{-4}, \\
 \Delta \alpha_{\rm had}(\psi^\prime) &=& (2.25 \pm 0.19 )\times 10^{-4}, \\
 \Delta \alpha_{\rm had}(\Upsilon(1S))    
    &=& (0.54 \pm 0.02 )\times 10^{-4}, \\
 \Delta \alpha_{\rm had}(\Upsilon(2S-6S)) 
    &=& (0.62 \pm 0.03 )\times 10^{-4}.
\end{eqnarray}

\subsection{Inclusive hadronic data contribution 
($\sqrt s < 11.09~\GeV$)}

\begin{figure}
\begin{center}
{\epsfxsize=8.2cm \leavevmode \epsfbox[138 270 432 550]{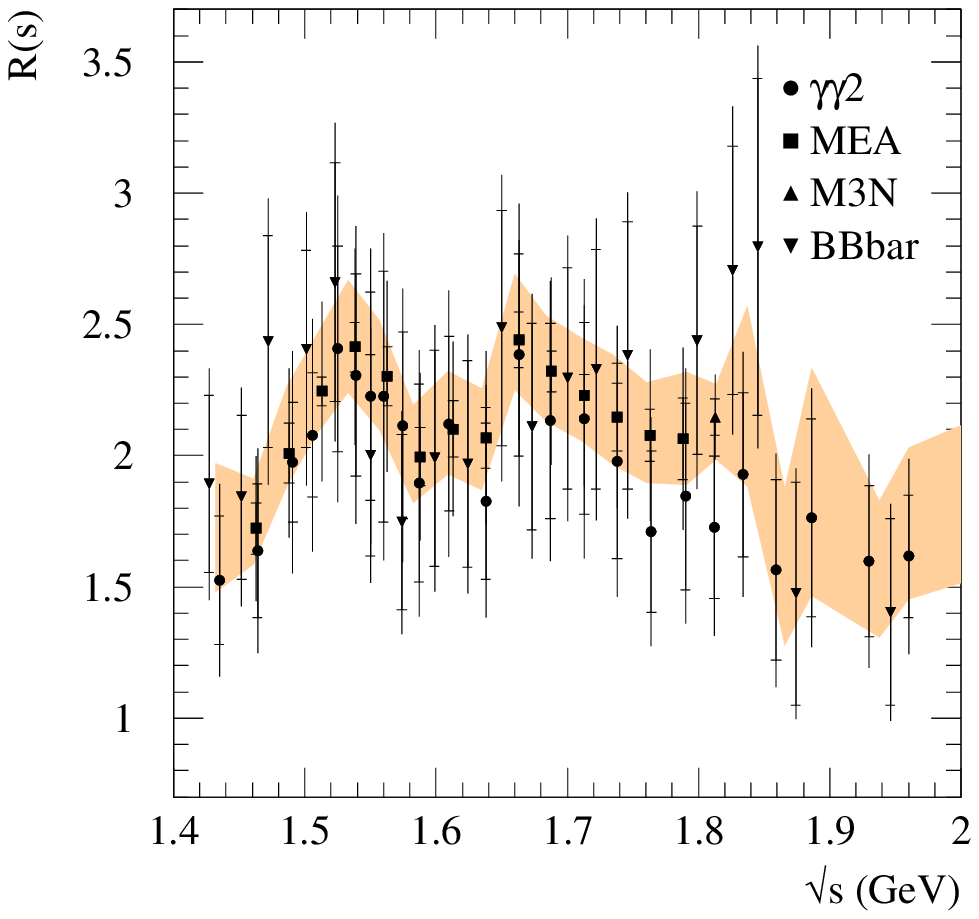}}
{\epsfxsize=8.2cm \leavevmode \epsfbox[138 270 432 550]{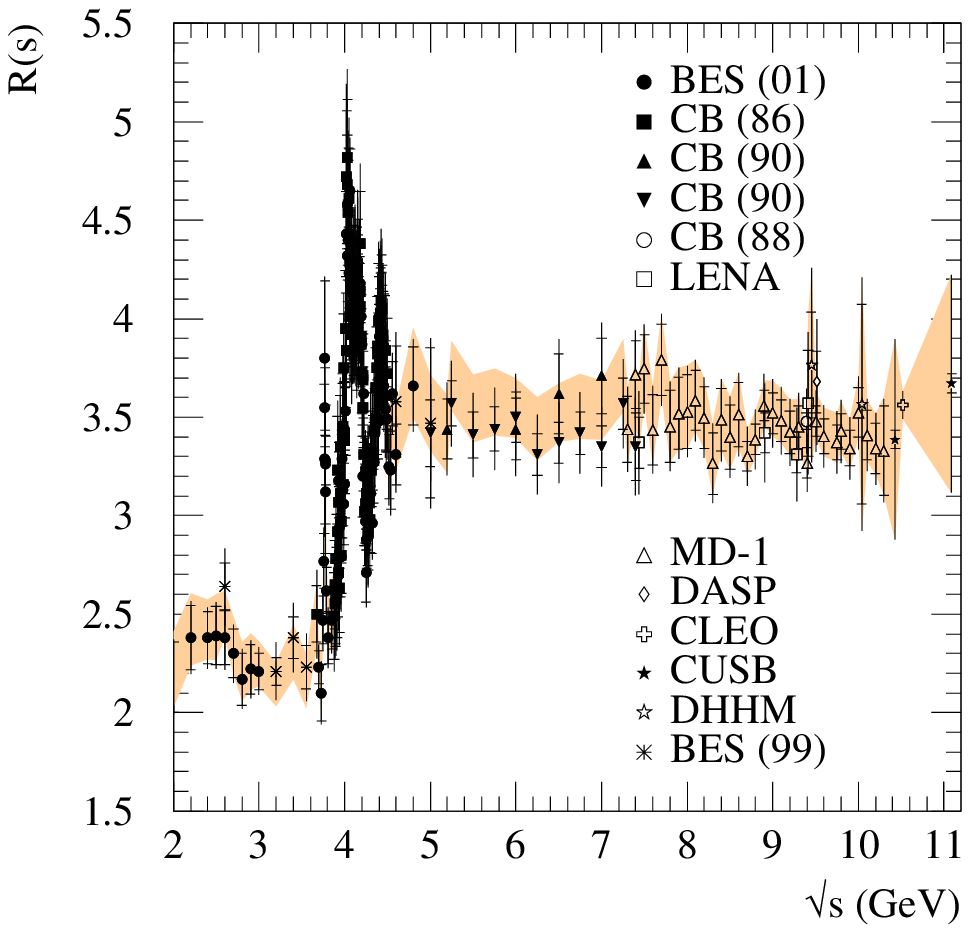}}
\vspace{-3.0ex}
\end{center}
\caption{Data for the measurement of the inclusive hadronic
  cross section below 2 GeV (left) and above 2 GeV (right).
  The shaded band shows the behaviour of the hadronic $R$ ratio
  after clustering and fitting the data.}
\label{fig:incl}
\end{figure}

We use four data sets below 
2~GeV~\cite{Bacci:1979ab}--\cite{Ambrosio:1980mf}, and twelve data 
sets above 2~GeV~\cite{Bai:1999pk}--\cite{Bock:1980ag} 
(see Fig.~\ref{fig:incl}). Below 2~GeV, we correct for the 
unaccounted modes.  Namely, we add the contributions from 
the $\omega (\to \pi^0\gamma) \pi^0$ and $K_S^0(\to 2\pi^0) K_L^0 \pi^0$
channels to the experimentally observed $R$-ratio, since the final 
states of these channels consist only of
electrically neutral particles, which are hard to see experimentally.
They shift the $R$ values by roughly 1\%, depending on $\sqrt{s}$.
In addition we correct some experiments for the contributions from 
missing two-body final states, as discussed at the end of Section 2.
We have also checked that corrections for $\gamma - Z$ interference
effects are completely negligible in the energy range below 11.09 GeV
where we use data.

The contributions to the muon $g-2$ and $\Delta\alpha_{\rm had}$ are,
from $1.43<\sqrt{s}<2$~GeV, 
\begin{eqnarray}
  a_{\mu}({\rm inclusive}, \sqrt{s}< 2~{\rm GeV}) &=&
 ( 31.91 \pm 2.42 )  \times 10^{-10}, \\
 \Delta \alpha_{\rm had} ({\rm inclusive}, \sqrt{s}< 2~{\rm GeV})
&=& (10.78 \pm 0.81) \times 10^{-4},
\end{eqnarray}
and from $2<\sqrt{s}<11.09$~GeV,
\begin{eqnarray}
  a_{\mu}({\rm inclusive}, 2<\sqrt{s}< 11.09~{\rm GeV}) &=&
 (42.05 \pm 1.14 )  \times 10^{-10}, \\
 \Delta \alpha_{\rm had} ({\rm inclusive}, 2<\sqrt{s}< 11.09~{\rm GeV})
&=& (81.97 \pm 1.53) \times 10^{-4},
\end{eqnarray}
respectively.

\subsection{Inclusive pQCD contribution ($\sqrt s > 11.09~\GeV$)}

Above 11 GeV we use perturbative QCD to evaluate the contributions 
to $a^{\rm had,LO}_\mu$ and $\Delta\alpha(M_Z^2)$. 
We incorporate ${\cal O}(\alpha_S^3)$ massless quark contributions, 
and the ${\cal O}(\alpha_S^2)$ massive quark 
contributions~\cite{CKK,massless,CHKST2,CKS,HJKT2}.  
We have checked that our code agrees very well with the 
code {\tt rhad} written by Harlander and Steinhauser~\cite{HAR}.
As input parameters, we use
\begin{eqnarray}
 \alpha_S(M_Z^2) = 0.1172 \pm 0.002, ~~~
 m_t = 174.3 \pm 5.1~{\rm GeV}, ~~~  m_b = 4.85 \pm 0.25~{\rm GeV},
\end{eqnarray}
and allow for an uncertainty in the renormalization scale 
of $\sqrt{s}/2 < \mu < 2 \sqrt{s}$. Here $m_t$ and $m_b$
are the pole masses of the top and bottom quark.  We obtain
\begin{eqnarray}
 a_\mu({\rm pQCD}, \sqrt{s}>11.09~{\rm GeV}) 
 &=& (2.11 \pm 0.00) \times  10^{-10},
\end{eqnarray}
where the uncertainty from $\alpha_S(M_Z^2)$
is dominant, which is less than $1\times 10^{-12}$.
Similarly, for $\Delta\alpha_{\rm had}$ we find
\begin{eqnarray}
 \Delta\alpha_{\rm had}({\rm pQCD}, \sqrt{s}>11.09~{\rm GeV}) 
 &=& (125.32 \pm 0.14 \pm 0.02 \pm 0.01) \times 10^{-4} \\
 &=& (125.32 \pm 0.15) \times  10^{-4},
\end{eqnarray}
where the first error comes from the uncertainty in
$\alpha_S(M_Z^2)$, the second from the renormalization scale
$\mu$, and the third from that on the mass of the bottom quark.

\subsection{Total contribution to the dispersion integrals}

\begin{figure}
\begin{center}
\psfig{file=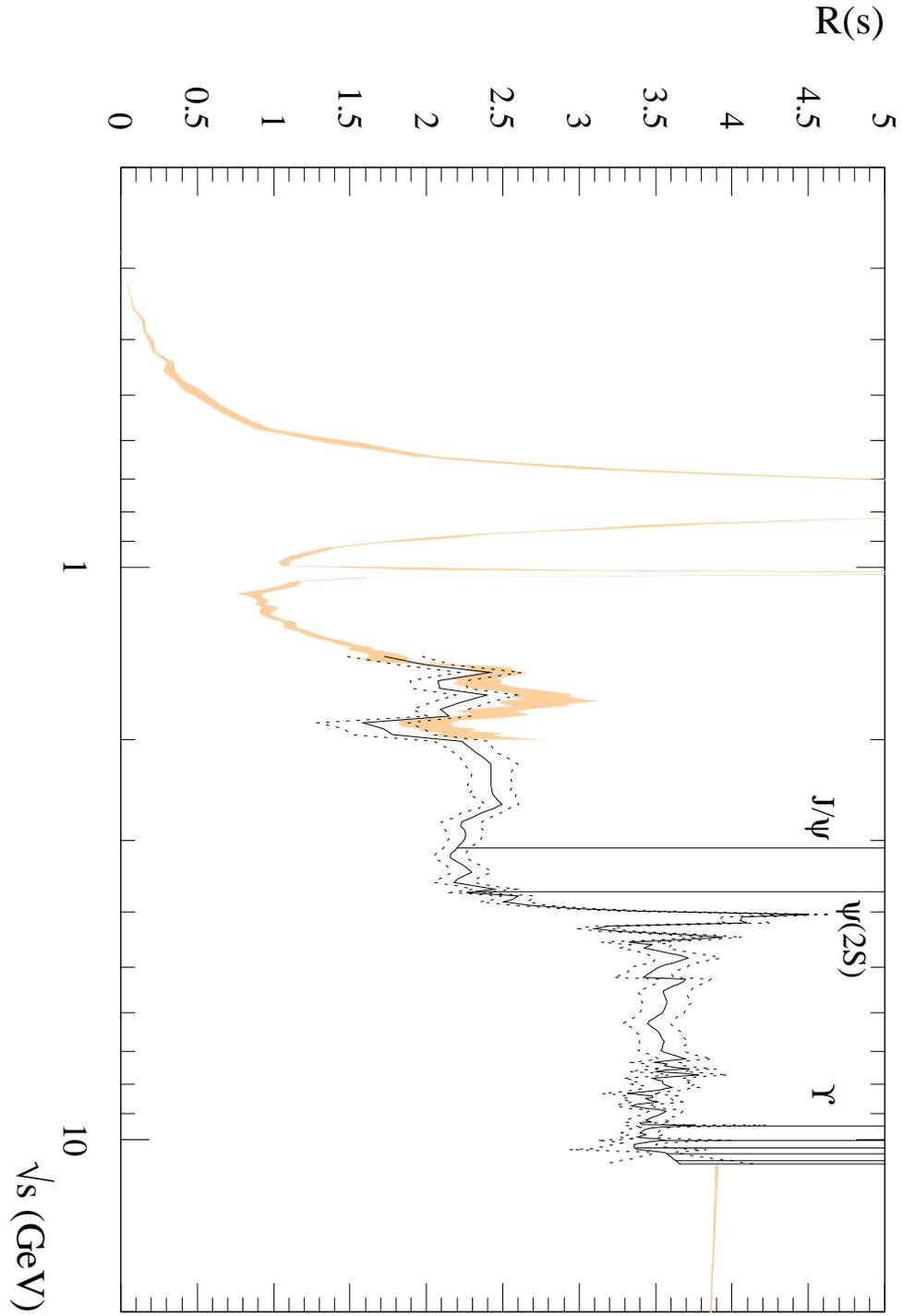, width=22cm, angle=270}
\end{center}
\vspace{-8.0ex}
\caption{The hadronic $R$ ratio as a function of $\sqrt{s}$.
  Note that the values of $R$ obtained from the sum of the
  exclusive channels and from the inclusive data overlap in the
  region $1.4 \lsim \sqrt{s} \lsim 2$  GeV.}
\label{fig:Rhad}
\end{figure}

To summarize, Table~\ref{tab:A} shows the values obtained 
for $a_\mu^{\rm had,LO}$ and $\Delta\alpha_{\rm had}$,
as well as showing the contributions of the individual channels. 
Summing all the contributions we obtain
\begin{eqnarray}
 a_\mu^{\rm had,LO}({\rm incl.}) 
    &=& (692.38 \pm 5.88_{\rm exp})\times 10^{-10},\label{eq:amuhadloincl}\\
 a_\mu^{\rm had,LO}({\rm excl.}) 
    &=& (696.15 \pm 5.68_{\rm exp})\times 10^{-10},\label{eq:amuhadloexcl}
\end{eqnarray}
where ``incl.'' means that we have used the inclusive data sets
for $1.43<\sqrt{s}<2~\GeV$, while ``excl.'' means that we used the 
exclusive data at the same interval.  ``exp.'' means that
the errors are from the experimental uncertainty. 
The corresponding results for $\Delta\alpha_{\rm had}$ are
\begin{eqnarray}
 \Delta\alpha_{\rm had}({\rm incl.}) &=& 
                (275.52 \pm 1.85_{\rm exp})\times 10^{-4},\\
 \Delta\alpha_{\rm had}({\rm excl.}) &=& 
                (276.90 \pm 1.77_{\rm exp})\times 10^{-4}.
\end{eqnarray}
We see that using the sum of the data for
exclusive channels to determine $R(s)$, in the intermediate energy 
interval $1.43<\sqrt{s}<2~\GeV$, yields values for
$a_\mu^{\rm had,LO}$ and $\Delta\alpha_{\rm had}$ which 
significantly exceed the values obtained using the
inclusive data for $R(s)$.  The mean values differ by about
$2/3$ of the total experimental error.  In Fig.\ \ref{fig:Rhad}
we show the hadronic $R$ ratio as a function of $\sqrt{s}$.
A careful inspection of the figure shows the discrepancy between 
the inclusive and exclusive data sets in the interval 
$1.43<\sqrt{s}<2~\GeV$, see Fig.~\ref{fig:incl-excl}.
The contribution from this region alone is
\begin{eqnarray}
 a_\mu^{\rm had,LO}(1.43 < \sqrt{s}< 2~\GeV, {\rm ~incl.}) 
    &=& (31.91 \pm 2.42_{\rm exp})\times 10^{-10},\\
 a_\mu^{\rm had,LO}(1.43 < \sqrt{s}< 2~\GeV, {\rm ~excl.}) 
    &=& (35.68 \pm 1.71_{\rm exp})\times 10^{-10},
\end{eqnarray}
and for $\Delta\alpha_{\rm had}$,
\begin{eqnarray}
 \Delta\alpha_{\rm had}(1.43 < \sqrt{s} < 2~\GeV, {\rm ~incl.}) &=& 
                (10.78 \pm 0.81_{\rm exp})\times 10^{-4},\\
 \Delta\alpha_{\rm had}(1.43 < \sqrt{s} < 2~\GeV, {\rm ~excl.}) &=& 
                (12.17 \pm 0.59_{\rm exp})\times 10^{-4}.
\end{eqnarray}
In the next Section we introduce QCD sum 
rules that are able to determine which choice
of $R(s)$ is consistent. We find that the sum rules strongly 
favour the use of the inclusive data in the above  
intermediate energy interval.

Table~\ref{tab:hadroniccontr} shows the breakdown of the contributions 
versus energy. It is also useful to show
the breakdown visually in terms of `pie' diagrams.
\begin{table}[htb]
\begin{center}
\begin{tabular}{|c|c|c|c|}
\hline
energy range (GeV) & comments & $a^{\rm had, LO}_\mu \times 10^{10}$ 
\rule[-1.5ex]{0ex}{4.5ex} &
$\Delta\alpha_{\rm had}(M_Z^2)\times10^4$ \\
\hline
$m_{\pi}$--0.32  & ChPT & $   2.36 \pm 0.05 $ & $  0.04 \pm 0.00$ \\
0.32--1.43 & excl.\ only &$ 606.55 \pm 5.22 $ & $ 47.34 \pm 0.35$  \\
1.43--2 & incl.\ only &   $  31.91 \pm 2.42 $ & $ 10.78 \pm 0.81$  \\
           & (excl.\ only& $ 35.68 \pm 1.71 $ & $ 12.17 \pm 0.59$) \\
2--11.09 & incl.\ only &   $ 42.05 \pm 1.14 $ & $ 81.97 \pm 1.53$  \\
$J/\psi$ and $\psi'$ & narrow width & $ 7.30 \pm 0.43 $  & $8.90\pm 0.51$ \\
$\Upsilon(1S-6S) $ & narrow width & $ 0.10 \pm 0.00 $ & $1.16\pm 0.04$ \\
11.09--$\infty $ & pQCD & $ 2.11 \pm 0.00 $ & $125.32 \pm 0.15$\\
\hline
 Sum of all & incl. 1.43--2 & $ 692.38 \pm 5.88$ & $ 275.52 \pm 1.85$  \\
           & (excl. 1.43--2 & $ 696.15 \pm 5.68$ & $ 276.90 \pm 1.77$) \\
\hline
\end{tabular}
\caption{A breakdown of the contributions to different 
  intervals of the dispersion
  integrals for $a_\mu^{\rm had,LO}$ and $\Delta\alpha_{\rm had}(M_Z^2)$. 
  The alternative numbers for the interval
  $1.43<\sqrt{s}<2$~GeV correspond to using data for either the sum of 
  the exclusive channels or the inclusive
  measurements, see Fig.~\ref{fig:incl-excl}.}
\label{tab:hadroniccontr} 
\end{center}
\end{table}
\begin{figure} \begin{center}
{\psfig{file=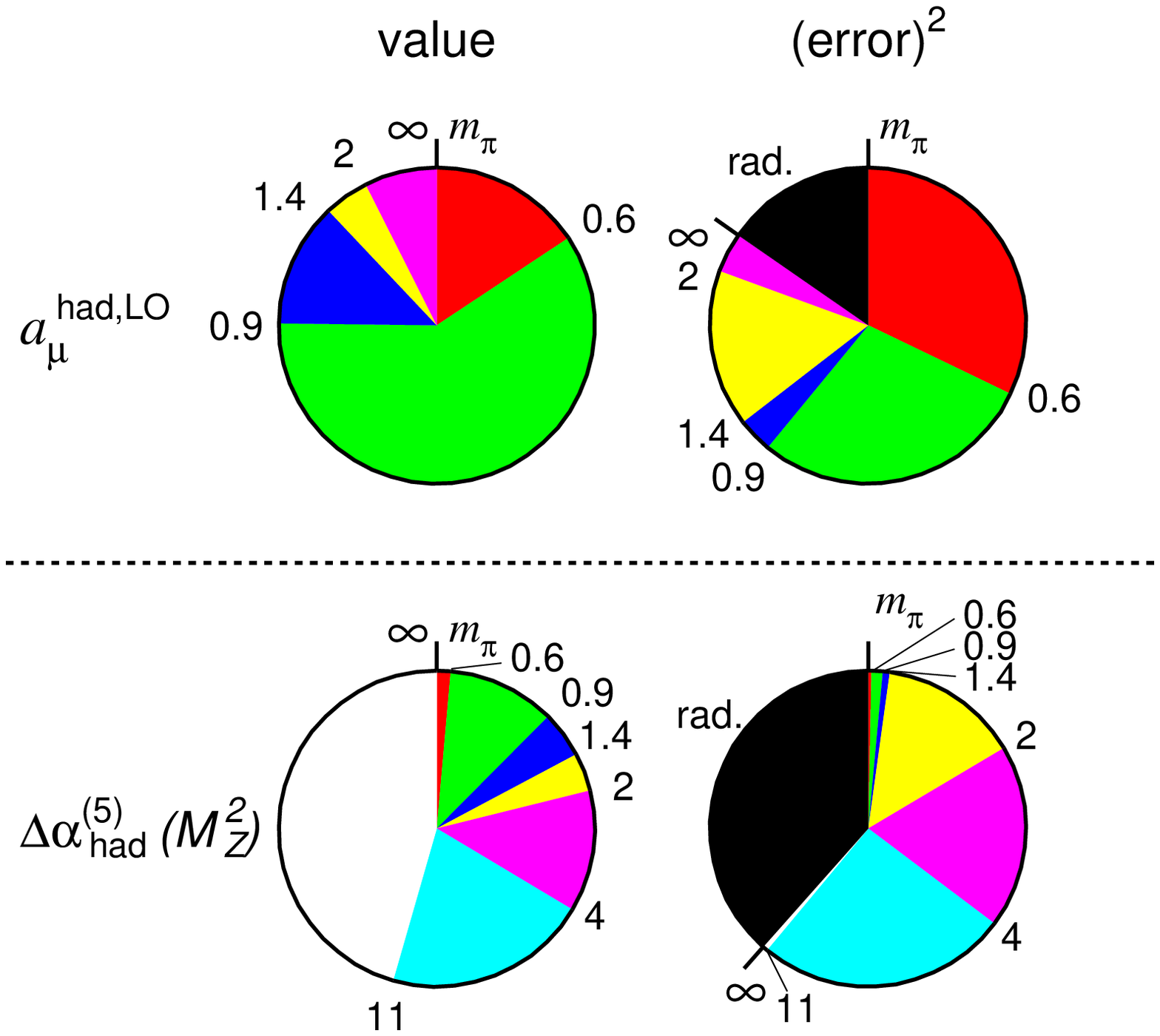,width=17cm}}
\end{center}   \vspace{-10.0ex}
\caption{The pie diagrams in the left- and right-hand columns show 
  the fractions of the total contributions and $({\rm errors})^2$,
  respectively, coming from various energy intervals in the dispersion 
  integrals (\ref{eq:disprel1}) and (\ref{eq:disprel2}). The pie 
  diagrams for the LO hadronic contribution to $g-2$, shown in 
  the first row, correspond to sub-contributions with energy boundaries at
  $m_\pi, 0.6, 0.9, 1.4, 2~\GeV {\rm ~and}~\infty$, whereas for the
   hadronic contribution to the QED coupling, shown in the second row, 
  the boundaries are at 
  $m_\pi, 0.6, 0.9, 1.4, 2, 4, 11.09~\GeV {\rm ~and}~\infty$.
  In the $({\rm error})^2$ pie diagrams we also included the
  $({\rm error})^2$ arising from the treatment of the radiative corrections
  to the data.}
\label{fig:pie}
\end{figure}
The `pie' diagrams on the left-hand side of Fig.~\ref{fig:pie} show the 
fraction of the total contributions to
$a_\mu^{\rm had,LO}$ and $\Delta \alpha_{\rm had}$ coming from 
various energy intervals of the dispersion
integrals (\ref{eq:disprel1}) and (\ref{eq:disprel2}). The plots 
on the right-hand-side indicate the fractional contributions to 
the square of the total error,
including the error due to the treatment of radiative corrections.
The values shown for $a_\mu^{\rm had,LO}$ in these plots
correspond to using the inclusive data in the intermediate energy interval.

In Section~\ref{sec:calculation} we use the value of 
$a_\mu^{\rm had,LO}$, along with the QED, weak and other
hadronic contributions, to predict the value of $g-2$ of the muon. 
In Section~\ref{sec:determination} we use the value of 
$\Delta\alpha_{\rm had}(M_Z^2)$ to predict the value of the QED 
coupling on the $Z$ pole, $\alpha(M_Z^2)$.

\section{Resolution of the ambiguity: QCD sum rules}
\label{sec:QCDsumrules}

To decide between the exclusive and inclusive data in the energy 
range $1.43<\sqrt{s}<2$~GeV (see Fig.~\ref{fig:incl-excl}), we make 
use of QCD sum rules~\cite{SVZ}, see also the review~\cite{SHIF}. 
The sum rules are based on the analyticity of the vacuum polarization 
function $\Pi(q^2)$, from which it follows that a relation of the form
\begin{eqnarray}
  \int_{s_{\rm th}}^{s_0}\ {\rm d}s\ R(s)f(s)
\ =\ \int_C\ {\rm d}s\ D(s)g(s)
  \label{eq:QCDsumrule}
\end{eqnarray}
must be satisfied for a non-singular function $f(s)$. $C$ is a
circular contour of radius $s_0$ and $g(s)$ is a known function
once $f(s)$ is given.  The lower limit of integration, $s_{\rm
th}$, is $4m_{\pi}^2$, except for a small $e^+e^-\to\pi^0\gamma$
contribution.  $D(s)$ is the Adler $D$ function,
\begin{eqnarray}
  D(s)\ \equiv\
  -12\pi^2s\frac{{\rm d}}{{\rm d}s}\left(\frac{\Pi(s)}{s}\right),
  \qquad {\rm where} \quad R(s)\ =\ \frac{12\pi}{s}\,{\rm Im}\, \Pi(s).
  \label{eq:AdlerDfunction}
\end{eqnarray}
Provided that $s_0$ is chosen sufficiently large for $D(s)$ to be
evaluated from QCD, the sum rules allow consistency checks of the
behaviour of the data for $R(s)$ for $s<s_0$. Indeed, by choosing
an appropriate form of the function $f(s)$ we can highlight the
average behaviour of $R(s)$ over a particular energy domain. To be
specific, we take $s_0$ just below the open charm threshold (say
$\sqrt s_0 = 3.7\ \GeV$) and choose forms for $f(s)$ which
emphasize the most ambiguous range
($1.5\lesim\sqrt{s}\lesim2$~GeV) of $R(s)$, so that the
discriminating power of the sum rules is maximized. We therefore
use the three flavour ($n_f=3$) QCD expressions for $D(s)$, and
omit the $J/\psi$ and $\psi(2S)$ $c\bar c$ resonance contributions
to $R(s)$.

To evaluate the function $D(s)$ from QCD, it is convenient to
express it as the sum of three contributions,
\begin{eqnarray}
  D(s)\ =\ D_0(s) + D_{\rm m}(s) + D_{\rm np}(s),
\end{eqnarray}
where $D_0$ is the $O(\alpha_S^3)$ massless, three-flavour QCD
prediction, $D_{\rm m}$ is the (small) quark mass correction and
$D_{\rm np}$ is a (very small) contribution estimated using
knowledge of the condensates.
$D_0$ is given by~\cite{massless}
\begin{eqnarray}
  D_0 (-s) =
3 \sum_f Q_f^2
\left\{
 1 +     \frac{\alpha_S(s)}{\pi}
   + d_1 \left(\frac{\alpha_S(s)}{\pi}\right)^2
   + \tilde{d}_2 \left(\frac{\alpha_S(s)}{\pi}\right)^3
   + {\cal O}\left( \alpha_S^4(s) \right)
\right\} ,
\end{eqnarray}
with
\begin{eqnarray}
 d_1 &=& 1.9857 - 0.1153 n_f, \\
 \tilde{d}_2 &=& d_2 + \frac{\beta_0^2 \pi^2}{48} ~~~~~
 \left({\rm with\ }\beta_0 = 11- \frac{2 n_f}{3} \right), \\
 d_2 &=& - 6.6368 - 1.2001 n_f - 0.0052 n_f^2
         - 1.2395 \frac{(\sum_f Q_f)^2}{3 \sum_f Q_f^2} ,
\end{eqnarray}
where the sum $f$ runs over $u, d$ and $s$ flavours.  $Q_f$ is the
electric charge of quark $f$, which takes the values 2/3, $-1/3$,
and $-1/3$ for $u, d$ and $s$, respectively.
The quark mass correction $D_{\rm m}$ reads~\cite{Chetyrkin:1990kr}
\begin{eqnarray}
 D_{\rm m}(-s) =
- 3 \sum_f Q_f^2 \frac{m_f^2(s)}{s}
 \left(
    6 + 28 \frac{\alpha_S(s)}{\pi}
      + (294.8 - 12.3 n_f) \left( \frac{\alpha_S(s)}{\pi} \right)^2
 \right) .
\end{eqnarray}
We take the $\overline{\rm MS}$ $s$-quark mass at 2~GeV 
$m_s(4~\GeV^2)$ to be $120 \pm 40$~MeV, and we neglect the $u$ and $d$
quark masses. The contribution from condensates, $D_{\rm np}$, is
given by
\begin{eqnarray}
 D_{\rm np}(-s) &=&
3 \sum_f Q_f^2
 \left\{
\frac{2\pi^2}{3}
 \left( 1 - \frac{11}{18} \frac{\alpha_S(s)}{\pi} \right)
\frac{\langle (\alpha_S/\pi) GG \rangle}{s^2}
\right.       \nonumber\\
&&
\phantom{3 \sum_f Q_f^2}
+ 8\pi^2
 \left( 1 - \frac{\alpha_S(s)}{\pi} \right)
 \frac{\langle m_f \overline{q}_f q_f \rangle}{s^2}
+ \frac{32\pi^2}{27} \frac{\alpha_S(s)}{\pi}
  \sum_k \frac{\langle m_k \overline{q}_k q_k \rangle}{s^2}    \nonumber\\
&& \left. \phantom{3 \sum_f Q_f^2} + 12\pi^2  \frac{\langle {\cal
O}_6 \rangle}{s^3} + 16\pi^2  \frac{\langle {\cal O}_8
\rangle}{s^4} \right\} ,
\end{eqnarray}
where, following~\cite{DH98a}, we take
\begin{eqnarray}
 \left\langle \frac{\alpha_S}{\pi} GG \right\rangle
&=& 0.037 \pm 0.019\ ({\rm GeV}^4),
 \nonumber\\
 \langle m_s \bar{s} s \rangle &=& - f_\pi^2  m_K^2.
\end{eqnarray}
Here $f_\pi\simeq 92$~MeV is the pion decay constant, and $m_K$ is
the kaon mass. As we will see later, the quark mass corrections and
the condensate contributions are very tiny---typically at most a
few percent of the whole QCD contribution. Hence we neglect the
higher dimensional condensates, $\langle {\cal O}_6 \rangle$ and
$\langle {\cal O}_8 \rangle$.

As for the weight function $f(s)$, we take it to be of the form
$(1-s/s_0)^m(s/s_0)^n$ with $n+m=0,1$ or $2$. For these six
choices of $f(s)$, the function $g(s)$ may be readily evaluated,
and the sum rules, (\ref{eq:QCDsumrule}), become
\begin{eqnarray}
%
%
 \int_{s_{\rm th}}^{s_0} {\rm d}s  R(s)
&=&
  \frac{i}{2\pi} \int_C  {\rm d}s
  \left\{
      1 - \frac{s_0}{s}
  \right\}  D(s)  ,           \label{eq:m=0n=0sumrule} \\
%
%
 \int_{s_{\rm th}}^{s_0} {\rm d}s  R(s) \frac{s}{s_0}
&=&
  \frac{i}{2\pi} \int_C  {\rm d}s
  \frac12
  \left\{
      \frac{s}{s_0} - \frac{s_0}{s}
  \right\}  D(s)  ,             \label{eq:m=0n=1sumrule} \\
%
%
 \int_{s_{\rm th}}^{s_0} {\rm d}s  R(s) \left( 1 - \frac{s}{s_0} \right)
&=&
  \frac{i}{2\pi} \int_C  {\rm d}s
  \left\{
     - \frac12 \frac{s}{s_0}
     + 1
     - \frac12 \frac{s_0}{s}
  \right\} D(s)    ,         \label{eq:m=1n=0sumrule} \\
%
%
 \int_{s_{\rm th}}^{s_0} {\rm d}s  R(s) \left( \frac{s}{s_0} \right)^2
&=&
  \frac{i}{2\pi} \int_C  {\rm d}s
  \frac13
  \left\{
      \left( \frac{s}{s_0} \right)^2 - \frac{s_0}{s}
  \right\}   D(s)   ,        \label{eq:m=0n=2sumrule} \\
%
%
 \int_{s_{\rm th}}^{s_0} {\rm d}s  R(s)
     \left( 1 - \frac{s}{s_0} \right) \frac{s}{s_0}
&=&
  \frac{i}{2\pi} \int_C  {\rm d}s
  \left\{
     - \frac13 \left( \frac{s}{s_0} \right)^2
     + \frac12 \frac{s}{s_0}
     - \frac16 \frac{s_0}{s}
  \right\} D(s)      ,    \label{eq:m=1n=1sumrule} \\
%
%
 \int_{s_{\rm th}}^{s_0} {\rm d}s  R(s)
     \left( 1 - \frac{s}{s_0} \right)^2
&=&
  \frac{i}{2\pi} \int_C  {\rm d}s
  \left\{
       \frac13 \left( \frac{s}{s_0} \right)^2
     - \frac{s}{s_0}
     + 1
     - \frac13 \frac{s_0}{s}
  \right\} D(s)       .  \label{eq:m=2n=0sumrule}
\end{eqnarray}

\begin{figure}[ht]
\begin{center}
{\psfig{file=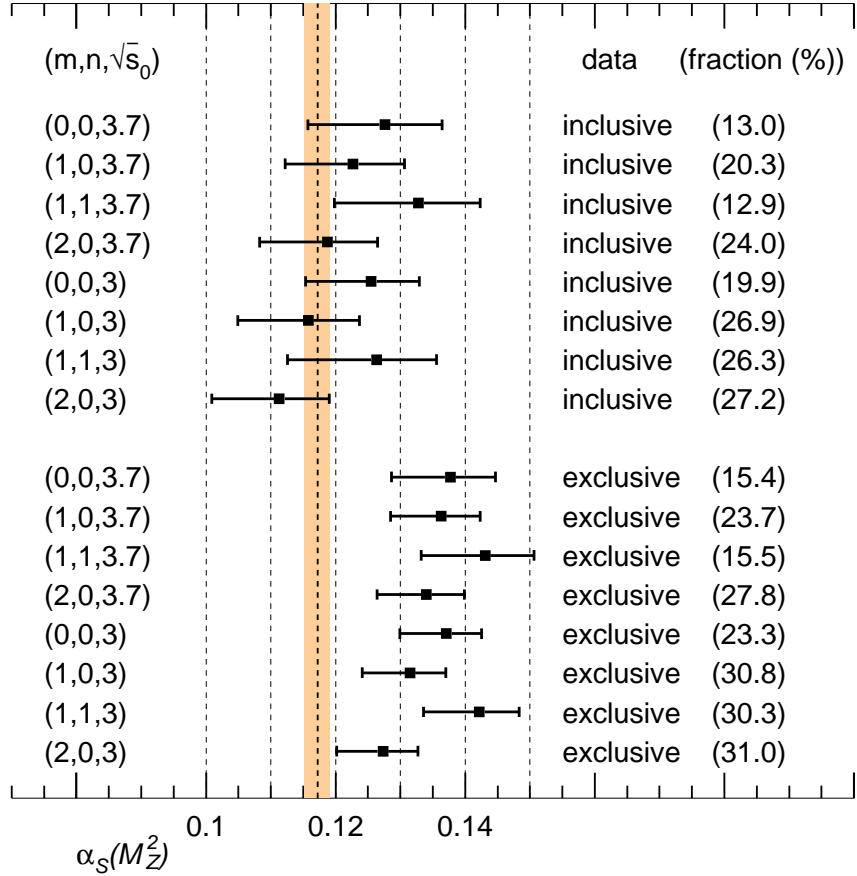,width=14cm}}
\end{center}
\vspace{-7.5ex}
\caption{The QCD sum rule predictions for $\alpha_S(M_Z^2)$
  compared with the world average value~\cite{PDG2002}.
  The results for the four sum rules for two values of $\sqrt{s_0}$
  are shown.  In each case we show results for the inclusive
  and the exclusive measurement of $R(s)$ in the intermediate
  energy region.  We also give in brackets the fractional
  contribution to the sum rule coming from the $1.43<\sqrt{s}<2$~GeV
  interval.}
\label{fig:sumrules}
\end{figure}

We evaluate each of these sum rules for $\sqrt{s_0}=3.7$~GeV using 
the clustered data values of $R(s)$ of Section~2 on the
l.h.s., and QCD for $D(s)$ (with
$\alpha_S=0.1172\pm0.0020$~\cite{PDG2002}) 
on the r.h.s.  We find, as
anticipated, that the sum rules with $m=0$ and $n=1$ or 2 have very 
small contributions from the disputed
1.43---2~GeV region. Indeed, this region contributes only about 
5\% and 2\%, respectively, of the total
contribution to the l.h.s. of (\ref{eq:m=0n=1sumrule}) and 
(\ref{eq:m=0n=2sumrule}). They emphasize the region
$s\lesim s_0$ and so essentially test data against perturbative QCD 
in this small domain. They are not useful for our purpose. The
results for the remaining four sum  rules are shown by the
numbers in brackets in Fig.~\ref{fig:sumrules}. For this choice of
$s_0$, the sum rules with $m=1$ or 2 and $n=0$ are found to maximize 
the fractional contribution to the sum rule 
coming from the $1.43<\sqrt{s}<2$~GeV interval. These two sum 
rules clearly favour the inclusive over the exclusive data.

\begin{table}[htb]
\begin{center}
\begin{tabular}{| c | c | c |}
\multicolumn{3}{l}{(a)~Breakdown of contributions to l.h.s. of sum
rules}
\\[1.5ex]\hline
energy range (GeV)    & contribution ($m=2, n=0$) & contribution
($m=n=0$) \rule[-1.5ex]{0ex}{4.5ex}
\\\hline
$2m_\pi - 0.32$ (ChPT) & 0.00 $\pm$ 0.00 &  0.00 $\pm$ 0.00 \\
$0.32 - 1.43$ (excl)   & 3.92 $\pm$ 0.03 &  4.49 $\pm$ 0.04 \\
$1.43 - 2.00$ (excl)   & 3.02 $\pm$ 0.26 &  4.93 $\pm$ 0.43 \\
$1.43 - 2.00$ (incl)   & 2.48 $\pm$ 0.19 &  4.03 $\pm$ 0.30 \\
$2.00 - 3.73$ (incl)   & 3.94 $\pm$ 0.14 & 22.56 $\pm$ 0.70
\\ \hline
sum (excl)            & 10.87 $\pm$ 0.30 & 31.98 $\pm$ 0.82 \\
sum (incl)            & 10.34 $\pm$ 0.24 & 31.08 $\pm$ 0.76 \\ \hline
\end{tabular}
\end{center}
\begin{center}
\begin{tabular}{| c | c | c |}\multicolumn{3}{l}{(b)~Breakdown of contributions to r.h.s. of sum
rules}
\\[1.5ex]\hline
origin    & contribution ($m=2, n=0$)  & contribution ($m=n=0$)
\rule[-1.5ex]{0ex}{4.5ex}
\\\hline
massless QCD                  & 10.31  $\pm$ 0.05 &  30.43 $\pm$ 0.11 \\
correction from finite $m_s$  &$-0.03$ $\pm$ 0.02 & $-0.03  \pm  0.02$ \\
quark and gluon condensates   &  0.03  $\pm$ 0.02 &   0.00 $\pm$ 0.00
\\\hline
prediction from QCD (total)   & 10.30  $\pm$ 0.06 & 30.40 $\pm$ 0.12 \\\hline
\end{tabular}
\caption{The breakdown of the sum
  rules for $\sqrt{s_0}=3.7$~GeV, for the choices $m=2, n=0$ and
  $m=n=0$. The contributions to the left-hand-side (data) are shown
  in the upper table, and the QCD contributions are given in the
  lower table.}
\label{tab:sumrule-breakdown} 
\end{center}
\end{table}

The comparison between the data and QCD can be translated into
another form.  We can treat $\alpha_S(M_Z^2)$ as a free parameter,
and calculate the value which makes the r.h.s. of a sum rule
exactly balance the l.h.s. The results are shown 
in Fig.~\ref{fig:sumrules}. 
We can see that in
this comparison the determination from the inclusive data is more
consistent with the world-average value,
$\alpha_S(M_Z^2)=0.1172 \pm 0.0020$~\cite{PDG2002}.

For illustration, we show in Table~\ref{tab:sumrule-breakdown} a
detailed breakdown of the contributions to both sides of the
sum rule for the cases of $m=2, n=0$ and $m=n=0$.   If we compare
the breakdown of the contribution from the data in both cases, we
can see that the weight function $f(s)=(1-s/s_0)^2$ highlights the
most ambiguous region of $R(s)$ very well. When we look into the
breakdown in the QCD part, we can see that the QCD contribution is
dominated by the massless part.

We repeated the sum rule analysis for $\sqrt{s_0}=3.0$~GeV,
see Fig.~\ref{fig:sumrules}.  The lower value of $s_0$ means 
that more weight is given to the disputed
1.43---2~GeV region. Taken together, we see that the sum rules
strongly favour the behaviour of $R(s)$ from the {\em inclusive}
measurements. Indeed, the overall consistency in this case is
remarkable.  This result can also be clearly seen from Fig.\ 
\ref{fig:sumrules}, which compares the world average value of 
$\alpha_S(M_Z^2)$ with the predictions of the individual sum
rules for, first $\sqrt{s_0}=3.7$ GeV and then for $\sqrt{s_0}=3$ GeV.
Again the consistency with the inclusive measurements of $R(s)$ 
is apparent.

The same conclusion with regard to the resolution of the
inclusive/exclusive ambiguity in the $1.43<\sqrt{s}<2$~GeV
interval was reached in an independent analysis~\cite{MOR}.

In an attempt to understand the origin of the 
discrepancy, we have studied the effect of possibly missing
(purely neutral) modes in the inclusive data, but found that these cannot
explain the difference.  One should, however, keep in mind that the
precision of both the (old) inclusive and the exclusive data in this
energy regime is quite poor.  We expect that future measurements 
at $B$-factories (via radiative return) and at the upgraded
machine VEPP-2000 in Novosibirsk will improve the situation in the future.

\section{Comparison with other predictions of $g-2$} 
\label{sec:comparison}

\begin{figure}[th]
\begin{center}
{\psfig{file=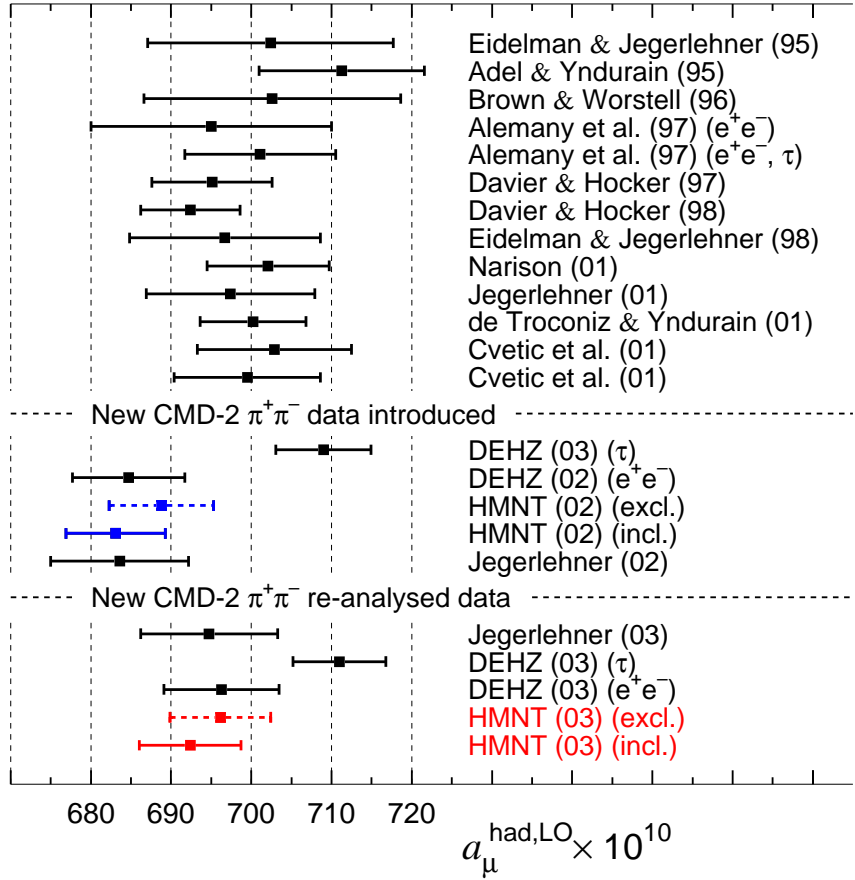,width=14cm}}
\end{center}
\vspace{-7.5ex}
\caption{Recent evaluations of 
  $a_\mu^{\rm had,LO}$~\cite{DEHZ,DEHZ03,HMNT,ADH,DH98a,f4+7,f4,JEG,
  Jegerlehner:2003qp}. 
  The entries below the first dashed line include the new CMD-2 $\pi^+\pi^-$ 
  data~\cite{Akhmetshin:2001ig}, and the values below the second dashed 
  line include the re-analysed CMD-2 $\pi^+\pi^-$ data~\cite{CMD2new}.} 
\label{fig:Z}
\end{figure}

Fig.~\ref{fig:Z} shows other determinations of $a^{\rm had,LO}_\mu$, 
together with the values (HMNT(03)) obtained in this work. The 
values listed below the first dashed line incorporate the new more 
precise data on $e^+e^-\to\pi^+\pi^-$~\cite{Akhmetshin:2001ig} into 
the analysis.  These data play a dominant role, and, as can be
seen from the Figure, significantly decrease the value of 
$a_\mu^{\rm had}$. However, very recently, the CMD-2 Collaboration 
have re-analysed their data and found that they should be increased 
by approximately 2\%, depending on $\sqrt s$. The new 
data~\cite{CMD2new} are included in our analysis. Inspection of 
Fig.~\ref{fig:Z} shows that the re-analysis of the CMD-2 data has 
led to an increase of $a_\mu^{\rm had,LO}\times10^{10}$ by about 10.

The entries denoted by ``DEHZ ($\tau$)'' also used information from 
hadronic $\tau$ decays~\cite{DEHZ,DEHZ03}, which through CVC give 
independent information on the $e^+e^-\to2\pi$ and $4\pi$ channels for 
$\sqrt s \lesim m_\tau$.  The apparent discrepancy between the 
prediction from this analysis and the pure $e^+e^-$ analyses is not yet
totally understood, however see the remarks in the Introduction.

\subsection{Comparison with the DEHZ evaluation}

It is particularly informative to compare the individual 
contributions to $a^{\rm had,LO}_\mu$ obtained in the
present analysis with those listed in the recent study of Davier 
et~al. (DEHZ03)~\cite{DEHZ03}, which used essentially the same 
$e^+e^-\to{\rm hadrons}$ data.  Such a comparison highlights regions 
of uncertainty, and indicates areas where further data and study 
could significantly improve the theoretical determination of $g-2$.
DEHZ provided a detailed breakdown of their contributions to 
$a^{\rm had,LO}_\mu$, and so, to facilitate the comparison, we have 
broken down our contributions into the energy intervals that they use. 
Table~\ref{tab:t7} shows the two sets of contributions of the 
individual $e^+e^-$ channels to dispersion relation 
(\ref{eq:dispersion_rel}) in the crucial low energy region with 
$\sqrt s < 1.8$~GeV.

\begin{figure}[tbp!]
\begin{center}
{\epsfxsize=13cm \leavevmode \epsffile[80 270 490 550]{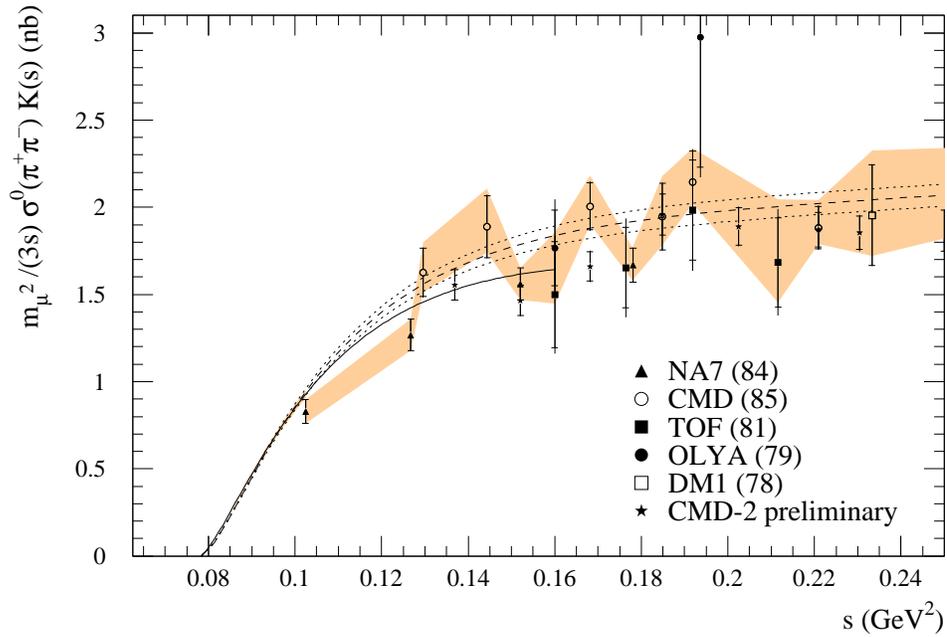}}
\end{center}
\caption{The $\pi^+\pi^-$ data just above threshold, plotted so that
  the area gives the contribution to dispersion relation
  (\ref{eq:dispersion_rel}) for $a_\mu^{\rm had,LO}$. The dashed curve
  is used by DEHZ~\cite{DEHZ,DEHZ03}, whereas the continuous curve 
  up to $\sqrt s = 0.32$~GeV ($s=0.10$~GeV$^2$) and data band are used 
  in this analysis; see text. We also show, but do not use, the 
  preliminary low energy CMD-2
  data, which were read off Fig.~3 of~\cite{EID}. These points,
  particularly the first, are subject to `reading-off' errors.} 
\label{fig:5}
\end{figure}

The last column of Table~\ref{tab:t7} shows the discrepancy between
the two analyses. The biggest difference occurs in the $\pi^+\pi^-$
channel, which gives the main contribution to $a_\mu^{\rm had,LO}$,
and the improvement in the SM prediction essentially comes from the
recent higher precision CMD-2 data in the region $0.6<\sqrt s < 0.9$~GeV 
(see the remarks in Chapter~\ref{pipichannel}).  We find that
this difference, $2.6\times10^{-10}$, appears to come from the region
just above the $\pi^+\pi^-$ threshold, especially in the region
$\sqrt{s} \sim 0.4$~GeV, see Fig.~\ref{fig:5}.
The figure shows the $\pi^+\pi^-$ contribution plotted in such a way
that the area under the curves (or data band) gives the contribution
to dispersion relation (\ref{eq:dispersion_rel}) for $a_\mu^{\rm
  had,LO}$. To determine the low energy $\pi^+\pi^-$ contribution,
DEHZ~\cite{DEHZ} first perform a three-parameter fit to 
$\pi^+\pi^-$ data for $\sqrt s < 0.6$~GeV, and obtain the dashed curve in
Fig.~\ref{fig:5}. This is then used to compute the $\pi^+\pi^-$
contribution of $(58.04\pm2.06)\times10^{-10}$ for $\sqrt s <
0.5$~GeV. They do not use either the NA7\footnote{
However, recently it has turned out that earlier worries about a
systematic bias in the NA7 data as mentioned in \cite{DEHZ} are not
justified and that there is no reason to neglect these important
data~\cite{Tenchini}.}
\cite{Amendolia:1984di} or the preliminary CMD-2 data.
On the other hand we use the chiral 
description~\cite{Gasser:1990bv},
shown by the continuous curve, only as far as $\sqrt s = 0.32$~GeV;
and then use the band obtained from our clustered data, which include
data from OLYA~\cite{Vasserman:1979hw}, TOF~\cite{Vasserman:1981xq}, 
NA7~\cite{Amendolia:1984di}, CMD~\cite{Barkov:1985ac} and 
DM1~\cite{Quenzer:1978qt} in this energy region.  In this
way we obtain a $\pi^+\pi^-$ contribution for $\sqrt s < 0.5$~GeV of
$(55.7\pm 1.9)\times10^{-10}$. We also show on Fig.~\ref{fig:5} the
preliminary CMD-2 data, obtained from Fig.~3 of Ref.~\cite{EID}. These
data were used in neither analysis, but do seem to favour the lower
$\pi^+\pi^-$ contribution. It is also interesting to note that
DEHZ~\cite{DEHZ,DEHZ03} obtain the low value of 
$(56.0\pm1.6)\times10^{-10}$ if $\tau$ decay and CVC are used in 
this region.

Other significant, with respect to the errors, discrepancies arise in
the $\pi^0 \gamma + \eta \gamma$ and the $K^0_S K^0_L$ channels, where
the treatment is different: DEHZ integrate over Breit--Wigner resonance
parametrizations (assuming that the $KK$ channels are saturated by the
$\phi$ decay), while we are integrating the available data in these
channels directly.  In our method there is no danger to omit or
double-count interference effects and resonance contributions from
tails still present at continuum energies, and the error estimate is
straight forward.  As a check, we made fits to Breit--Wigner-type
resonance forms and studied the possibility that trapezoidal
integration of concave structures overestimate the resonance
contributions.  We found the possible effects are negligible compared to the
uncertainties in the parametrization coming from poor quality data in
the tail regions.  The one exception is the $\phi\to K_S^0 K_L^0$
contribution.  Here the lack of data in certain regions of the
resonance tails (see Fig.\ \ref{fig:kskl}) has caused us to increase 
the uncertainty on this contribution to $a_\mu$, see Section 
\ref{subsec:2K}.

Apart from these channels, it is only the two four-pion channels which
show uncomfortably large and relevant discrepancies.  Here, the data
input is different between DEHZ and our analysis.  We use, in addition
to DEHZ, also data from $\gamma\gamma 2$~\cite{Bacci:1981zs,Bacci:1980ru} 
and ORSAY-ACO~\cite{Cosme:1976tf} for both $4\pi$
channels, and data from M3N~\cite{Paulot:1979ep} and two more 
data sets from CMD-2 \cite{Akhmetshin:1999ty,Akhmetshin:2000it} for the
$\pi^+\pi^-\pi^+\pi^-$ channel.  However, it should be noted, that the
available data are not entirely consistent, a fact reflected in the
poor $\chi^2_{\rm min}/{\rm d.o.f.}$ of our fits resulting in the need of
error inflation\footnote{
If for a given channel $\chi^2_{\rm min}/{\rm d.o.f.} > 1.2$, 
then we enlarge the error by $\sqrt{\chi^2_{\rm min}/{\rm d.o.f.}}$. 
This was necessary for three channels, see Section \ref{sec:combdatasets}.
}.  
Clearly, in these channels, new and better data is
required.  As mentioned already in Section \ref{subsec:npi}, the
situation is expected to improve as soon as the announced re-analysis 
from CMD-2 will become available.

There are no data available for some of the exclusive channels. Their
contribution to the dispersion relation is computed using isospin
relations. The corresponding entries in Table~\ref{tab:t7} have been
marked by the word ``isospin''.

\subsection{Possible contribution of the $\sigma(600)$ resonance to
$g-2$} 
\label{sec:poss_cont}

This subsection is motivated by the claim~\cite{Narison} that the 
isosinglet scalar boson\footnote{$\sigma(600)$ is denoted by $f_0(600)$ 
in the Review of Particle Physics~\cite{PDG2002}.} $\sigma(600)$ 
can have a non-negligible contribution to the muon $g-2$. Here, we 
evaluate its contribution and find that it is at most of order 
$0.1 \times 10^{-10}$. This is negligible as compared to the 
uncertainty of the hadronic vacuum polarization contribution of 
$6 \times 10^{-10}$, and hence we can safely neglect it.

The argument presented in Ref.~\cite{Narison} is twofold. First $\sigma$ may contribute to the muon $g-2$ through
unaccounted decay modes of the narrow spin~1 resonances into the $\sigma \gamma$ channel.  The second possibility,
considered in \cite{Narison}, is that $\sigma$ may contribute directly to the muon $g-2$ through its coupling to
the muon pair. We estimate the two contributions below.

In the zero-width limit, narrow spin~1 resonances, $V$, contribute 
to the muon $g-2$ as
\be  a_\mu^V = (3/\pi) K(m_V^2) \Gamma(V\to ee)/m_V,  \label{eq:hag1} \ee
where $K(m_V^2)$ is the kernel function (\ref{eq:K}) at $s=m_V^2$.  
We find, for example\footnote{We take vector mesons,
$V$, which, according to~\cite{Narison}, may have significant 
contributions.},
\be a_\mu^\omega = 391 \times 10^{-10},  \label{eq:hag2a} \ee
\be a_\mu^\phi   =  39 \times 10^{-10},  \label{eq:hag2b} \ee
\be a_\mu^{\phi(1.68)} = 3.4 \times 10^{-10},  \label{eq:hag2c} \ee
where, in (\ref{eq:hag2c}), we have used 
$\Gamma(\phi(1.68)\to ee)=0.48$~keV \cite{PDG2002} to give a rough
estimate. If the decays $V \to \sigma \gamma$ of the above vector 
bosons escape detection, a fraction of the above
contributions up to $B(V \to \sigma \gamma)$ may have been missed. 
On the other hand we find that 99.8\% of 
$\phi$ decays has been accounted for in the five decay channels 
explicitly included in our analysis
hence $B(\phi \to \sigma \gamma)<0.002$. This severely constrains 
the $\sigma\gamma\gamma$ coupling. Hence we
can use the Vector Meson Dominance (VMD) approximation to show that 
the other branching fractions satisfy
$B(\omega \to \sigma \gamma) < 7.2\times10^{-5}$ and 
$B(\phi(1.68) \to \sigma \gamma) < 3.5\times10^{-5}$, see
Appendix~B. By inserting these constraints into the estimates 
(\ref{eq:hag2a})--(\ref{eq:hag2c}), we find that the
unaccounted $V\to\sigma\gamma$ contributions to $g-2$ of the muon 
are less than $(2.8,\, 7.8,\, 0.01)\times10^{-12}$ for 
$V=\omega,\ \phi,\ \phi(1.68)$ respectively, assuming 
$m_\sigma=600$~MeV. These estimates are much smaller than those 
presented in \cite{Narison}. It is clear that the total contribution 
of unaccounted $\sigma \gamma$ modes through narrow resonance decays 
is negligibly small.  It is also worth pointing out here
that the unaccounted fraction $0.2\%$ of the $\phi$ contribution
($7.8\times 10^{-12}$ above) has been taken into account
in our analysis, whether it is $\phi\to \sigma\gamma$ or not.

We now turn to the $\sigma$ contribution to the muon $g-2$ through 
its direct coupling to a muon-pair. To evaluate
this, it is essential to estimate the magnitude of the $\sigma\mu\mu$ 
coupling.  Since the coupling through the
$\sigma$-Higgs boson mixing is negligibly small, the leading 
contribution should come from two-photon exchange. In
this regard, the effective coupling strength should be of the same 
order as that of the $\eta$ isoscalar
pseudoscalar meson, which should also be dominated by the two-photon 
exchange.  By using the observed width
$\Gamma(\eta\to\mu\mu)$, and by neglecting the form factor 
suppression, we find that the point-like $\eta$
contribution to the muon $g-2$ is
\be a_\mu^\eta = - 3 \times 10^{-13}, \label{eq:eta_cont} \ee
which is negligibly small. It follows that this implies that 
$a_\mu^\sigma$ is also negligibly small, see
(\ref{eq:210703-12}) below. However, the discussion can be made far 
more general. It is presented in the next Section.

\section{Internal light-by-light contributions} 
\label{sec:internal}

In this section we present a very primitive discussion of the hadronic
contribution to the internal light-by-light amplitudes,
motivated by the study of the direct $\sigma$ and $\eta$ 
contributions to the muon $g-2$.

\begin{figure}[htbp!]
\begin{center}
{\psfig{file=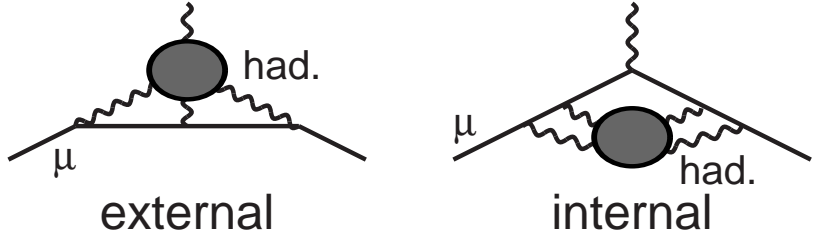,width=12cm}}
\end{center}
\caption{{\em External} and {\em internal} light-by-light contributions 
  to $g-2$.  The former is an ${\cal O}(\alpha^3)$ and the latter is
  an ${\cal O}(\alpha^4)$ contribution.  In this paper we compute the 
  contribution of the {\em internal} diagram.  In 
  Section~\ref{sec:int_mes_con} we take the shaded blob to be scalar 
  ($\sigma$) or pseudoscalar ($\pi^0,\eta$) mesons, whereas in 
  Section~\ref{sec:int_lep_con} we take it to be a light ($u,d,s$) 
  quark loop, using the result for lepton ($e,\mu$) loops as a guide.} 
\label{fig:lbyl-inex}
\end{figure}

The meaning of `internal' can be seen from Fig.~\ref{fig:lbyl-inex}. 
We call the diagram on the right `internal' to distinguish it from the 
left diagram, that is the familiar light-by-light contribution which, 
here, we call `external'.  We should note that the external light-by-light
diagram is of ${\cal O}(\alpha^3)$ and the internal light-by-light
diagram is an ${\cal O}(\alpha^4)$ contribution.

\subsection{Internal meson contributions}  
\label{sec:int_mes_con}

Just like the external light-by-light amplitude is dominated by
a single pseudoscalar meson contribution~\cite{KNO}--\cite{HayK},
it is likely that the hadronic contribution to the internal
light-by-light amplitudes is dominated by a single meson exchange
contribution.

In general, we can estimate the internal contribution to $a_\mu$ from 
arbitrary scalar and pseudoscalar mesons.  Using the effective coupling
\be {\cal L} = \bar\psi_\mu (g_S S + ig_P \gamma_5 P)\psi_\mu,
\label{eq:210703-2} 
\ee
we find~\cite{Mery:1989dx}
\begin{equation} 
  a_\mu^S = 
  \frac{g_S^2}{48\pi^2} \frac{r}{(1-r)^4}
  \left[6(1-2r)\ln\frac{1}{r} - 7 + 24r - 21r^2 + 4r^3\right],
\label{eq:210703-3} 
\end{equation}
\begin{equation} 
 a_\mu^P = 
  \frac{g_P^2}{48\pi^2} \frac{r}{(1-r)^4} 
  \left[ -6\ln\frac{1}{r} + 11 - 18r + 9r^2 -2r^3\right],
\label{eq:210703-4} 
\end{equation}
where $r \equiv m_\mu^2/m_h^2$, with $h=S$ and $P$ in
(\ref{eq:210703-3}) and (\ref{eq:210703-4}) respectively.
The scalar contribution is positive definite and the pseudoscalar 
contribution is negative definite. In the large mass limit ($r\ll1$) 
we have
\begin{equation} 
 a_\mu^S = 
 \frac{g_S^2}{8\pi^2} 
 r\left[\ln\frac{1}{r} - \frac{7}{6} + O(r)\right], 
\label{eq:210703-5} 
\end{equation}
\begin{equation} 
 a_\mu^P = 
 \frac{g_P^2}{8\pi^2} 
 r\left[-\ln\frac{1}{r} + \frac{11}{6} + O(r)\right]. 
\label{eq:210703-6} 
\end{equation}
Further, in the parity-doublet limit of $g_S=g_P$ and $m_S=m_P$, the 
leading terms cancel~\cite{Peccei83} and only a tiny positive 
contribution remains. The effective couplings in (\ref{eq:210703-2}) 
can be extracted from the leptonic widths
\begin{equation} 
 \Gamma(h\to\mu^+\mu^-) 
 = \frac{g_h^2}{8\pi} m_h\left(1-\frac{4m_\mu^2}{m_h^2}\right)^{n/2},
\label{eq:210703-7} 
\end{equation}
where $n=3,1$ for $h=S,P$ respectively.

Let us estimate the pseudoscalar $\pi^0$ contribution. We use
\begin{equation} 
 \Gamma(\pi^0 \to e^+e^-) \simeq 5\times10^{-7}~\eV. 
\label{eq:210703-8} 
\end{equation}
After we allow for the helicity suppression factor of $m_e/m_\mu$
for the $\pi^0ee$ coupling, this gives a $\pi^0\mu\mu$ coupling
\begin{equation} 
 \frac{g_\pi^2}{8\pi} 
 \simeq \left(\frac{m_\mu}{m_e}\right)^2
        \frac{\Gamma(\pi^0\to e^+e^-)}{m_\pi} 
 \simeq 1.6\times10^{-10} 
\label{eq:210703-9} 
\end{equation}
and hence, from (\ref{eq:210703-4}), a contribution
\be a_\mu^{\pi^0} \simeq -6\times10^{-11}. \label{eq:210703-10} \ee
Although this contribution is not completely negligible, we expect 
a form factor suppression of the effective couplings and so the pion 
structure effects should suppress the magnitude significantly.

In the scalar sector, we do not find a particle with significant 
leptonic width. Although the $\sigma$ leptonic width is unknown, 
we find no reason to expect that its coupling is bigger than the 
$\eta\mu\mu$ coupling. If we use
\begin{equation} 
 \frac{g_\sigma^2}{8\pi} \simeq \frac{g_\eta^2}{8\pi} 
= \frac{\Gamma(\eta\to \mu^+\mu^-)}{m_\eta \sqrt{1-4m_\mu^2/m_\eta^2}} 
\simeq  1.4 \times 10^{-11} 
\label{eq:210703-11} 
\end{equation}
we find that $\Gamma(\sigma\to\mu^+\mu^-) \simeq 
7 \times 10^{-3}{\rm eV}$, and hence
\be a_\mu^\sigma = 7\times 10^{-13}    \label{eq:210703-12} \ee
for $m_\sigma = 600$~MeV.  Again we should expect form-factor 
suppressions. Because pseudoscalar mesons are lighter than the 
scalars, there is a tendency that the total contribution is negative 
rather than positive.

\subsection{Internal lepton or quark contributions} 
\label{sec:int_lep_con}

The internal light-by-light scattering contributions in the 4-loop 
order have been evaluated in QED. The electron-loop contribution 
is~\cite{KNIO}
\be 
a_\mu^{\rm int.\ l{\mbox -}b{\mbox -}l} (e{\rm {\mbox -}loop}) 
 = -4.43243(58)\left(\frac{\alpha}{\pi}\right)^4 
 \simeq 1.29\times10^{-10}, 
\label{eq:210703-13} 
\ee
whereas the muon-loop contribution is
\be 
  a_\mu^{\rm int.\ l{\mbox -}b{\mbox -}l} (\mu{\rm {\mbox -}loop}) 
  = -0.99072(10)\left(\frac{\alpha}{\pi}\right)^4 
  \simeq -0.29\times10^{-10}. 
\label{eq:210703-14} \ee
The $\mu$-loop contributions to the electron anomalous moment has 
also been estimated~\cite{TK}
\be 
 a_e^{\rm int.\ l{\mbox -}b{\mbox -}l}(\mu{\rm {\mbox -}loop})  
 = -0.000184(14) \left(\frac{\alpha}{\pi}\right)^4. 
\label{eq:210703-15}
\ee
If we interpolate between (\ref{eq:210703-14}) and (\ref{eq:210703-15}) 
by assuming the form
$(m_\mu^2/m_l^2)[A\ln(m_l^2/m_\mu^2)+B]$, we obtain the estimate
\be 
 a_\mu^{\rm int.\ l{\mbox -}b{\mbox -}l}(l{\rm {\mbox -}loop}) 
 \simeq -\left[0.65\ln\frac{m_l^2}{m_\mu^2} + 1\right]
        \frac{m_\mu^2}{m_l^2}\left(\frac{\alpha}{\pi}\right)^4, 
\label{eq:210703-16} 
\ee
which may be valid for an arbitrary lepton mass in the range
$m_\mu<m_l<m_\mu^2/m_e \sim 20~\GeV$.  
For a $\tau$-loop internal light-by-light contribution to $a_\mu$,
the relation (\ref{eq:210703-16}) gives
\begin{eqnarray}
  a_\mu^{\rm int.\ l{\mbox -}b{\mbox -}l}(\tau{\rm {\mbox -}loop}) = 
   - 0.0165  \left( \frac{\alpha}{\pi} \right)^4,
\end{eqnarray}
which agrees with the actual numerical result 
\begin{eqnarray}
  a_\mu^{\rm int.\ l{\mbox -}b{\mbox -}l}(\tau{\rm {\mbox -}loop}) =  
 -0.01570 (49) \left( \frac{\alpha}{\pi} \right)^4
\end{eqnarray}
within 10\%.  We can now
estimate the hadronic contribution by using the constituent quark model
\be 
a_\mu^{\rm int.\ l{\mbox -}b{\mbox -}l}(u,d,s{\rm {\mbox -}loop}) 
\simeq -\frac{2}{3}
  \left[0.65\ln\frac{m_q^2}{m_\mu^2} + 1\right]
\frac{m_\mu^2}{m_q^2}\left(\frac{\alpha}{\pi}\right)^4, 
\label{eq:210703-17} 
\ee
where we use $m_u = m_d = m_s = m_q$ to set the scale, and where 
$\frac{2}{3} = 3\left( \left(\frac{2}{3}\right)^4
+ 2\left(\frac{1}{3}\right)^4\right)$ is the charge factor. 
For $m_q = 300~\MeV$, (\ref{eq:210703-17}) gives
\be a_\mu^{\rm int.\ l{\mbox -}b{\mbox -}l} (u,d,s{\rm {\mbox -}loop}) 
 \simeq -6\times10^{-12}. \label{eq:210703-18} 
\ee

\subsection{Quark loop estimates of the hadronic light-by-light contributions}

If the same massive quark loop estimate is made for the 3-loop 
(external) hadronic light-by-light scattering
contribution, it is found that~\cite{Laporta:1992pa}
\be a_\mu^{\rm ext.\ l{\mbox -}b{\mbox -}l}(u,d,s{\rm {\mbox -}loop}) 
  \simeq \frac{2}{3}\times0.615
\left(\frac{m_\mu}{m_q}\right)^2\left(\frac{\alpha}{\pi}\right)^3 
\simeq 6\times10^{-10}. \label{eq:210703-19} 
\ee
As we shall see later, this estimate is in reasonably good agreement 
with the present estimate of the total contribution of
$(8\pm4)\times10^{-10}$ of (\ref{eq:hadlbyl}), and of its sign.

The above well-known result has been regarded as an accident,
because in the small quark mass limit the quark-loop contribution
to the external light-by-light amplitude diverges.  The light-meson
contributions could only be estimated by adopting the effective 
light-meson description of low-energy QCD.  Although the same may well 
apply for the internal light-by-light amplitudes, we note here that the 
quark-loop contributions to the internal light-by-light amplitudes 
remain finite in
the massless quark limit because of the cancellation of mass
singularities \cite{Kinoshita:ur,Lee:is}.  We find no strong reason
to discredit the order of magnitude estimate based on (\ref{eq:210703-18})
against the successful one of (\ref{eq:210703-19}) for the external 
light-by-light amplitudes.  Although the point-like $\pi$ 
contribution of (\ref{eq:210703-10})
is a factor of ten larger than the estimate (\ref{eq:210703-18}), 
the corresponding point-like $\pi$ contribution to the external 
light-by-light amplitudes diverges.  We can expect that the form 
factor suppression of the effective
vertices should significantly reduce its contribution.
Also, since these mesons are lighter than the scalar
mesons, we expect the sign of the total meson
contribution to be negative, in agreement with the quark loop estimate 
of (\ref{eq:210703-18}). In conclusion, we
use (\ref{eq:210703-18}) to estimate that the hadronic {\em internal} 
light-by-light contribution is given by
\be a_\mu^{\rm int.\ l{\mbox -}b{\mbox -}l}({\rm hadrons}) 
 = -(0.6\pm0.6)\times10^{-11}, \label{eq:210703-20} \ee
which is totally negligible.  We do not take this contribution 
into account in our final results.

\section{Calculation of $a_\mu^{\rm had}$ and $g-2$ of the muon} 
\label{sec:calculation}

\subsection{Results on $a_\mu^{\rm had, LO}$}

We calculated the LO hadronic contribution $a_\mu^{\rm had,LO}$ 
in Section~\ref{sec:evaluation}.  We found
\begin{eqnarray}
 a_\mu^{\rm had, LO} 
&=& (692.4 \pm 5.9_{\rm exp} \pm 1.4_{\rm rad, VP} 
           \pm 1.9_{\rm rad, FSR}) \times 10^{-10}
\\
&=& (692.4 \pm 5.9_{\rm exp} \pm 2.4_{\rm rad})   \times 10^{-10},
 \label{eq:hadLO_byHMNT03}
\end{eqnarray}
where the first error comes from the systematic and statistic errors 
in the hadronic data which we included in the clustering
algorithm, and the second error is from the uncertainties
in the radiative correction in the experimental data.  Below
we explain this in more detail.

We add the VP error from the experiments and the narrow
resonances linearly.  Out of $1.4\times10^{-10}$, $1.2\times10^{-10}$
is from the data, and $0.2\times10^{-10}$ is from the narrow
resonances.
For the errors from the final state radiation we assign
$1.9\times10^{-10}$, which is the sum of 
the errors,
$\delta a_{\mu}^{{\rm fsr},\pi^+\pi^-}=0.68\times 10^{-10}$,
$\delta a_{\mu}^{{\rm fsr},K^+K^-}=0.42\times 10^{-10}$ and  
$\delta a_{\mu}^{\rm fsr, ~other~excl.}=0.81\times 10^{-10}$.
We added the errors from the VP and the FSR in
quadrature, which is the second error in (\ref{eq:hadLO_byHMNT03}).

\subsection{Calculation of the NLO hadronic 
contributions to $g-2$ of the muon}

\begin{figure}[!htb]
\begin{center}
{\psfig{file=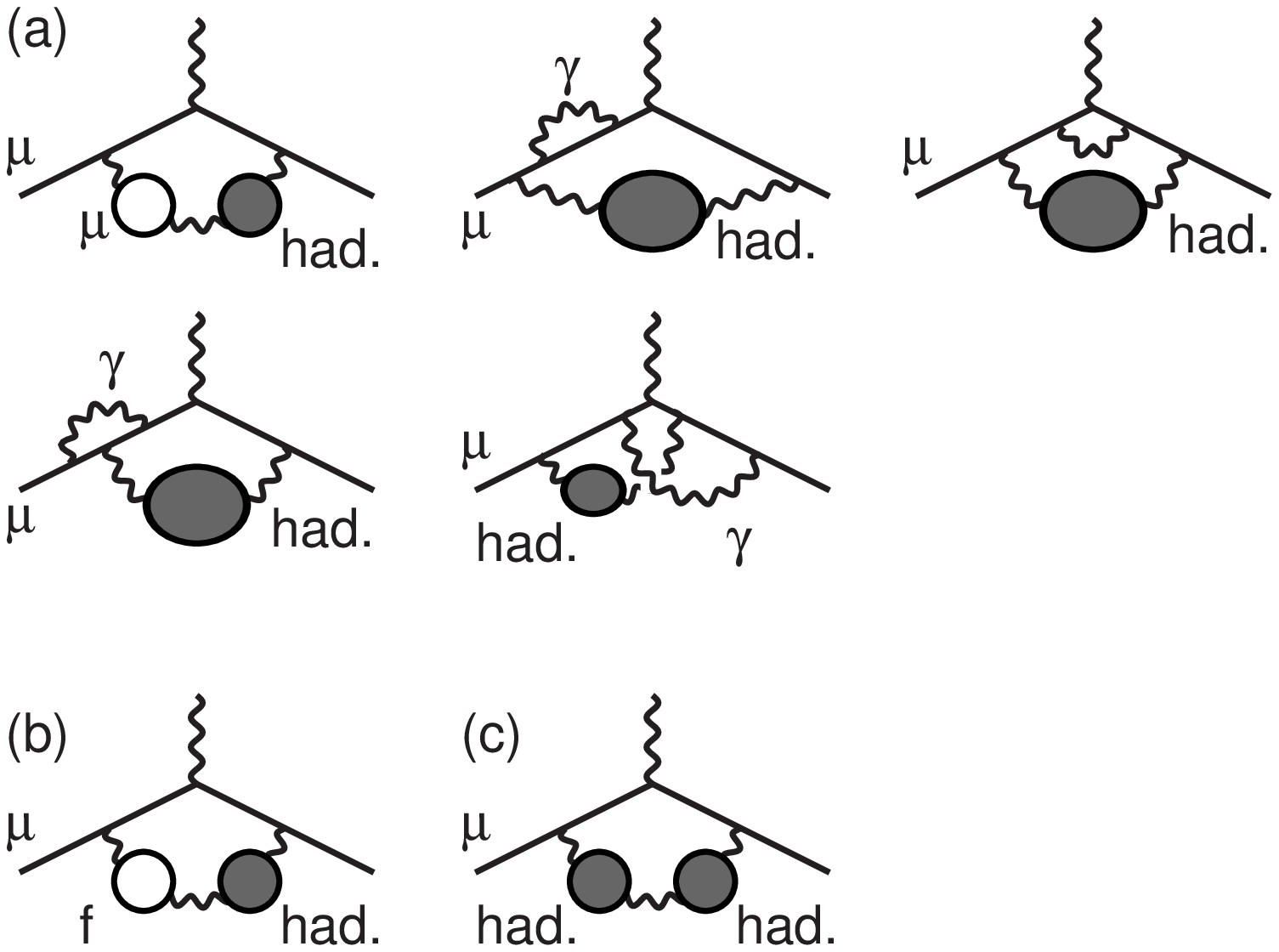,width=12cm}}
\end{center}
\caption{The three classes of diagrams (a,b,c) which contribute 
to $a_\mu^{\rm had,NLO}$. Class~(a) contains the
first five diagrams. In the class~(b) diagram, $f=e$ or $\tau$, 
but not $\mu$. Mirror counterparts and diagrams
with an interchange between the massless photon and the 
``massive photon'' propagators should be understood.}
\label{fig:NLOhadg-2}
\end{figure}

In this subsection we update the computation of the NLO hadronic 
contribution to $g-2$ of the muon.  It proceeds in a similar way to 
that for the LO contribution,
but now the kernel of the dispersion relation is a little more complicated.  
There are three types of NLO contributions, which were denoted (2a),
(2b) and (2c) by Krause \cite{Krause}:  (2a) consists of the
diagrams which contain one hadronic bubble and which do not
involve leptons other than the muon, (2b) is the 
diagram which has one hadronic bubble and one electron (or tau) loop, 
and, finally,  (2c) is the diagram which has two hadronic bubbles.
The three different classes of NLO contributions correspond to 
the diagrams which are denoted (a,b,c) respectively 
in Fig.~\ref{fig:NLOhadg-2}.

The contributions from (2a), (2b), and (2c) can be written as
\begin{eqnarray} 
a_\mu^{\rm had,NLO (2a)} &=&
\frac{\alpha}{4\pi^4} \int_{s_{\rm th}}^\infty {\rm d}s\, 
  \sigma_{\rm had}^0 (s) K^{(2a)}(s),
\label{eq:BB}  \\
a_\mu^{\rm had,NLO (2b)} &=&
\frac{\alpha}{4\pi^4} \int_{s_{\rm th}}^\infty {\rm d}s\, 
  \sigma_{\rm had}^0 (s) K^{(2b)}(s),  
\label{eq:BB2b}  \\
a_\mu^{\rm had,NLO (2c)} &=&
\frac1{16 \pi^5 \alpha} 
\int_{s_{\rm th}}^\infty {\rm d}s\, 
\int_{s_{\rm th}}^\infty {\rm d}s^\prime\, 
  \sigma_{\rm had}^0 (s) \sigma_{\rm had}^0 (s^\prime) 
  K^{(2c)}(s, s^\prime),
\label{eq:BB2c}  
\end{eqnarray}
where the analytic expressions for $K^{(2a)}, K^{(2b)}$ and $K^{(2c)}$
are given in Ref.~\cite{Krause}.  We use the clustered data for the
cross section for $e^+e^-\to{\rm hadrons}$, $\sigma_{\rm had}^0$ of
(\ref{eq:sigma_had}), with the choice of inclusive data in the regime
above 1.43 GeV to compute the
contributions of the three different classes of NLO diagrams. We find
\begin{eqnarray}
  a_\mu^{\rm had,NLO(2a)} & = & 
( -20.73 \pm 0.18_{\rm exp} \pm 0.07_{\rm rad} )\times10^{-10} , \\
  a_\mu^{\rm had,NLO(2b)} & = & 
(  10.60 \pm 0.09_{\rm exp} \pm 0.04_{\rm rad} ) \times10^{-10} ,
\label{eq:CC} \\
  a_\mu^{\rm had,NLO(2c)} & = & 
( 0.34 \pm 0.01_{\rm exp} \pm 0.00_{\rm rad}) \times10^{-10} ,
\end{eqnarray}
where we have assigned the uncertainty from the radiative 
correction similarly to the LO hadronic contribution.
When summed,
\begin{eqnarray}
  a_\mu^{\rm had,NLO} 
 & = & (-9.79 \pm 0.09_{\rm exp} \pm 0.03_{\rm rad})
 \times 10^{-10}, \label{eq:DD}
\end{eqnarray}
which may be compared to the original calculation of 
Krause~\cite{Krause},  $$a_\mu^{\rm had,NLO} =  
( -21.1 (0.5) + 10.7 (0.2) + 0.27(0.01)) \times 10^{-10}
= -10.1(0.6)\times 10^{-10}.$$

In (\ref{eq:DD}) we added the error linearly 
with an opposite relative sign since the errors in (2a) and (2b) 
are nearly 100\% correlated in the opposite directions.  Hence
the total error is the difference of the two.  In combining 
the errors we neglected the errors from (2c) since it is
negligibly small compared to the other errors.  
 
Note that the contribution of diagram (2c) does not agree with
the result given by Krause, when account is taken of the
small error on this contribution.  We have therefore performed 
two checks of our numerical programme.
First we replaced the two hadronic blobs of the diagram (2c) with
two muon loops, since the contribution from such a diagram is 
known analytically \cite{Mignaco-Remiddi}
as a part of the QED contribution.  It is 
\begin{eqnarray}
\lefteqn{
  a_\mu({\rm two~muon~loops~along~one~photon~propagator})} \nonumber\\
     &=& \left(\frac{\alpha}{\pi} \right)^3 
        \left( 
     -\frac{943}{324} - \frac{8}{45} \zeta(2) + \frac83 \zeta(3) 
        \right) \\
     &=& \left(\frac{\alpha}{\pi} \right)^3 \times 0.002558 \ldots \\
     &=& 0.3206 \ldots \times 10^{-10} .
\label{eq:K2ctest1}
\end{eqnarray}
Our programme reproduced $0.321\times 10^{-10}$,
which agrees with (\ref{eq:K2ctest1}) within an accuracy of 
$10^{-12}$, which is the accuracy of the 
calculation throughout this paper.

As a second check, we have taken $R(s)$ to be a step function.
In the first line of Eq.\ (13) of the paper by Krause, the contribution
from the diagram (2c) is written as a triple integral over $s$,
$s^\prime$ and $x$, where $s$ and $s^\prime$ are "mass-squared" of 
the hadronic blobs, and $x$ is a Feynman parameter.
By explicitly integrating over $x$, Krause obtained the second
line of Eq.\ (13), which is a double integral over $s$ and $s^\prime$.  
We are using this expression to integrate over the hadronic data.
If $R(s)$ is a constant, we can explicitly integrate over
$s$ and $s^\prime$, instead of $x$.  Then we are left with one 
dimensional integral over $x$, which is much more tractable than 
the double integral over $s$ and $s^\prime$.  We compared the 
result obtained from this integral over $x$ with the double 
integral over $s$ and $s^\prime$.  Below are the numerical results.

When $R(s)$ is a constant (more rigorously, when $R(s)$ is a step 
function with $R(s)=1$ for $s>4m_\pi^2$, otherwise $R(s)=0$), the 
result from the double integral is
\begin{eqnarray}
  a_\mu = 0.21 \times 10^{-10},
\end{eqnarray}
(which has only two significant digits) and the result from the integral
over $x$ is
\begin{eqnarray}
  a_\mu = 0.2109\ldots \times 10^{-10} .
\end{eqnarray}
The agreement is very good. From the above two checks we 
believe our result for diagram (2c) is correct.

\subsection{Hadronic contribution to $g-2$ of the muon}

The hadronic contribution $a_\mu^{\rm had}$ has been divided 
into three pieces,
\begin{eqnarray}
 a_\mu^{\rm had} = a_\mu^{\rm had,LO}
                 + a_\mu^{\rm had,NLO}
                 + a_\mu^{\rm had,l \mbox{-} b \mbox{-} l}.
\end{eqnarray}
The lowest-order (vacuum polarisation) hadronic contribution, 
$a_\mu^{\rm had,LO}$, was calculated in
Section~\ref{sec:evaluation}. There we found
\begin{eqnarray}
 a_\mu^{\rm had, LO} = (692.4 \pm 5.9_{\rm exp} \pm 2.4_{\rm rad})
                       \times 10^{-10},
\label{eq:hadLO}
\end{eqnarray}
where we have used the QCD sum rule analysis to resolve the discrepancy in
favour of the inclusive $e^+e^-\to{\rm hadrons}$ data in the
region $1.4\lesim\sqrt s\lesim 2$~GeV.
The value of the next-to-leading order hadronic contribution, 
$a_\mu^{\rm had,NLO}$, 
was updated by the calculation described in the previous subsection. 
We obtained
\begin{eqnarray}
  a_\mu^{\rm had,NLO} 
 & = & (-9.79 \pm 0.09_{\rm exp} \pm 0.03_{\rm rad} )\times 10^{-10}.
\label{eq:hadNLO_byHMNT03}
\end{eqnarray}
Finally, we must include the hadronic light-by-light scattering
contribution $a_\mu^{\rm had, l\mbox{-}b\mbox{-}l}$.  It has attracted
much study. Recent re-evaluations can be found, for example, in 
Refs.~\cite{KN}--\cite{RMW}.
Here we take the representative value\footnote{However, see the note added
  in proof.} 
\begin{eqnarray}
a_\mu^{\rm had, l\mbox{-}b\mbox{-}l} = (8.0 \pm 4.0) \times 10^{-10} ,
  \label{eq:hadlbyl}
\end{eqnarray}
as given in Ref.~\cite{Nyffeler}. From
Eqs.~(\ref{eq:hadLO_byHMNT03}), (\ref{eq:hadNLO_byHMNT03}) and
(\ref{eq:hadlbyl}), we can see that $a_\mu^{\rm had,LO}$ has the
largest uncertainty, although the uncertainty in the
light-by-light contribution $a_\mu^{\rm had,l \mbox{-} b \mbox{-} l}$
is also large.

When we combine all the three contributions to the hadronic
contribution, we find 
\begin{eqnarray}
 a_\mu^{\rm had} = (690.6 \pm 7.4) \times 10^{-10}.
\label{eq:hadtot}
\end{eqnarray}
To calculate the number above, we first added the uncertainties
associated with the LO and NLO diagrams linearly, and then added the
uncertainty in the light-by-light contribution quadratically. 
We did so since the errors in the LO and the NLO contributions
are nearly 100\% correlated.

\subsection{SM prediction of $g-2$ of the muon}

The SM value of the anomalous magnetic moment of the muon,
$a_\mu$, may be written as the sum of three terms,
\begin{eqnarray}
 a_\mu^{\rm SM} = a_\mu^{\rm QED} + a_\mu^{\rm EW} + a_\mu^{\rm had} .
\label{eq:amusm}
\end{eqnarray}
The QED contribution, $a_\mu^{\rm QED}$, 
has been calculated up to and including estimates of the 5-loop
contribution, see reviews ~\cite{KNIO,HuK,CM,Nuff},
\begin{eqnarray}
 a_\mu^{\rm QED} = 116~584~703.5 (2.8) \times 10^{-11} .
 \label{eq:QED}
\end{eqnarray}
This value~\cite{Nuff} includes the recent update from \cite{KNIO}.
In comparison with the experimental error in Eq.~(\ref{eq:BNL2001}), 
and the error of the hadronic contribution,
the uncertainty in $a_\mu^{\rm QED}$ is much less important than the 
other sources of uncertainty.
The electroweak contribution $a_\mu^{\rm EW}$ is calculated through 
second order to be~\cite{CKM}--\cite{CMV}
\begin{eqnarray}
 a_\mu^{\rm EW} = 154 (2) \times 10^{-11} .
 \label{eq:EW}
\end{eqnarray}
Here we quote the result of \cite{CMV}. Although some discrepancies on
conceptual questions remain, this result agrees numerically 
with the one of \cite{KPPdR}, and here again the error is negligibly small.

\begin{figure}[ht]
\begin{center}
{\psfig{file=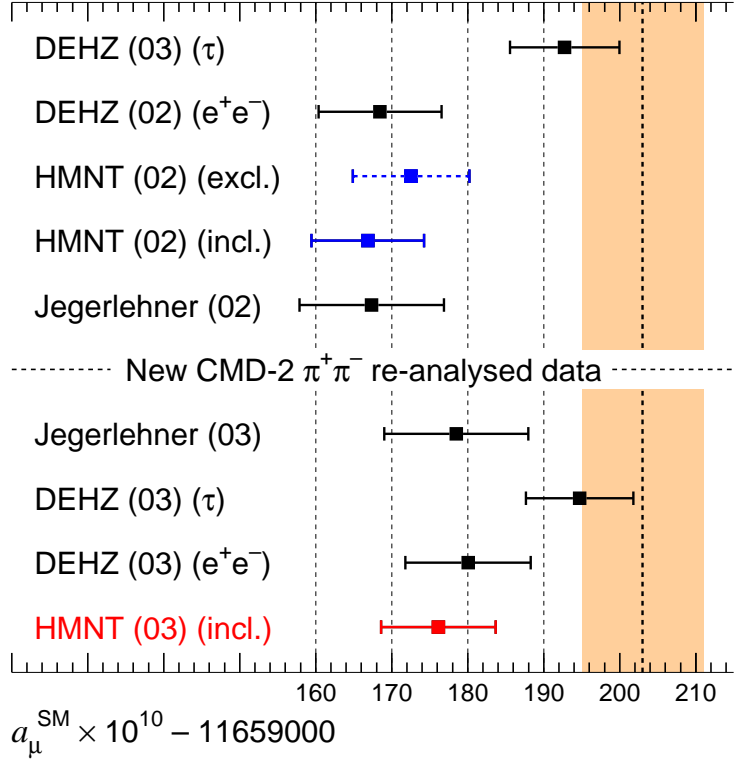,width=12cm}}
\end{center}
\vspace{-5.5ex}
\caption{Recent evaluations of $a_\mu^{\rm SM}$ and the current 
  world-average of the measured value
  (shown as a band). The band corresponds to a 1-$\sigma$ range. 
  The final values, HMNT(03), are the predictions of
  this work, and include the recently re-analysed CMD-2 $\pi^+\pi^-$ 
  data~\cite{CMD2new} in our analyses.}
\label{fig:Zsm}
\end{figure}

Summing up the SM contributions to $a_{\mu}^{\rm SM}$, as given in
(\ref{eq:hadtot}), (\ref{eq:QED}) and (\ref{eq:EW}), we conclude
that
\begin{equation}
a_\mu^{\rm SM}=(11659176.3\pm7.4)\times10^{-10},
\end{equation}
which is $26.7\times10^{-10}$ ($2.4\sigma$) below the world average
experimental measurement. If, on the other hand, we were to take,
instead of (\ref{eq:hadLO}), the value of
$a_\mu^{\rm had,LO}$ obtained using the sum of the exclusive data
in the interval $1.43<\sqrt{s}<2$~GeV, then we would find
$a_\mu^{\rm SM}=(11659180.1 \pm 7.4)\times10^{-10}$, which is
$22.9\times10^{-10}$ ($2.1\sigma$) below $a_\mu^{\rm exp}$.
The above values of $a_\mu^{SM}$ is compared 
with other determinations in Fig.~\ref{fig:Zsm}.

\section{Determination of $\alpha_{\rm QED}(M_Z^2)$}
\label{sec:determination}

As mentioned in the Introduction, the value of the QED coupling at
the $Z$ boson mass is the least well known of the three input
parameters ($G_\mu$, $M_Z$ and $\alpha(M_Z^2)$) which are the 
three most fundamental inputs of the standard electroweak model. 
Its uncertainty is
therefore the limiting factor for precision electroweak physics.
It is clearly important to determine $\alpha(M_Z^2)$ as accurately
as possible.

The value of $\alpha(M_Z^2)$ is obtained from \cite{PDG2002}
\be
\alpha^{-1}\equiv \alpha(0)^{-1} = 137.03599976(50)
\label{eq:alpha^{-1}}
\ee
using the relation
\be 
\alpha(s)^{-1} = 
 \left( 1 - \Delta\alpha_{\rm lep}(s)
          - \Delta\alpha_{\rm had}^{(5)}(s) 
          - \Delta\alpha^{\rm top}(s)   
 \right) \alpha^{-1}, 
\ee
where the leptonic contribution to the running of $\alpha$ is
known to three loops \cite{STEINH},
\be \Delta\alpha_{\rm lep}(M_Z^2) = 0.03149769. \ee
The evaluation of the hadronic contribution, 
$\Delta\alpha_{\rm had}^{(5)}(M_Z^2)$, is described below.

\subsection{The hadronic contribution to the running of 
$\alpha$ up to $s=M_Z^2$}

It is conventional to determine the contribution from 5 quark
flavours, $\Delta\alpha_{\rm had}^{(5)}$, and to include the
contribution of the sixth flavour~\cite{KuhnStein},
\be \Delta\alpha^{\rm top}(M_Z^2) = -0.000070 (05), \ee
at the end. The quark contribution cannot be calculated just 
from perturbative QCD because of low energy strong
interaction effects. Rather we determined the contribution, 
$\Delta\alpha_{\rm had}^{(5)}(M_Z^2)$, by evaluating
the dispersion relation (\ref{eq:210703-1}). 
The results were shown in Table~\ref{tab:hadroniccontr}. We found
\begin{eqnarray}
 \Delta\alpha_{\rm had}^{(5)}(M_Z^2) 
&=& 0.02755\pm 0.00019_{\rm exp}\pm 0.00013_{\rm rad, VP}
                                \pm 0.000019_{\rm rad, FSR}     \\
&=& 0.02755\pm 0.00019_{\rm exp}\pm 0.00013_{\rm rad} \\
&=& 0.02755\pm 0.00023, 
\end{eqnarray}
if we use the {\em inclusive} measurements of $R(s)$ in the 
interval $1.43<\sqrt{s}<2\ \GeV$.  The corresponding
value of the QED coupling is given by
\begin{equation}
 \alpha(M_Z^2)^{-1}=128.954 \pm 0.031 . \label{eq:QEDcoupling}
\end{equation}
If, on the other hand, we were to use the sum of the {\em exclusive} 
data for the various $e^+e^-\to {\rm hadron}$
channels, then the result would become 
$0.02769 \pm0.00018_{\rm exp} \pm 0.00013_{\rm rad}$
and $\alpha(M_Z^2)^{-1}=128.935 \pm 0.030$. 
Table~\ref{tab:hadroniccontr} shows the contributions
to $\Delta\alpha_{\rm had}^{(5)}(M_Z^2)$ from the different energy
intervals of the dispersion integral,
(\ref{eq:210703-1}), together with the sum.  An alternative view may 
be obtained from the (lower) pie diagrams of Fig.~\ref{fig:pie}. They 
display the fractions of the
total contribution and error coming from various energy
intervals in the dispersion integral.   As anticipated, 
both Table~\ref{tab:hadroniccontr} and the pie diagrams, 
show that the hadronic contributions to $\alpha(M_Z^2)$
are more weighted to higher $s$ values in the dispersion integral
for $\Delta\alpha_{\rm had}^{(5)}(M_Z^2)$, than those in the integral 
for $a^{\rm had}_\mu$ needed to predict $g-2$ of the muon.

\begin{figure}[tbp!]
\begin{center}
{\psfig{file=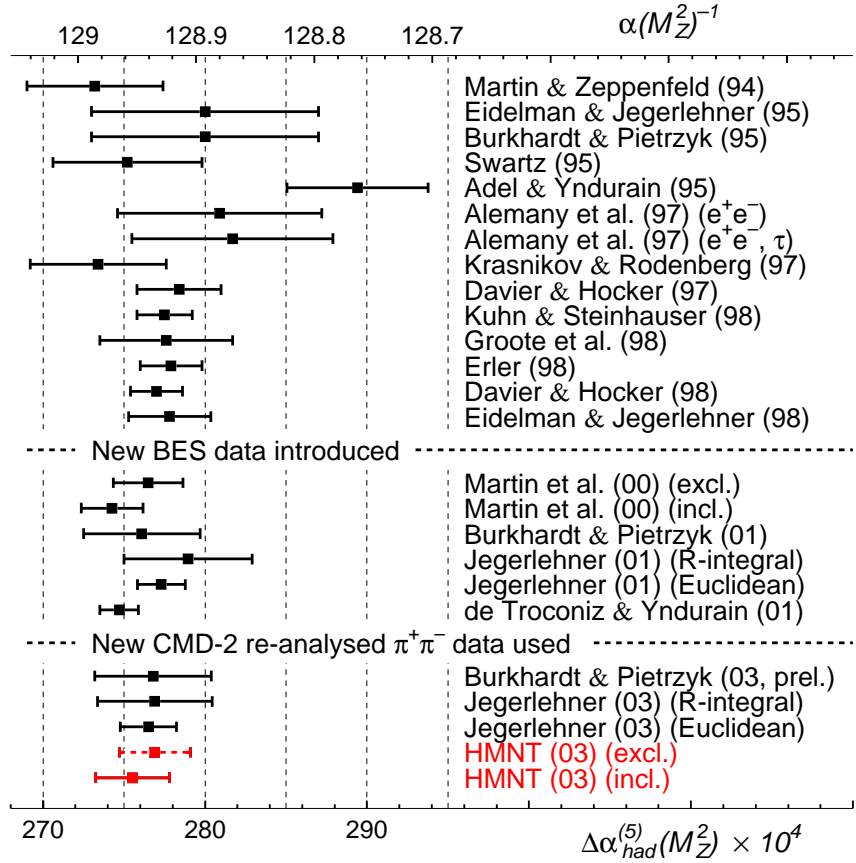,width=14cm}}
\end{center}
\vspace{-7.5ex}
\caption{
  Recent 
  determinations~\cite{SWARTZ,Burkhardt:2001xp,ADH,DH98a,MOR,f4+7,JEG,
  Jegerlehner:2003qp,KuhnStein,f7,BP03} 
  of $\Delta\alpha_{\rm had}^{(5)}(M_Z^2)$ (lower scale) with the 
  corresponding value of $\alpha(M_Z^2)^{-1}$ at the $Z$ boson mass shown
  on the upper scale. The last two entries, HMNT(03), are the values 
  obtained in this work, and include the recent CMD-2 
  (re-analysed) data~\cite{CMD2new} in the evaluation.} 
\label{fig:A}
\end{figure}

The above values of $\Delta\alpha_{\rm had}^{(5)}(M_Z^2)$, and the
corresponding values of $\alpha^{-1}$ at $s=M_Z^2$, are compared 
with other determinations in Fig.~\ref{fig:A}.
The BES data~\cite{Bai:2001ct} became available for the analyses 
from \cite{MOR} onwards.  In Table~\ref{tab:compBP01} we compare 
contributions to the dispersion relation (\ref{eq:disprel2}) for
$\Delta \alpha_{\rm had}^{(5)}(M_Z^2)$ obtained in this work with 
those found by
Burkhardt and Pietrzyk~\cite{Burkhardt:2001xp}.
Since new $e^+e^-$ data became available for the former analysis, the
comparison is only meaningful in the higher energy intervals.
Nevertheless, although the agreement in the size of the contributions is
good, we see that the latter analysis has considerably larger uncertainties
in some energy intervals, which
explains, in part, the difference in the size of the overall error 
shown in Fig.~\ref{fig:A}.

\begin{table}[htbp]
\begin{center}
\epsfig{file=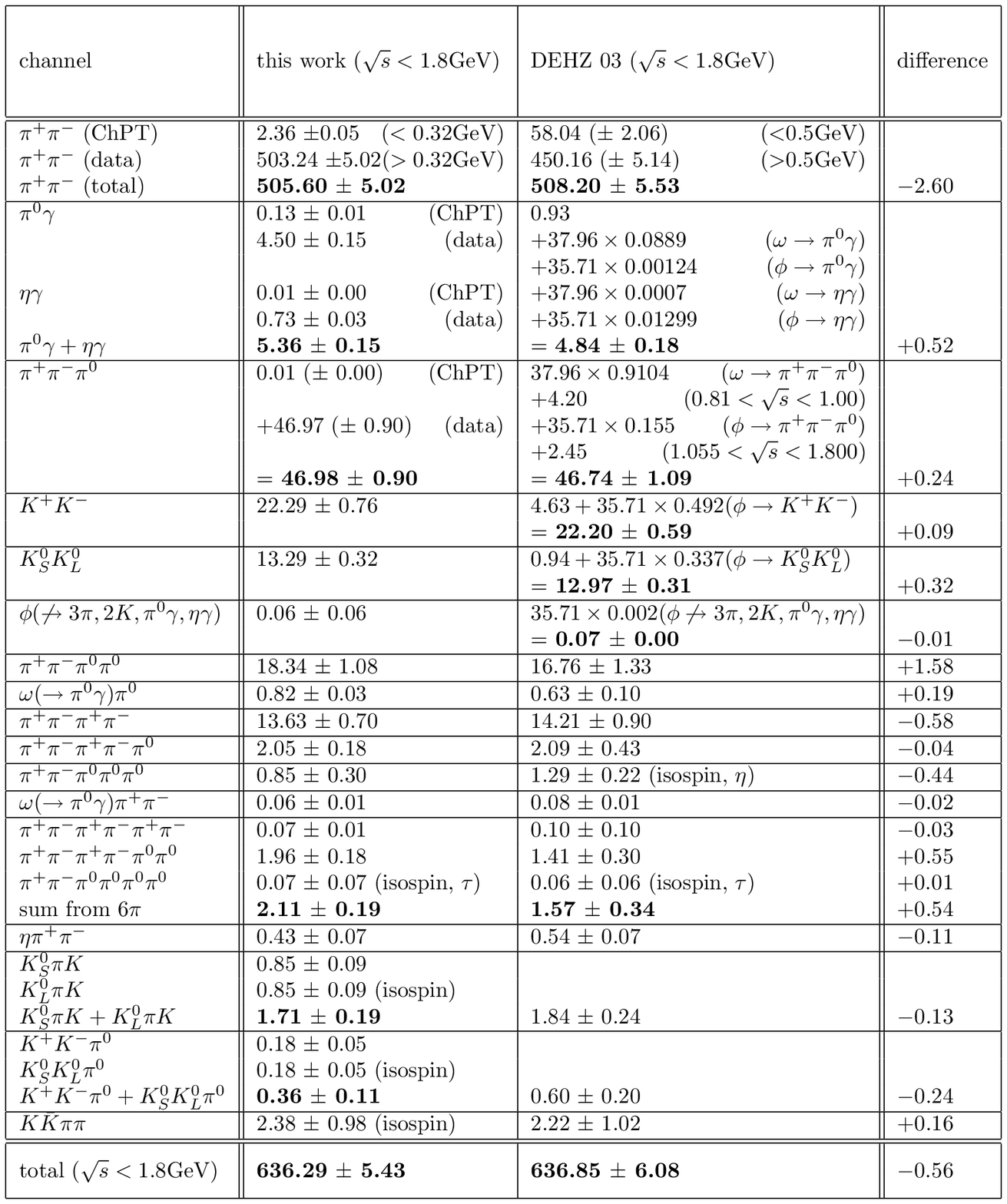, width=16cm} 
\caption{
  The contributions of the individual $e^+e^-$
  channels, up to $\sqrt s = 1.8$~GeV, to dispersion relation 
  (\ref{eq:dispersion_rel}) for $a_\mu^{\rm had,LO}\
  (\times 10^{10})$ that were obtained in this analysis and in the 
  DEHZ03 study~\cite{DEHZ03}. The last column shows the difference.
  ``Isospin'' denotes channels for which no data exist, and
  for which isospin relations or bounds are used. We have divided 
  the DEHZ $\omega$ contribution into the respective channels according 
  to their branching fractions~\cite{PDG2002}, with their sum 
  normalized to unity.}
\label{tab:t7} 
\end{center}
\end{table}

\begin{table}[htb]
\begin{center}
\begin{tabular}{|c|c|c|}
\hline
energy range (GeV) & HMNT 03 & BP 01 \\
\hline
1.05--2.0  & $ 16.34 \pm 0.82$\hfill (excl+incl) & 
             $ 15.6 \pm 2.3 $\hfill (excl)  \\
     &      ($  5.56 \pm 0.13$\hfill ~~(1.05--1.43 GeV, excl)) & \\ 
     &      ($ 10.78 \pm 0.81$\hfill ~~(1.43--2.0 GeV, incl)) & \\ \hline
 2.0--5.0  & $ 38.13 \pm 1.10$\hfill (incl) & 
             $ 38.1 \pm 2.2$ \hfill (incl)\\ \hline
 5.0--7.0  & $ 18.52 \pm 0.64$\hfill (incl) & 
           $ 18.3 \pm 1.1$\hfill (incl) \\ \hline
 7.0--12 & $ 30.16 \pm 0.61$\hfill (incl+pQCD) 
           & $ 30.4 \pm 0.4$\hfill (incl)  \\
     &      ( $ 25.32 \pm 0.61$\hfill ~~(7.0--11 GeV, incl)) & \\ 
     &      ( $  4.84 \pm 0.02$\hfill ~~(11--12 GeV, pQCD)) & \\ \hline
 12--$\infty$ & $120.48 \pm 0.13$ \hfill (pQCD) & 
                $ 120.3 \pm 0.2 $ \hfill (pQCD) \\
\hline
\end{tabular}
\caption{Comparison 
  of the contributions to $\Delta\alpha_{\rm had}(M_Z^2) \times 10^4$ 
  with the analysis of BP 01~\cite{Burkhardt:2001xp}.}
\label{tab:compBP01} 
\end{center}
\end{table}

\subsection{Implications for the global fit to electroweak data}

\begin{figure}[htbp]
\begin{center}
{\psfig{file=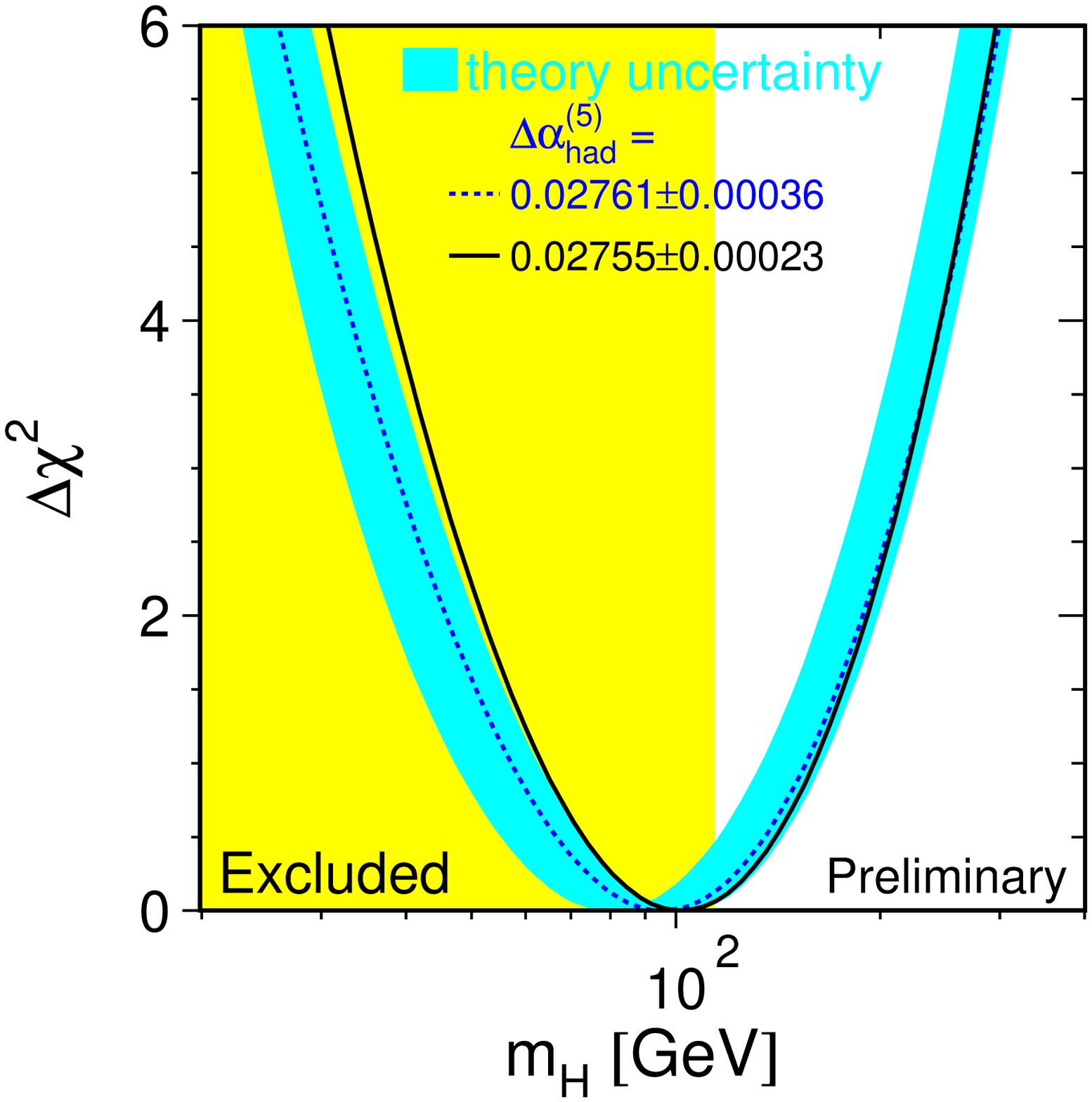,width=8cm}}
\end{center}
\vspace{-5.5ex}
\caption{The $\chi^2$ profile versus the mass of a
  Standard Model Higgs boson obtained in a global analysis of
  electroweak data. 
  The solid curve is obtained using the value we found in this work, 
  and the dotted curve is obtained using the value 
  in \cite{Burkhardt:2001xp}.
  We thank Martin Gr\"unewald for making this plot.} 
\label{fig:C}
\end{figure}
\noindent
The value of the QED coupling on the Z pole is an important
ingredient in the global fit of all the precise electroweak data.  
The continuous curve in Fig.~\ref{fig:C} shows the $\chi^2$ profile 
as a function of $\ln m_H$ obtained in the global analysis if our value
of $\Delta\alpha_{\rm had}$ is used.  (whereas the dashed shows the profile
that would result from the BP01~\cite{Burkhardt:2001xp} 
determination of the QED coupling).   The measured value of $m_t$ 
has been included in the analysis.  When our new determination is 
taken, the fit predicts that a Standard Model Higgs has a mass
\be 
  m_H =  102 \,^{\mbox{\normalsize +58}}_{-\mbox{\normalsize 38}}\ \GeV 
\ee
or $m_H<221\ \GeV$ at the 95\% confidence level.

\section{Conclusions} \label{sec:conclusions}

The anomalous magnetic moment of the muon, $(g-2)/2$, and the QED 
coupling at the $Z$ boson mass, $\alpha(M_Z^2)$, are two important 
quantities in particle physics. At present, the accuracy of the 
theoretical predictions is limited by the uncertainty of the hadronic 
vacuum polarization contributions. Here we use all the available data
on $e^+e^-\to{\rm hadrons}$ to achieve the best presently possible 
data-driven determination of these contributions.  In this way, we
obtain a Standard Model prediction of the muon anomalous magnetic 
moment of
\be a_\mu^{\rm SM}  = 0.00116591763 (74), \label{eq:amm1} \ee
to be compared with the present experimental value of
\be a_\mu^{\rm exp} = 0.0011659203 (8), \label{eq:amm2} \ee
which shows a $2.4 \sigma$ difference. As this comparison of the 
measurement and prediction becomes more and more
precise, we will obtain an increasingly powerful constraint on physics 
beyond the Standard Model.

We have also used our optimal compilation of the available
$e^+e^-\to{\rm hadrons}$ data to predict
\be \alpha(M_Z^2)^{-1} = 128.954 \pm 0.031. \ee
The accuracy is now $24\times 10^{-5}$. 
This again is an important quantity. It is the most poorly-determined 
of the three parameters which specify the electroweak model.
Although significantly improved from the error of 
Burkhardt and Pietrzyk's preliminary result~\cite{BP03}, 
it is still the least accurately determined of the three 
fundamental parameters of the electroweak theory;
$\Delta G_\mu/G_\mu= 1\times 10^{-5}$  and 
$\Delta M_Z/M_Z= 2\times 10^{-5}$.

\subsection{Future prospects for reducing the error on $g-2$}

We have stressed that the comparison of the measurement and the Standard Model
prediction of the muon anomalous magnetic moment,
$a_\mu\equiv(g_\mu-2)/2$, is very important.
It provides a valuable constraint on, or an indicator of, new physics
beyond the Standard Model.  From the above discussion, we see that the 
present uncertainties on the measurement and the prediction are 8 and 
7 $\times 10^{-10}$ respectively.  How realistic is it to improve the 
accuracy in the future?   On the experimental side, the accuracy is 
dominated at present by the BNL measurement.  We can expect a further 
improvement in the BNL measurement of $(g-2)$, since the collaboration 
are at present analysing 3.7~billion $\mu^-$ events which should 
give a total relative error of about 0.8~ppm. As a
consequence, the $\pm8 \times 10^{-10}$ uncertainty in (\ref{eq:amm2}) 
should be improved\footnote{See the note added in proof.} to about $\pm6
\times 10^{-10}$. 
If the error on the theory prediction can be improved beyond this
value then the case for another dedicated experiment with even more 
precision is considerably enhanced.    

The error attributed to the
theoretical prediction  of $a_\mu$ is dominated by the uncertainties in the
computation of the hadronic contribution, $a_\mu^{\rm had}$; in particular
on the calculation of $a_\mu^{\rm had,LO}$ and 
$a_\mu^{\rm had,l \mbox{-} b \mbox{-} l}$,
which at present have uncertainties of about 6 and 4 $\times 10^{-10}$ 
respectively.  The latter error, on the light-by-light contribution, 
is generally believed to be able to be improved to 2 $\times 10^{-10}$ 
(25\% error); and, optimistically, it is perhaps not
hopeless to envisage an eventual accuracy of about 1 $\times 10^{-10}$ 
(10\% error), but this would require a breakthrough in the understanding 
of this contribution.  We are left to consider how much the error
on $a_\mu^{\rm had,LO}$ could be improved.  Already we are claiming 
a 1\% accuracy.  To reduce the error from the present 6 $\times
10^{-10}$ to 1 $\times 10^{-10}$ is not realistic.  However we should 
note (see, for example, Ref.~\cite{ZHAO}) there will be 
progress from all experiments that are measuring $R$. Indeed, with the 
improvements, already in progress or planned, of the BES, CMD-3 + SND 
at VEPP-2000, BaBar, Belle, CLEO-C and KLOE experiments, we may anticipate 
an eventual accuracy of 0.5\% in the crucial $\rho$ domain and 
1-2\% in the region above 1 GeV.  It will be challenging, but not 
impossible.  This statement also applies to improving the accuracy
of the radiative corrections.

In this connection, note that the measurements of the radiative return 
experiments are just becoming available.  From these experiments 
we may anticipate low energy data for a variety of $e^+e^-$ channels, 
produced via initial state radiation, at the $\phi$-factory 
DA$\Phi$NE \cite{Kloe,DENIG} and at the $B$-factories, BaBar and Belle,
see, for example,~\cite{SOL}. For instance, by detecting the 
$\pi^+\pi^-\gamma$ channel, it may be possible to measure the vital 
$e^+e^-\to\pi^+\pi^-$ cross section in the threshold region.
For the radiative return experiments there is no 
problem with statistics, and the accuracy is at present due to 
systematics, which come mainly from theory.  These new experiments are 
motivating much theoretical work to improve their accuracy.  Already, 
today, it is claimed to be 2\% in the $\rho$ region.

In summary, we may hope for an improvement in accuracy down to about 
3 $\times 10^{-10}$ in the theoretical prediction of $a_\mu$ in the 
foreseeable future, which in turn emphasizes the need for an 
experimental measurement with improved precision.\\

\noindent
{\bf Note added in proof}\\

\begin{figure}[htb]
\begin{center}
{\psfig{file=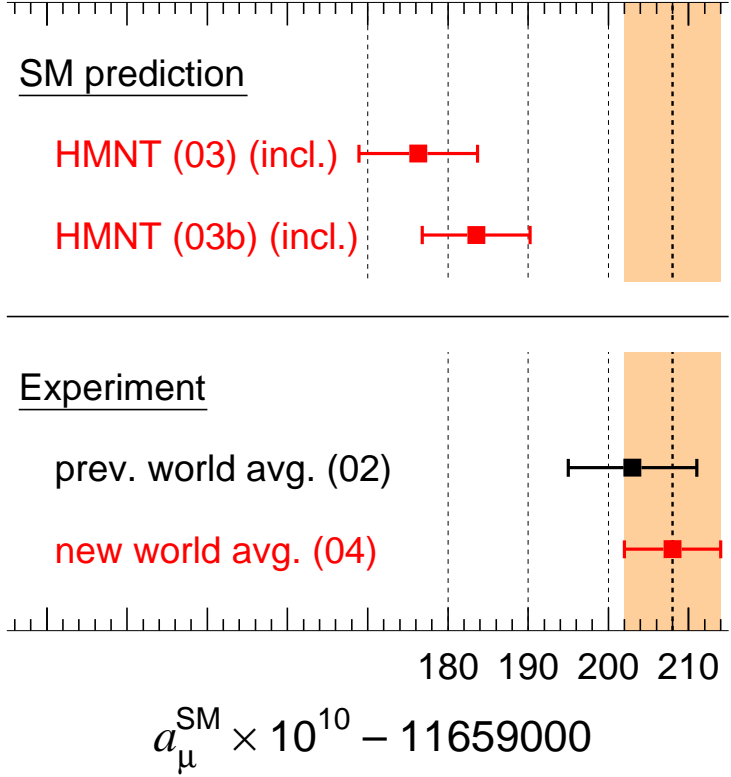,width=12cm}}
\end{center}
\vspace{-5.5ex}
\caption{The new (world average) experimental value of the muon 
  $(g_{\mu}-2)/2 \equiv a_\mu$ given in Ref.~\cite{Bennett:2004pv}, compared
  with the SM prediction as given in the text, HMNT (03), and with the 
  value, HMNT (03b), which is obtained using the hadronic light-by-light 
  contribution as recently calculated in Ref.~\cite{Melnikov:2003xd} 
  and the updated QED contribution given in Ref.~\cite{Kinoshita:2004wi}.}
\label{fig:Zsmnew}
\end{figure}

The BNL muon $g-2$ collaboration have just published
\cite{Bennett:2004pv} the results of 
their analysis of the $\mu^-$ data which updates their experimental 
determination of $a_{\mu}$.
As a result they now obtain a new world average
\begin{equation}
a_{\mu}^{\rm exp} = 0.0011659208(6)\,.
\end{equation}
Comparing this value with our SM prediction of Eq.~(\ref{eq:amm1}) we find
a $3.3 \sigma$ discrepancy, as shown by the HMNT (03) (incl.) error bar
in Fig.~\ref{fig:Zsmnew}.  That is the discrepancy is $\delta a_{\mu} =
(31.7 \pm 9.5) \times 10^{-10}$.

Also, very recently, the hadronic light-by-light contribution has been 
recalculated, paying particular attention to the matching between the 
short- and long-distance behaviour~\cite{Melnikov:2003xd}.  The 
contribution is found to be $a_\mu^{\rm had,l \mbox{-} b \mbox{-} l} =
(13.6 \pm 2.5) \times 10^{-10}$.  In addition Kinoshita and Nio have 
updated the calculation of the $\alpha^4$ QED contribution and 
find~\cite{Kinoshita:2004wi}
\begin{equation}
  a_\mu^{\rm QED} = 116 584 719.35 (1.43) \times 10^{-11},
\end{equation}
which should be compared with the value (\ref{eq:QED}) we have used.  
If we use these new $a_\mu^{\rm had,l-b-l}$ and $a_\mu^{\rm QED}$ 
values, then our prediction is given by the HMNT (03b) (incl.) error 
bar in Fig.~\ref{fig:Zsmnew}, and corresponds to 
\begin{equation}
     a_\mu^{\rm SM} = (11 659 183.53 \pm 6.73) \times 10^{-10}
\end{equation}
in the place of Eq.~(\ref{eq:amm1}).  If this prediction is compared 
with the new BNL result above, then there is a discrepancy of 
$2.7\sigma$, that is $\delta a_{\mu} = ( 24.5 \pm 9.0 ) \times 10^{-10}$.

\newpage
\section*{Appendix A: Threshold behaviour of
$\pi^0\gamma$ and $\eta\gamma$ production}
\addcontentsline{toc}{section}{Appendix A: Threshold behaviour of
$\pi^0\gamma$ and $\eta\gamma$ production}

We take the Wess-Zumino-Witten (WZW) local interaction as
\begin{eqnarray}
 {\cal L}_{WZW} = 
 - \frac{\alpha}{8\pi f_\pi} c_P P F_{\mu\nu} \widetilde{F}^{\mu\nu},
 \label{eq:appendix:WZW}
\end{eqnarray}
where $f_\pi \simeq 93$ MeV, and $P$ denotes the electrically neutral 
members, $\pi^0$ or $\eta_8$, of the $SU(3)$ 
pseudoscalar octet.  The $c_P$ coefficients are $c_{\pi^0}=1$ and 
$c_{\eta_8}=1/\sqrt3$.
We may extend the multiplet to include the $SU(3)$ singlet, $\eta_1$, 
for which the coefficient is $c_{\eta_1}=2\sqrt2/\sqrt3$.
As usual, $F_{\mu\nu}$ is the QED field strength tensor, and 
$\widetilde{F}_{\mu\nu}$ is its dual, 
\begin{eqnarray}
 \widetilde{F}_{\mu\nu} 
 \equiv \epsilon_{\mu\nu\rho\sigma}{F}^{\rho\sigma},
\end{eqnarray}
where $\epsilon_{\mu\nu\rho\sigma}$ is a totally
antisymmetric tensor with $\epsilon_{0123}=1$.

\subsection*{A.1: $\pi^0 \to 2\gamma$ decay and
$e^+e^-\to \pi^0\gamma$}

The WZW interaction, (\ref{eq:appendix:WZW}), is responsible for the 
$\pi^0\to 2\gamma$ decay.   The lowest-order amplitude ${\cal M}$ is
\begin{eqnarray}
 {\cal M} = 
   \frac{\alpha}{\pi f_\pi}~ 
   \epsilon^{\mu\nu\lambda\sigma} ~ p_{1\mu} ~ p_{2\lambda} ~
   \epsilon^{\ast}_\nu (p_1) ~ \epsilon^{\ast}_\sigma (p_2),
\end{eqnarray}
which results in the partial decay width
\begin{eqnarray}
 \Gamma(\pi^0\to 2\gamma) = 
 \frac{\alpha^2 m_{\pi^0}^3}{64 \pi^3 f_\pi^2},
\end{eqnarray} 
when summed over
the polarization of the final state photons.
If we take $f_\pi = (130 \pm 5) /\sqrt2$ MeV and 
$m_{\pi^0} = 134.9766 \pm 0.0006$ MeV~\cite{PDG2002},
then this gives
\begin{eqnarray}
  \Gamma(\pi^0\to 2\gamma) = 7.81 \pm 0.60 ~{\rm eV},
\end{eqnarray}
which is in good agreement with the experimental 
value~\cite{PDG2002},
\begin{eqnarray}
 \left. \Gamma(\pi^0\to 2\gamma) \right|_{\rm exp.} 
= 7.7 \pm 0.6 ~{\rm eV} .
\end{eqnarray}

The cross section of $e^+e^- \to \pi^0\gamma$ can be written in 
terms of the $\pi^0\to 2\gamma$ width as
\begin{eqnarray}
 \sigma(e^+e^-\to \pi^0\gamma) 
 = \sigma_{\rm pt}(e^+e^-\to \pi^0\gamma)
 \equiv
  \frac{8\alpha\pi\Gamma(\pi^0\to2\gamma)}{3m_\pi^3} 
  \left( 1 - \frac{m_\pi^2}{s} \right)^3.
\label{eq:pi0gammaLO1_app}
\end{eqnarray}
We can further improve the behaviour of the cross section 
by assuming vector meson dominance:
\begin{eqnarray}
 \sigma_{\rm VMD}(e^+e^-\to \pi^0\gamma) 
= \sigma_{\rm pt} (e^+e^-\to \pi^0\gamma)
  \left(\frac{m_\omega^2}{m_\omega^2 - s}\right)^2.
\label{eq:pi0gammaLO2_app}
\end{eqnarray}
We use the equation above in calculating the $\pi^0\gamma$
contribution from the threshold region in 
Section \ref{subsec:eval_pi0gamma}.

\subsection*{A.2: $\eta\to 2\gamma$ decay and
$e^+e^-\to \eta\gamma$}

If we neglect $\eta_8$-$\eta_1$ mixing and identify
$\eta_8$ as $\eta$, then the $\eta\to 2\gamma$ decay is 
dictated by the WZW interaction,
\begin{eqnarray}
 {\cal L}_{WZW} = 
 - ~ \frac{\alpha}{8 \sqrt3 \pi f_\pi} ~
  \eta_8 ~ F_{\mu\nu} \widetilde{F}^{\mu\nu},
 \label{eq:appendix:WZW:eta}
\end{eqnarray}
which contains an extra factor of $1/\sqrt3$ as compared with the
$\pi^0 \gamma\gamma$ coupling term.
The calculation of the decay rate is exactly analogous to
that of $\pi^0$ decay.  The result is
\begin{eqnarray}
 \Gamma(\eta \to 2\gamma) 
= 
\frac{\alpha^2 m_\eta^3}{192 \pi^3 f_\pi^2} 
~~~~~ ({\mbox{LO ChPT without $\eta_1$-$\eta_8$ mixing}}).
\end{eqnarray}
Taking $f_\pi = (130\pm 5)/\sqrt2$ MeV 
and $m_\eta = 547.30 \pm 0.12$ MeV~\cite{PDG2002}, we obtain
\begin{eqnarray}
 \Gamma(\eta \to 2\gamma) 
= 
 0.174 \pm 0.013~{\rm keV}  
~~~~~ ({\mbox{LO ChPT without $\eta_1$-$\eta_8$ mixing}}),
\end{eqnarray}
which differs from the observed value~\cite{PDG2002} by
about a factor of 3,
\begin{eqnarray}
  \left. \Gamma(\eta \to 2\gamma) \right|_{\rm exp.} 
= 
  0.46 \pm 0.04 ~{\rm keV} . 
\label{eq:eta2gamma:exp}
\end{eqnarray}

The agreement becomes better when we allow for the mixing between 
the $\eta$ and $\eta^\prime$ states.  
Following Ref.\ \cite{QuarkModelRev}, we define
the mixing angle $\theta_P$ by 
\begin{eqnarray}
\left(  \begin{array}{c} \eta \\ \eta^\prime \end{array} \right)
=
\left(       \begin{array}{rr} 
 \cos\theta_P &  - \sin\theta_P \\
 \sin\theta_P &    \cos\theta_P 
\end{array}  \right)
\left( \begin{array}{c}  \eta_8 \\ \eta_1 \end{array}  \right).
\end{eqnarray}
The Lagrangian now becomes
\begin{eqnarray}
 {\cal L} =
 - \frac{\alpha}{8\pi f_\pi} 
   \left( c_{\eta_8} \cos\theta_P - c_{\eta_1} \sin\theta_P \right)
      \eta F_{\mu\nu} \widetilde{F}^{\mu\nu}
 -\frac{\alpha}{8\pi f_\pi} 
   \left( c_{\eta_8} \sin\theta_P + c_{\eta_1} \cos\theta_P \right)
      \eta^\prime F_{\mu\nu} \widetilde{F}^{\mu\nu}.
\end{eqnarray}
If we take $\theta_P \approx -20^\circ$ \cite{QuarkModelRev}, 
then the coefficient of the $\eta F \widetilde{F}$ term is
\begin{eqnarray}
 c_{\eta_8} \cos\theta_P - c_{\eta_1} \sin\theta_P 
 ~=~ 1.91 \times c_{\eta_8}  ~=~ 1.10,
\end{eqnarray}
and the predicted decay width is 
\begin{eqnarray}
  \Gamma(\eta \to 2\gamma) 
&=&
 \frac{\alpha^2 m_\eta^3}{64 \pi^3 f_\pi^2} 
 (c_{\eta_8} \cos\theta_P - c_{\eta_1} \sin\theta_P)^2    \nonumber\\
&\simeq& 
 0.63 ~{\rm keV}  
~~~~~ ({\mbox{LO ChPT with $\eta_1$-$\eta_8$ mixing}}) .
\end{eqnarray}

We find the residual discrepancy with the observed rate is removed 
when we introduce the higher-order effect,
$f_1 \ne f_8 \ne f_\pi$.
In this case,
\begin{eqnarray}
 {\cal L} = 
  - \frac{\alpha}{8\pi} 
         \left(
            \frac{c_{\eta_8}}{f_8} \cos\theta_P 
          - \frac{c_{\eta_1}}{f_1} \sin\theta_P 
         \right)
         \eta F_{\mu\nu} \widetilde{F}^{\mu\nu}
 - \frac{\alpha}{8\pi} 
         \left(
            \frac{c_{\eta_8}}{f_8} \sin\theta_P 
          + \frac{c_{\eta_1}}{f_1} \cos\theta_P 
         \right)
         \eta^\prime F_{\mu\nu} \widetilde{F}^{\mu\nu}.
\end{eqnarray}
If we take $f_8\approx 1.3 f_\pi$,  $f_1\approx 1.1 f_\pi$,
as given by Eqs.~(162) and (163) of Ref.~\cite{Bijnens:xi},     
and $\theta_P \approx -20^\circ$, then the Lagrangian becomes 
\begin{eqnarray}
 {\cal L} \simeq 
- 1.60 \times \frac{\alpha}{8\pi f_\pi} c_{\eta_8}  
  \eta F_{\mu\nu} \widetilde{F}^{\mu\nu}, 
 \label{eq:imprvd_eta2gammaLag}
\end{eqnarray}
and the predicted decay rate is 
\begin{eqnarray}
  \Gamma(\eta \to 2\gamma) 
&=& 
 \frac{\alpha^2 m_\eta^3}{64 \pi^3 f_\pi^2} 
 \left(   \frac{f_\pi}{f_8} c_{\eta_8} \cos\theta_P 
        - \frac{f_\pi}{f_1} c_{\eta_1} \sin\theta_P \right)^2  \nonumber\\
&\simeq& 
 0.45 ~{\rm keV}  
~~~~~ ({\mbox{NLO ChPT with $\eta_1$-$\eta_8$ mixing}}),
\end{eqnarray}
which is now in excellent agreement with the observed value,
(\ref{eq:eta2gamma:exp}).

Similarly to the $e^+e^-\to \pi^0\gamma$ case, we can 
use the VMD approach to predict the cross section of
$e^+e^-\to \eta\gamma$.
\begin{eqnarray}
 \sigma_{\rm VMD}(e^+e^-\to \eta\gamma) 
= \sigma_{\rm pt} (e^+e^-\to \eta\gamma)
  \left(\frac{m_\omega^2}{m_\omega^2 - s}\right)^2,
\label{eq:e+e-_etagamma_VMD_app}
\end{eqnarray}
where 
\begin{eqnarray}
 \sigma_{\rm pt}(e^+e^-\to \eta\gamma) \equiv
  \frac{\alpha^3}{24\pi^2} 
         \left(
            \frac{c_{\eta_8}}{f_8} \cos\theta_P 
          - \frac{c_{\eta_1}}{f_1} \sin\theta_P 
         \right)^2
  \left( 1 - \frac{m_\eta^2}{s} \right)^3.
\label{eq:etagamma_pt_app}
\end{eqnarray}
We take the parametrization (\ref{eq:e+e-_etagamma_VMD_app})
in calculating the $e^+e^-\to \eta\gamma$ cross section near
the threshold region in Section \ref{subsec:eval_etagamma}.

\section*{Appendix B: Constraints on $V\to\sigma\gamma$ decay
branching fractions}
\addcontentsline{toc}{section}{Appendix B: Constraints on
  $V\to\sigma\gamma$ decay branching fractions} 
\begin{figure}[htbp]
\begin{center}
{\psfig{file=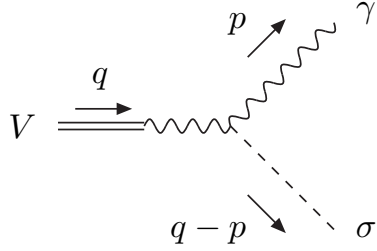,width=5cm}}
\end{center}
\vspace{-3.0ex}
\caption{The $V\to \gamma\sigma$ decay in the VMD approach.} 
\label{fig:V_gamma_sigma}
\end{figure}
\noindent
Here we calculate the $V\to \gamma\sigma$ decay of a vector meson 
using the Vector Meson Dominance (VMD) model.
To calculate the amplitude, we have used the VMD
Lagrangian~\cite{VMD}
\begin{eqnarray}
 {\cal L}_{\rm VMD} =
     - \frac14 F_{\mu\nu} F^{\mu\nu}
     - \frac14 V_{\mu\nu} V^{\mu\nu}
     + \frac12 m_V^2 V_{\mu} V^{\mu}
     - g_{V\pi\pi} V^\mu J_\mu
     - e J_\mu A^\mu
     - \frac{e}{2g_V} F_{\mu\nu} V^{\mu\nu},
\end{eqnarray}
where $J_\mu$ is the electromagnetic current.
$V_{\mu\nu}$ is defined by
\begin{eqnarray}
 V_{\mu\nu} \equiv \partial_\mu V_\nu - \partial_\nu V_\mu.
\end{eqnarray}
Here $V^\mu$ describes the neutral vector meson 
($V=\rho,\omega,\phi,\cdots$). We take $e$ to be
positive.  
The diagram which contributes to the decay is shown in 
Fig.~\ref{fig:V_gamma_sigma}. The amplitude ${\cal M}$ is given by
\begin{eqnarray}
 i {\cal M} &=&
 4 i g_{\sigma\gamma\gamma}
  ( (p\cdot q) g_{\alpha\beta} - p_\alpha q_\beta )
  \left( -i \frac{e\qsq}{g_V} \right) \frac{-i}{\qsq}
  \epsilon_\gamma^{\beta\ast}(p)   \epsilon_V^\alpha(q)
\nonumber\\
&=&
 - 4 ig_{\sigma\gamma\gamma} \frac{e}{g_V}
 ( (p\cdot q) g_{\alpha\beta} - p_\alpha q_\beta )
  \epsilon_\gamma^{\beta\ast}(p) \epsilon_V^\alpha(q), \label{eq:VMD_iM}
\end{eqnarray}
where $\epsilon_V$ and $\epsilon_\gamma$ are the polarization 
vectors of $V$ and the photon, respectively.  We have assumed that 
the interaction between the $\sigma$ meson and photon is given by
\begin{eqnarray}
 {\cal L} = g_{\sigma\gamma\gamma} \sigma F_{\mu\nu} F^{\mu\nu},
\end{eqnarray}
where $g_{\sigma\gamma\gamma}$ is a coupling constant. From the 
amplitude of (\ref{eq:VMD_iM}) 
we can readily calculate the required partial decay width
\begin{eqnarray}
 \Gamma(V\to \gamma\sigma)
 = \frac{m_V^3}{6\pi}
   \left( \frac{eg_{\sigma\gamma\gamma}}{g_V}\right)^2
   \left( 1 - \frac{m_\sigma^2}{m_V^2} \right)^3.
\label{eq:G_V_gammasigma}
\end{eqnarray}

If we use the parameters~\cite{PDG2002}
\begin{eqnarray}
& m_\phi = 1019 {\rm ~MeV},     ~~~~~
 \Gamma_\phi = 4.26 {\rm ~MeV},& \nonumber \\
&
 g_\phi^2/\pi = 14.4, ~~~~~
 B(\phi \to \gamma \sigma) < 0.002,&
\end{eqnarray}
and assume $m_\sigma = 600 {\rm ~MeV}$,
then the coupling constant $g_{\sigma\gamma\gamma}$ is constrained to be
\begin{eqnarray}
 g_{\sigma\gamma\gamma} < 5.2 \times 10^{-4} ({\rm MeV}^{-1}).
\end{eqnarray}
This bound gives constraints on 
$B(\omega\to \sigma\gamma)$ and $B(\phi(1.68) \to \sigma\gamma)$. From
(\ref{eq:G_V_gammasigma}), the branching ratio $B(V \to \sigma\gamma)$ is
\begin{eqnarray}
 B(V \to\sigma\gamma) =
\frac{2\alpha m_V^3}{3\Gamma_V}
 \left( 1 - \frac{m_\sigma^2}{m_V^2} \right)^3
 \frac{g_{\sigma\gamma\gamma}^2}{g_V^2}.
\end{eqnarray}
For the $\omega$ decay, we use the parameters
\begin{eqnarray}
 m_\omega = 783 {\rm ~MeV},     ~~~~~
 \Gamma_\omega = 8.44 {\rm ~MeV},~~~~~
 g_\omega^2/\pi = 23.2,  ~~~~~
 m_\sigma = 600 {\rm ~MeV},
\end{eqnarray}
to obtain the constraint
\begin{eqnarray}
 B(\omega\to \sigma\gamma) < 7.2 \times 10^{-5}.
\end{eqnarray}
Similarly, for the $\phi(1.68)\to \sigma\gamma$ decay, we have
\begin{eqnarray}
 B(\phi(1.68) \to \sigma\gamma) < 3.5 \times 10^{-5},
\end{eqnarray}
using the parameters
\begin{eqnarray}
 m_{\phi(1.68)} = 1680 {\rm ~MeV},     ~~~~~
 \Gamma_{\phi(1.68)} = 150 {\rm ~MeV},~~~~~
 g_{\phi(1.68)}^2/\pi = 249, ~~~~~
 m_\sigma = 600 {\rm ~MeV}.
\end{eqnarray}
These constraints are used in Section~\ref{sec:poss_cont}.

\newpage
\section*{Acknowledgements}
We thank Simon Eidelman for numerous helpful discussions concerning 
the data. We also thank M.~Fukugita, M.~Gr\"unewald, M.~Hayakawa, 
F.~Jegerlehner, T.~Kinoshita, V.~A.~Khoze, M.~Nio and M.~Whalley
for stimulating discussions and the UK Particle Physics and Astronomy 
Research Council for financial support.

The work of KH is supported in part by Grant-in-Aid for Scientific 
Research from MEXT, Ministry of Education, Culture, Science
and Technology of Japan.  ADM thanks the Leverhulme trust for an
Emeritus Fellowship.


\vfill

\end{document}